\documentclass[12pt]{article}
\usepackage{a4wide}
\usepackage{amssymb,amsmath}
\usepackage{graphicx}
\begin{document}
{\renewcommand{\thefootnote}{\fnsymbol{footnote}}
\begin{center}
{\LARGE  Quasiclassical model of inhomogeneous cosmology}\\
\vspace{1.5em}
Martin Bojowald\footnote{e-mail address: {\tt bojowald@psu.edu}}
and Freddy Hancock\footnote{e-mail address: {\tt fuh109@psu.edu}}\\
\vspace{0.5em}
Department of Physics,
The Pennsylvania State
University,\\
104 Davey Lab, University Park, PA 16802, USA
\end{center}
}

\vspace{1.5em}

\setcounter{footnote}{0}

\begin{abstract}
  Fluctuation terms and higher moments of a quantum state imply corrections to
  the classical equations of motion that may have implications in
  early-universe cosmology, for instance in the state-dependent form of
  effective potentials. In addition, space-time properties are relevant in
  cosmology, in particular when combined with quantum corrections 
  required to maintain general covariance in a consistent way. Here, an
  extension of previous investigations of static quasiclassical space-time
  models to dynamical ones is presented, describing the evolution of
  1-dimensional space as in the classical Lemaitre--Tolman--Bondi models. The
  corresponding spatial metric has two independent components, both of which
  are in general subject to quantum fluctuations. The main result is that
  individual moments from both components are indeed required for general
  covariance to be maintained at a semiclassical level, while quantum
  correlations between the components are less relevant.

\end{abstract}

\section{Introduction}

If gravity is quantized, the geometry of space-time is subject to quantum
fluctuations that modify the dynamics of phenomena such as black holes or
structure formation in the expanding universe. Traditionally, fluctuation
effects in gravity have been treated in an indirect way by higher-curvature
corrections to the Einstein--Hilbert action, implied for instance by loop
diagrams of perturbative quantum gravity. Such treatments are indirect because
they implicitly replace dynamical quantum fluctuations (or higher moments of a
state) with geometrical interaction terms. At a technical level, a derivative
expansion in space and time is included, at least implicitly, in standard
methods of effective field theory, expressing non-local quantum effects in a
local manner. However, there may be regimes in which non-local effects may be
significant, in particular when fluctuations are dominant. The recent results
of \cite{QuantumHiggsInflation,EffPotInflation}, using similar methods as
employed below but in a spatially homogeneous setting, have confirmed this
expectation by showing how Higgs inflation can be made consistent by including
non-adiabatic quasiclassical effects in an effective potential. The topic of
the present paper is a discussion of basic conditions required for an
extension of effective potentials to inhomogeneous configurations.

The methods used, given by canonical effective theory, provide the possibility
to generalize standard effective field theories by avoiding derivative
expansions. Instead, quantum fluctuations are initially kept as dynamical
fields, coupled to the classical fields which can be interpreted as field
expectation values in a quasiclassical state. Again at a technical level,
non-locality is described not by restricting it by a derivative expansion, but
rather by reformulating it as a local theory with auxiliary fields. These
auxiliary fields are nothing but quantum fluctuations, and therefore have an
important physical interpretation as well.

Systematic formulations of moments of a state in quantum field theory face
several challenges and are still being developed. Nevertheless, recent
progress allows us to capture some of the fluctuation degrees of freedom in
tractable systems and to explore their physical implications. Independent
fields for quantum corrections have been used in similar models analyzed in
\cite{CQC,CQCFields,CQCFieldsHom,CQCHawking}. The relationship between these
fields and moments of an evolving state follows from the canonical effective
theory of \cite{EffAc,Karpacz}, written in terms of Casimir--Darboux variables
that bring the phase-space structure of moments to canonical form
\cite{Bosonize,EffPotRealize}. In general, deriving Casimir--Darboux variables
that coordinatize phase space by pairs of position and momentum coordinates as
well as conserved quantities (the Casimir variables) is a challenging task in
higher-dimensional cases as presented by moment systems in particular in a
case of field theory. For tractability, it is then important to know whether
certain moments, such as quantum correlations, can be ignored under certain
conditions, allowing one to use lower-dimensional phase spaces. In a field
theory, a natural candidate for moments to ignore is correlations between
field values at different points. (Such correlations may, however, be relevant
at early stages of structure formation in the universe \cite{Backreaction}.)
At each point, single-particle quasiclassical descriptions can then be used.

In \cite{SphSymmMoments}, the case of static spherically symmetric space-times
has been considered in this way with applications to quantum black holes. Here, we drop
the staticity assumption (while maintaining spherical symmetry) and thereby
open up investigations of fluctuations effects in inhomogeneous quantum
cosmology. Our main focus is on the covariance problem, which turns out to be
more challenging in non-static space-times compared with static ones. Our
combination of analytical and numerical studies will show that it is important
to include moments of both fields present in the model (two independent
components of a spherically symmetric spatial metric) while correlations
between the two fields can be ignored without significant violations of
general covariance. Tractable models of quasiclassical cosmological dynamics
can therefore be constructed, up to technical problems in the current
numerical implementation.

\section{Quasiclassical dynamics}
\label{s:QuasiClass}

We assume that a system described by canonical variables
$(q_i,p_j)$ with Hamiltonian $H(q_i,p_j)$ has been quantized by the
assignment of an operator algebra with generators $(\hat{q}_i,\hat{p}_j)$ obeying
basic commutation relations
\begin{equation}\label{eq:q_canonical_pois}
  [\hat{q}_i,\hat{q}_j] = 0\quad,\quad
  [\hat{p}_i,\hat{p}_j] = 0\quad,\quad    
  [\hat{q}_i,\hat{p}_j] = i\hbar\delta_{ij}\,.
\end{equation}
States of the quantum system are positive linear maps from the algebra to the
complex numbers, denoted here as expectation value functionals
$\langle\cdot\rangle\colon \hat{A}\mapsto \langle\hat{A}\rangle$. The specific
definition of a state can be defined through wave functions or density
matrices, depending on the action of the algebra on a Hilbert space. States in
this formulation may therefore be pure or mixed.
  
The Hamilton operator $\hat{H}$ is constructed from the basic operators
$(\hat{q}_i,\hat{p}_j)$ and defines equations of motion
\begin{equation}
  \frac{{\rm d} \langle\hat{f}\rangle}{{\rm d}t}= \frac{\langle[\hat{f},\hat{H}]\rangle}{i\hbar}
  \label{eq:quantum_hamiltonian}
\end{equation}
for expectation values. Depending on the underlying formulation, the time
dependence may reside in the states (Schr\"odinger picture) or in the
operators (Heisenberg picture). To be specific, we assume the former
interpretation. The Hamiltonian flow (\ref{eq:quantum_hamiltonian}) then
defines a dynamical system on the infinite-dimensional space of states.

\subsection{Quasiclassical formulation for a single degree of freedom}
\label{s:Single}

In order to parameterize this flow, it is useful to introduce central moments
\begin{equation}
  \Delta(q^np^m) = \langle{(\hat{q}-\langle{\hat{q}\rangle})^n(\hat{p}-\langle{\hat{p}})^m\rangle}_{\text{symm}}\, ,
\end{equation}
where ``symm'' denotes completely symmetric (Weyl) ordering. The set of these
moments, together with the basic expectation values $\langle\hat{q}_i\rangle$
and $\langle\hat{p}_j\rangle$, can be seen as a decomposition of a full
quantum state. A fully quantum system could in principle be reproduced by
considering the moments corresponding to every positive integer value of $m$
and $n$. Our quasiclassical description, resulting in an effective field
theory approach to canonical systems, is implemented by truncating the moment
expansion at a finite order. 

The quasiclassical nature of the dynamics follows from a Poisson structure on
the space of states, defined by 
\begin{equation}
    \label{eq:quasiclassical_poisson}
    \{\langle{\hat{f}\rangle},\langle{\hat{g}\rangle}\} = \frac{1}{i\hbar}\langle{[\hat{f},\hat{g}]\rangle}\,,
  \end{equation}
  combined with the Leibniz rule for products of expectation values as they
  appear in central moments. Using
(\ref{eq:quasiclassical_poisson}) and (\ref{eq:quantum_hamiltonian}), it
immediately follows that
\begin{equation}
  \label{eq:quasiclassical_eom_gen}
  \frac{{\rm d}\langle{\hat{f}\rangle}}{{\rm d}t}=  \{\langle{\hat{f}\rangle},H_{\text{eff}}\} 
\end{equation}
with the effective Hamiltonian $H_{\text{eff}}=\langle{\hat{H}\rangle}$,
considered for a given Hamilton operator as a function on the space of states.
To find this effective Hamiltonian in terms of our coordinates and moments, we
make use of a Taylor expansion. For simplicity, we define
$\langle{\hat{q}\rangle} = q$ and $\langle{\hat{p}\rangle} = p$. We can expect
this expansion to take the form
\begin{equation}
    H_{\text{eff}} = H + H_2\, ,
\end{equation}
where $H$ is the classical Hamiltonian and $H_2$ is a quantum correction. To
perform this expansion, we make the alteration
\begin{equation}
    H_{\text{eff}} = \langle{\hat{H}(\hat{q},\hat{p})\rangle} =
    \langle{\hat{H}(q + (\hat{q}-q),
      p + (\hat{p}-p))\rangle}\, ,
\end{equation}
apply a Taylor expansion in $(\hat{q}-q)$ and $(\hat{p}-p)$
and distribute the expectation value over the terms in the expansion. The
result is
\begin{equation}
    \label{eq:quasiclassical_hamiltonian}
    H_{\text{eff}} = H(q,p) + \sum_{n+m=2}^N
    \frac{1}{n!m!}\frac{\partial^{n+m}H(q,p)}{\partial q^n
      \partial p^m}\Delta(q^n p^m)\, ,
\end{equation}
where $N$ is any integer. The order of the expansion in $\hbar$ is given by
$N/2$. In this investigation, we truncate this series at second-order or
$N=2$.

An important aspect of these moments $\Delta(q^n p^m)$ is that they themselves
are not canonical coordinates, as their brackets do not follow
(\ref{eq:q_canonical_pois}). Rather, as was shown in
\cite{Bosonize,EffPotRealize}, second-order moments follow the brackets of the
Lie algebra $\text{sp}(2M,\mathbb{R})$, where $M$ is the number of classical
degrees of freedom. We can make a transformation from moments to canonical
coordinates $(s,p_s)$ by
\begin{equation}
    \label{eq:quantum_coords_def}
    \Delta(q^2) = s^2 \quad , \quad
    \Delta(qp) = sp_s \quad , \quad
    \Delta(p^2) = p_s^2 + \frac{U}{s^2}\, ,
\end{equation}
where $U=\Delta(q^2)\Delta(p^2)-\Delta(qp)^2$, a Casimir function of the
Poisson manifold such that $\{U,s\}=0=\{U,p_s\}$, is required by Heisenberg's
uncertainty relation to obey $U \geq \hbar^2/4$. This transformation
has been found independently several times and in various fields
\cite{VariationalEffAc,EnvQuantumChaos,GaussianDyn,QHDTunneling,CQC}, but without
a connection to Poisson geometry. The latter is important for a systematic
generalization to higher orders or multiple degrees of freedom
\cite{Bosonize,EffPotRealize}. A field-theory version related to the one to be
used here has been studied in a cosmological and black-hole context
\cite{CQCFields,CQCFieldsHom,CQCHawking}. Our new results consist mainly in an
extension to constrained dynamics and related consistency tests.

In (\ref{eq:quantum_coords_def}), it is important to include the variable $U$
in order to have a one-to-one mapping from the three second-order moments
$(\Delta(q^2),\Delta(qp),\Delta(p^2))$ to three independent canonical/Casimir
variables, $(s,p_s,U)$. As a Casimir variable, $U$ has vanishing Poisson
brackets with any function on this phase space, including any possible
Hamiltonian for the dynamics of second-order moments and basic expectation
values. The value of $U$ is therefore time-independent in this
truncation. Nevertheless, it has dynamical implications because it appears in
effective Hamiltonians derived from applying (\ref{eq:quantum_coords_def}) to
the second-order truncation of (\ref{eq:quasiclassical_hamiltonian}). For a
harmonic oscillator, for instance, we obtain
\begin{eqnarray}
  H_{\rm eff} &=& \left\langle \frac{\hat{p}^2}{2m}+\frac{1}{2}m\omega^2
                  \hat{q}^2\right\rangle\nonumber\\
  &=& \frac{\langle\hat{p}\rangle^2+\Delta(p^2)}{2m}+ \frac{1}{2}m\omega^2
      \left(\langle\hat{q}\rangle^2+\Delta(q^2)\right)\nonumber\\
  &=& \frac{\langle\hat{p}\rangle^2}{2m}+ \frac{p_s^2}{2m}+
      \frac{U}{2ms^2}+\frac{1}{2}m\omega^2\langle\hat{q}\rangle^2+
      \frac{1}{2}m\omega^2s^2\,.
\end{eqnarray}
The ground-state value is obtained by setting both momenta equal to zero and
minimizing the remaining effective potential in $\langle\hat{q}\rangle$ (which
also turns out to be zero) and $s$ as well as $U$. For $s$, which appears in
two terms, we obtain the value
\begin{equation}
  s_{\rm min} = \sqrt[4]{\frac{U}{m^2\omega^2}}= \sqrt{\frac{\hbar}{2m\omega}}
\end{equation}
where $U$ must take its boundary value for a minimum. The effective
Hamiltonian then takes the value $H_{\rm eff,min} = \frac{1}{2}\hbar\omega$,
the correct ground-state energy. In an extension to field theories, $U$ will,
in a similar fashion, provide zero-point energies. Since they would formally
diverge and must be subtracted by some regularization procedure, we will, for
the most part, ignore field-theory analogs of $U$ in our first detailed
analysis of evolving cosmological inhomogeneity. Finite remnants of $U$ in a
field theory context may nevertheless be of interest, and we will show that
they would have a noticeable effect.

\subsection{Quasiclassical constraints}

In gravitational systems, the dynamics is provided by constraints $\hat{C}_i$
that ensure time reparameterization and diffeomorphism covariance. Unless one
chooses a specific gauge, there is no fixed Hamiltonian $\hat{H}$ with respect
to an absolute time $t$ as used in the equations above.  Beginning with a
single constraint, $C=0$, in canonical quantizations it is usually imposed by
turning it into an operator $\hat{C}$ and requiring pure physical states
$|\psi\rangle$ to be annihilated by it:
\begin{equation} \label{DiracC}
    \hat{C}|\psi\rangle = 0\, .
\end{equation}
This equation immediately implies
\begin{equation}\label{CExp}
    \langle \hat{C} \rangle = \langle\psi|\hat{C}|\psi\rangle = 0\,,
\end{equation}
which can be
written as a constraint on basic expectation values and moments by an
expansion analogous to (\ref{eq:quasiclassical_hamiltonian}).

However, the single quasiclassical constraint (\ref{CExp}) does
not imply that a state solving this equation obeys (\ref{DiracC}). In particular,
for any other operator $\hat{f}$ acting on the Hilbert space, it must be true that 
\begin{equation}
    \hat{f}\hat{C}|\psi\rangle = 0\, .
\end{equation}
As such, the constraint  (\ref{CExp})
is clearly insufficient to capture the behavior of the constraint
operator on the space of states, parameterized by moments. We must additionally introduce
higher-order constraints of the form
\begin{equation} \label{fCExp}
    \langle \hat{f}\hat{C} \rangle = \langle\psi|\hat{f}\hat{C}|\psi\rangle = 0
\end{equation}
which account for every possible $\hat{f}$. There are therefore infinitely
many independent quasiclassical constraints for every classical constraint
\cite{EffCons,EffConsRel,EffConsComp}. Heuristically, this infinite number
makes sure that not only basic expectation values but also their fluctuations,
correlations and higher moments are suitably constrained. An application to
cosmological power spectra can be found in \cite{EffConsPower}.

It is convenient to write the higher-order constraints with
$\hat{f}\not\propto {\mathbb I}$ as
\begin{equation}
    \label{eq:high_ord_con}
    C_{q^n p^m} = \langle{\left((\hat{q}-\langle{\hat{q}}\rangle)^n(\hat{p}-
        \langle{\hat{p}\rangle})^m\right)_{\text{symm}}\hat{C}\rangle} = 0\, .
\end{equation}
This system of constraints is equivalent to (\ref{fCExp}) if polynomials in
$\hat{q}$ and $\hat{p}$ are used for the set of all $\hat{f}$, which is
suitable for an algebra generated by $\hat{q}$ and $\hat{p}$. At any fixed
order $n+m$, the constraints (\ref{eq:high_ord_con}) are then linear
combinations of the constraints (\ref{fCExp}) for $\hat{f}$ of polynomials
degree up to $n+m$ in $\hat{q}$ and $\hat{p}$. The constraints
(\ref{eq:high_ord_con}) are amenable to semiclassical truncations: Finitely
many constraints up to $n+m=N-1$ are sufficient to constrain a quasiclassical
system of moment order $N$.

If there are multiple constraint operators $\hat{C}_i$, each can be treated in
the way described for a single constraint. As shown in \cite{EffCons}, the
system of higher-order constraints is first class if the system of constraint
operators $\hat{C}_i$ is first class. In addition to constraining moments,
higher-order constraints then also generate gauge flows on them. In general,
some of the higher-order constraints are complex valued because the operators
in (\ref{eq:high_ord_con}) cannot always be arranged in symmetric ordering
while keeping the full constraint operator $\hat{C}$ acting on the state on
the right. Solving the constraints, some of the moments therefore take complex
values. Reality conditions must then be imposed to ensure that observable
moments, defined as functions of the original moments that are invariant under
the gauge flows, are real.

As a simple but instructive example, consider the constraint $C=p=0$ on a
single pair of canonical variables, $(q,p)$. The constraint expectation value
$\langle\hat{C}\rangle=\langle\hat{p}\rangle=0$ then implies that the momentum
expectation value vanishes, $\langle\hat{p}\rangle=0$, while the position
expectation value $\langle\hat{q}\rangle$ can be changed arbitrarily by a
gauge transformation and is therefore non-physical. Gauge-invariant
observables could only be provided by independent canonical pairs. For
higher-order constraints at the next order, $n+m=1$ in
(\ref{eq:high_ord_con}), we obtain
\begin{eqnarray}
  C_q&=&\langle(\hat{q}-\langle\hat{q}\rangle)\hat{p}\rangle=\langle\hat{q}\hat{p}\rangle-
         \langle\hat{q}\rangle\langle\hat{p}\rangle= 
  \Delta(qp)+\frac{1}{2}i\hbar\\ 
  C_p&=&
         \langle(\hat{p}-\langle\hat{p}\rangle)\hat{p}\rangle=\langle\hat{p}^2\rangle-\langle\hat{p}\rangle^2=\Delta(p^2)\,.
\end{eqnarray}
Introducing canonical variables for second-order moments, we have
\begin{equation} \label{CqCp}
  C_q=sp_s+\frac{1}{2}i\hbar=0 \quad,\quad C_p=  p_s^2+\frac{U}{s^2}=0\,.
\end{equation}
These constraints are first class because
\begin{equation}
  \{C_q,C_p\}= 2p_s^2+\frac{2U}{s^2}=2C_p
\end{equation}
while both constraints have vanishing Poisson brackets with
$\langle\hat{p}\rangle$.

Solving the constraints requires some care because they are defined on a
Poisson manifold with canonical coordinates $(s,p_s,U)$ that is not
symplectic. For this reason, it is possible to have two independent
constraints, $C_q$ and $C_p$, even though there is only one canonical pair,
$(s,p_s)$. We first see that $C_q=0$ implies $s\not=0$ and
$p_s\not=0$. Inserting $C_q$ in $C_p$ then shows that $U=\hbar^2/4$ has
the minimum value allowed by uncertainty relations. The remaining equation,
$C_q=0$, is then solved by $p_s=-i\hbar/(2s)$. This solution relates both the
real and the imaginary part of $p_s$ to the real and imaginary parts of
$s$. On the solution space, the gauge flows
\begin{equation}
  \delta_{C_q}s=\{s,C_q\}= s \quad,\quad \delta_{C_p}s=\{s,C_p\}=
  2p_s=\frac{i\hbar}{s}
\end{equation}
show that both the real and imaginary parts of $s$ can be changed arbitrarily
by gauge transformations. The Casimir variable $U$ has, by definition,
vanishing Poisson brackets with any phase-space function, including the
constraints, and is therefore an observable. Its value on the constraint
surface, $U=\hbar^2/4$, is guaranteed to be real, but it is directly related
to a fundamental constant and does not contain information about the system.
As seen for expectation values, other observable moments could only be
provided by independent canonical degrees of freedom, which can then be
required to be real.

In this example, the imaginary part of $C_q$ is important because it
eliminates the submanifolds $s=0$ or $p_s=0$ from the constraint surface on
which $C_q=0$, and at the same time allows the Casimir variable $U$ to stay
within its allowed range. Even though some of the kinematical moments are then
complex, uncertainty relations are formally satisfied. The cosmological
dynamics to which we turn now will be governed by more complicated
constraints. We will therefore focus on dynamical questions and downplay
formal questions by ignoring, in this first study, complex contributions to
the constraints.

\section{Cosmological dynamics}
\label{s:CosmoDyn}

In order to make the relevant equations tractable, we now analyze cosmological
dynamics in the restricted setting of spherically symmetric time-dependent
models of gravity, given by Lemaitre--Tolman--Bondi (LTB) space-times in the
classical setting. We begin with the most general spherically symmetric
line element
\begin{equation}
  \label{eq:spherically_symm_metric}
  {\rm d}s^2= g_{\mu\nu}{\rm d}x^\mu {\rm d}x^\nu = -N^2 {\rm d}t^2 +
  q_{xx}({\rm d}x + M {\rm d}t)^2 + q_{\phi\phi}{\rm d}\Omega^2\, .
\end{equation}
In general, the four functions $N$, $M$, $q_{xx}$ and $q_{\phi\phi}$ may
depend on the time coordinate $t$ and the radial coordinate $x$, while
$d\Omega^2 = d\theta^2 + \sin^2{\theta}d\phi^2$ is the line element on a
2-sphere.

We now transform the metric variables $q_{xx}$ and $q_{\phi\phi}$ into the
canonical coordinates used in \cite{SphSymmMoments}. The new variables are
motivated by a formulation in which spatial metrics are expressed by
densitized triads. One advantage of triad formulations in the present context
is that they allow classical phase spaces without boundaries because the
metric $q_{ab}=e^i_ae_{bi}$ constructed from a triad $e_a^i$, $i=1,2,3$, is
guaranteed to be positive semidefinite, $\det q\geq 0$. A densitized triad is
defined by multiplying a triad with its determinant. Such formulations are
useful in certain canonical quantizations such as loop quantum gravity
\cite{LoopRep,ALMMT}.  Our spatial metric components then read
\begin{equation}
    \label{eq:phi1_phi2_defs}
    q_{xx} = \frac{\phi_2^2}{|\phi_1|} \quad , \quad q_{\phi\phi} = |\phi_1|\, ,
\end{equation}
where $\phi_1$ and $\phi_2$, identified with the components of a spherically
symmetric densitized triad \cite{SymmRed,SphSymm}, are canonical coordinates
with conjugate momenta $p_1$ and $p_2$ respectively. The sign of $\phi_1$
determines the orientation of the triad, while the sign of $\phi_2$ is subject
to a residual gauge transformation that follows from triad rotations. In what
follows, we assume that both $\phi_1$ and $\phi_2$ are positive in order to
simplify our notation. Nevertheless, a full quantization of these variables
does not require an implementation of phase-space boundaries.

The gravitational dynamics is completely constrained: Instead of a
Hamiltonian, we have a Hamiltonian constraint which restricts the canonical
variables and generates gauge transformations that correspond to deformations
of a spatial slice in a normal direction in space-time
\cite{ADM,Regained}. The deformations are independent at different spatial
points, just as changes of time coordinates can be performed independently in
different regions. The Hamiltonian constraint therefore amounts to infinitely
many constraints, denoted as $H[N]$ with a free lapse function $N$. In
addition, there are infinitely many diffeomorphism constraints, $D[M]$ with a
free shift vector $M$ in the 1-dimensional spatial direction, which generate
tangential deformations of a spatial slice, corresponding to spatial
coordinate changes.  In units such that $2G=1$, the spherically symmetric
versions of the Hamiltonian and diffeomorphism constraints take the forms
\begin{equation}
    \label{eq:clas_hamil}
    H[N] = -\int {\rm d}x N \left(\frac{\phi_2p_2^2}{2\sqrt{\phi_1}} +
      2\sqrt{\phi_1}p_1p_2 + \left(1 -
        \left(\frac{\phi_1'}{\phi_2}\right)^2\right)\frac{\phi_2}{2\sqrt{\phi_1}}
      - 2\left(\frac{\phi_1'}{\phi_2}\right)'\sqrt{\phi_1}\right) 
\end{equation}
and
\begin{equation}
    \label{eq:clas_diff}
    D[M] = \int {\rm d}x M (-\phi_1'p_1 + p_2'\phi_2)\, ,
\end{equation}
where primes denote spatial derivatives. Evolution can be generated by the
constraints along any timelike vector field in space-time, whose components
depend on $N$ and $M$. (These functions are, however, not identical with the
vector components because the latter refer to coordinate directions while the
former refer to directions normal and tangential to spatial hypersurfaces. We
will not use the precise relationship in this paper.)  Given a timelike vector
field $t^a$ and applying a linear transformation of its components to a lapse
function and shift vector, the corresponding Hamiltonian is the sum of these
two constraints, $H_t[N,M]=H[N]+D[M]$. Since the Hamiltonian and
diffeomorphism constraints are known to form a first-class system, we can
follow the procedure for constrained quasiclassical quantization given in the
last section.

\subsection{Expansion in One Canonical Pair}

Expanding quasiclassical constraints in moments implies additional terms added
to them. As explicitly shown in \cite{SphSymmMomentsMasters}, the extended
system up to moments of second order remains first class if moments of all
variables as well as cross-moments such as $\Delta(\phi_1\phi_2)$ are
included. However, the resulting system contains a large number of independent
fields which, formulated as moments, do not immediately appear in canonical
form. For tractable equations we therefore aim to find further truncations in
order to reduce the number of independent fields and, at the same time, make
it possible to formulate canonical variables for them. Such additional
truncations should be motivated by physical assumptions, such as using states
with small cross-correlations compared with other moments. They should also be
subject to consistency conditions such as general covariance, which in formal
terms means that the truncated constraints should remain first class.

In a previous study \cite{SphSymmMoments}, in which the aim was to analyze
static black-hole models, a suitable truncation turned out to be given by
ignoring all moments related to the component $\phi_1$ and its momentum $p_1$,
including cross-moments with $\phi_2$ and $p_2$. The physical motivation in
this case was given by the fact that $\phi_1$, or $q_{\phi\phi}$ to which it
is directly related, can, in the static case, always be set equal to $x^2$ by
a suitable choice of the spatial coordinate. Once $\phi_1$ is identified with
a space-time coordinate, it should no longer be subject to fluctuations or
correlations. Only the three second-order moments for $\phi_2$ and $p_2$ then
remained in the system, which can be formulated canonically using an extension
of (\ref{eq:quantum_coords_def}) to fields. A further truncation, assuming
subdominant correlations between fields at different points, leads us to
\begin{equation}
    \label{eq:one_pair_quantum_vars_def}
    \Delta(\phi_2^2) = \phi_3^2 \quad , \quad
    \Delta(\phi_2p_2) = \phi_3p_3 \quad , \quad
    \Delta(p_2^2) = p_3^2 + \frac{U}{\phi_3^2}\, .
\end{equation}
The resulting system has only one additional canonical field, $\phi_3$,
together with its momentum $p_3$. In general, the Casimir variable $U$ may
also be a spatial function (of density weight one because it has the same
density weight as the product $\phi_3p_3$ of a scalar field times its
momentum). But, as discussed in Section~\ref{s:Single}, its precise form is
subject to regularization and need not obey the single-particle condition
$U\geq \hbar^2/4$. According to \cite{SphSymmMoments}, the truncated
quasiclassical constraints are first class in the static case, and consistent
solutions had indeed been found.

Such a strong truncation can no longer be applied in cosmological
models. Clearly, they cannot be static, complicating the consistency analysis
that requires first-class constraints. Moreover, the partial gauge fixing
$\phi_1 = x^2$ should be relaxed because it may no longer be possible if
$\phi_1$ is allowed to depend on $t$. (Upon trying this partial gauge fixing
in a time-dependent case, in which it could at least suggest partial
information about a subset of solutions, we found the resulting system unable
to yield solutions.) For these reasons, we will keep $\phi_1$ as a free
canonical field and will not apply any partial gauge fixing before solving the
extended constraints. In a first attempt of simplifying the equations, we
assume that $\phi_1$, even though it is a canonical field, has subdominant
moments compared with $\phi_2$.

\subsubsection{Constraints}

After performing the quasiclassical expansion, the effective versions of both
classical constraints are given by their classical form plus a quantum
correction, following the derivations in \cite{SphSymmMoments}. For an
effective Hamiltonian constraint defined by
\begin{equation}
    \bar{H}[N] = H[N] + H_2[N]\, ,
\end{equation}
we find
\begin{eqnarray}
  H[N] &=& -\int {\rm d}x N \left(\frac{\phi_2p_2^2}{2\sqrt{\phi_1}} +
          2\sqrt{\phi_1}p_1p_2 + \left(1 -
          \left(\frac{\phi_1'}{\phi_2}\right)^2\right)\frac{\phi_2}{2\sqrt{\phi_1}}
          - 2\left(\frac{\phi_1'}{\phi_2}\right)'\sqrt{\phi_1}\right) \\ 
    H_2[N] &=& -\int {\rm d}x N \biggl(\frac{\phi_2p_3^2}{2\sqrt{\phi_1}} +
              \frac{\phi_3p_2p_3}{\sqrt{\phi_1}} +
              \left(6\frac{\sqrt{\phi_1}\phi_1'\phi_2'}{\phi_2^4} -
              \frac{1}{2}\frac{(\phi_1')^2}{\sqrt{\phi_1}\phi_2^3} -
              2\frac{\phi_1''\sqrt{\phi_1}}{\phi_2^3}\right)\phi_3^2
  \\\nonumber 
       &&- 4\frac{\sqrt{\phi_1}\phi_1'\phi_3\phi_3'}{\phi_2^3}
 %         +\frac{U\phi_2}{2\sqrt{\phi_1}\phi_3^2}
       \biggr)\, .
\end{eqnarray}
The effective diffeomorphism constraint is given by 
\begin{equation}
    \bar{D}[M] = \int {\rm d}x M (-\phi_1'p_1 + p_2'\phi_2 + p_3'\phi_3)\\
    \label{eq:1exp_eff_diff_con}\, .
\end{equation}

In addition, we need a higher-order constraint corresponding to each classical
constraint for each fluctuating classical field. These are given by
\begin{eqnarray}
    H_{\phi_2}[L] &=& -\int {\rm d}x L
                     \biggl(\left(\frac{p_2^2}{2\sqrt{\phi_1}} +
                     \frac{1}{2\sqrt{\phi_1}} + \frac{(\phi_1')^2 +
                     4\phi_1\phi_1''}{2\sqrt{\phi_1}\phi_2^2} -
                     4\frac{\sqrt{\phi_1}\phi_1'\phi_2'}{\phi_2^3}\right)\phi_3^2
  \\\nonumber 
    &&+\frac{2\sqrt{\phi_1}\phi_1'}{\phi_2^2}\phi_3\phi_3' +
      \left(\frac{\phi_2p_2}{\sqrt{\phi_1}} +
      2\sqrt{\phi_1}p_1\right)\phi_3p_3\biggr)\\ 
    D_{\phi_2}[K] &=& \int {\rm d}x K (p_2'\phi_3^2 + \phi_2\phi_3 p_3')\\
    H_{p_2}[J] &=& -\int {\rm d}x J \biggl(\left(\frac{p_2^2}{2\sqrt{\phi_1}}
                  + \frac{1}{2\sqrt{\phi_1}} +
                  \frac{(\phi_1')^2}{2\sqrt{\phi_1}\phi_2^2} +
                  \frac{2\sqrt{\phi_1}\phi_1''}{\phi_2^2} -
                  \frac{4\sqrt{\phi_1}\phi_1'\phi_2'}{\phi_2^3}\right)p_3\phi_3
  \\\nonumber 
    && + \frac{2\sqrt{\phi_1}\phi_1'}{\phi_2^2}p_3\phi_3' + \left(\frac{\phi_2
      p_2}{\sqrt{\phi_1}} + 2\sqrt{\phi_1}p_1\right)p_3^2 \biggr)\\ 
    D_{p_2}[I] &=& \int{{\rm d}x I \left(p_2'p_3\phi_3 + \phi_2p_3p_3'\right)}\, .
\end{eqnarray}
The last three constraints are new here because they vanish identically in the
static case studied in \cite{SphSymmMoments}.

These higher-order constraints, in combination with the effective Hamiltonian and diffeomorphism
constraints, define the time evolution of this system. For a function
$f$ on this new phase space, its time evolution is given by
\begin{equation}
    \dot{f} = \{f,\bar{H}[N] + \bar{D}[M] + H_{\phi_2}[L] + D_{\phi_2}[K] + H_{p_2}[J] + D_{p_2}[J]\}\, .
\end{equation}
This definition allows us to derive an equation of motion for each
coordinate. From our definition of these brackets, we have that the equations
of motion corresponding to canonical pair $(\phi_i,p_i)$ are
\begin{eqnarray}
    \dot{\phi_i} &=& \frac{\delta}{\delta p_i}\left(\bar{H}[N] + \bar{D}[M] +
                    H_{\phi_2}[L] + D_{\phi_2}[K] + H_{p_2}[J] +
                    D_{p_2}[J]\right)\\ 
    \dot{p_i} &=& -\frac{\delta}{\delta \phi_i}\left(\bar{H}[N] + \bar{D}[M]
                 + H_{\phi_2}[L] + D_{\phi_2}[K] + H_{p_2}[J] +
                 D_{p_2}[J]\right)\, . 
\end{eqnarray}
The individual equations of motion for all fields are lengthy and therefore
reproduced in Appendix \ref{section:app_a}.

If the system is consistent, then for initial conditions satisfying the
constraints, these equations should produce solutions that continue to satisfy
them. The initial goal with this model will be to see if this is the
case. Since each constraint must vanish for all solutions, we can extract a
set of equations that all initial conditions must satisfy by setting each
constraint equal to zero. These are
\begin{eqnarray}
    0&=&p_2^2 \phi _3^2 \phi _2^5+p_3^2 \phi _3^2 \phi _2^5+2 p_2 p_3 \phi
         _3^3 \phi _2^4+4 p_1 p_2 \phi _1 \phi _3^2 \phi _2^4
%        +U \phi _2^5
         -\phi _3^2
    \phi _2^3 \left(\phi _1'\right){}^2+4 \phi _1 \phi _3^2 \phi _2^2 \phi _1'
    \phi _2'\\\nonumber 
    &&-\phi _3^4 \phi _2 \left(\phi _1'\right){}^2-8 \phi _1 \phi _3^3 \phi _2
    \phi _1' \phi _3'+12 \phi _1 \phi _3^4 \phi _1' \phi _2'-4 \phi _1 \phi
    _3^2 \phi _2^3 \phi _1''-4 \phi _1 \phi _3^4 \phi _2 \phi _1''+\phi _3^2
    \phi _2^5\\ 
    0&=&\phi _2 p_2'+\phi _3 p_3'-p_1 \phi _1'\\
    0&=&2 p_2 p_3 \phi _2^4+4 p_1 p_3 \phi _1 \phi _2^3+p_2^2 \phi _3 \phi
    _2^3+\phi _3 \phi _2 \left(\phi _1'\right){}^2+4 \phi _1 \phi _2 \phi _1'
    \phi _3'-8 \phi _1 \phi _3 \phi _1' \phi _2'\\\nonumber 
    &&+4 \phi _1 \phi _3 \phi _2 \phi _1''+\phi _3 \phi _2^3\\  
    0&=&\phi _3 p_2'+\phi _2 p_3'\, .
\end{eqnarray}
Note that the six initial constraints have been reduced to a system of four
equations. The first two constraints are quantum extensions of the original
Hamiltonian and diffeomorphism constraints, while the other two are
higher-order constaints for moments of $(\phi_2,p_2)$. The remaining two of
the original six constraints turn out to be redundant because we ignored
imaginary contributions and $U$-terms. For comparison, if we do the same in
(\ref{CqCp}), both constraints in this equation can be solved by a single
condition, $p_2=0$. Keeping imaginary parts and $U$-terms in (\ref{CqCp})
allowed us to solve also for $U$, a possibility that we forgo in the main part
of our present cosmological treatment.

\subsubsection{FRW Background and perturbations}

In order to simplify solving the equations governing this system, we will
proceed perturbatively. We will treat the classical fields as their classical
exact solutions plus small perturbations. The background solutions we will use
are those of the Friedmann--Robertson--Walker (FRW) cosmological model, an
exact solution of general relativity that assumes the universe is homogeneous,
isotropic, and filled with a ``perfect fluid.'' It results in a metric dependent
only on two parameters: the scale factor, $a(t)$, which gives the ``size'' of
the universe and the spatial curvature constant, $k$, which encodes the
large-scale curvature of the universe. Since this background is homogeneous,
the classical perturbations will supply perturbative inhomogeneity, a notable
increase in complexity from the classical model.

The FRW line element takes the form
\begin{equation}
    \label{eq:frw_metric}
    {\rm d}s^2=g_{\mu\nu}{\rm d}x^\mu {\rm d}x^\nu = -{\rm d}t^2 +
    a(t)^2\left(\frac{{\rm d}r^2}{1-kr^2} + r^2 {\rm d}\Omega^2\right)\, . 
\end{equation}
One can infer the corresponding values of $\phi_1$, $\phi_2$, $N$, and $M$ by
inspection using (\ref{eq:spherically_symm_metric}) and
(\ref{eq:phi1_phi2_defs}). The momenta, $p_1$ and $p_2$, can then be found by solving the
system using the classical brackets and the constraints
(\ref{eq:clas_hamil})--(\ref{eq:clas_diff}). In doing this, a few questions
arise. First, in order to find analytic solutions, assumptions about
$a(t)$ must be made. Since this investigation focuses on no era in particular,
we make the somewhat arbitrary choice
\begin{equation}
    a(t) = a_0 t^{1/2}\, ,
\end{equation}
which corresponds to a radiation-dominated universe. Here, $a_0$ is a scaling
constant which we will assume to equal one when finding solutions. The second
question is that of the value of the parameter $k$. There is no a priori
reason to assume any particular value. However, since modern measurements have
so far been unable to constrain this parameter to be nonzero, we will assume
$k=0$ for simplicity. Under these assumptions, our new definitions of the
classical variables take the form
\begin{eqnarray}
    \phi _1=a_0^2 t x^2+\delta \phi _1 \quad &,& \quad
    p_1=-\frac{a_0}{2 \sqrt{ t}} + \delta{p}_1\\
    \phi _2=2 a_0^2 t x+\delta \phi _2 \quad &,& \quad
    p_2=-\frac{a_0x}{2\sqrt{t}}+ \delta{p}_2\\
    N=1 \quad &,& \quad
    M=0\, .
\end{eqnarray}
We can now simplify the equations considerably. 

Imposing the background solutions, we can simplify the constraint and
evolution equations via Taylor expansion under the assumption that the
perturbations are small compared with the classical solutions. The new quantum
correction fields, related to the variances of the classical fields, should be
even smaller than these for the moment expansion to remain valid as well. We
will perform a Taylor expansion in all small quantities and remove any terms
of greater than second order in these quantities.  We will expand the vacuum
Hamiltonian constraint for this purpose, eliminating any pure background terms
because they would cancel out with a background matter Hamiltonian.

These conditions give the new set of constraint equations
\begin{eqnarray}
    \label{eq:one_pair_simp_con1}
    0&=& 16 p_3^2 t^{7/2} x^4-8 p_3 t^2 x^4 \phi _3+32
         \delta{p}_1 \delta{p}_2 t^{7/2} x^5+2 \delta \phi _2 t^{3/2} x^5+16
         \delta{p}_2^2 t^{7/2} x^4\\\nonumber 
    &&+16 t^{5/2} x^4 \delta \phi _2'-16 t^{5/2} x^4 \delta \phi _1''+2 \delta
       \phi _1 t^{3/2} x^4-16 \delta \phi _2 t^{3/2} x^3 \delta \phi _2'+8
       t^{3/2} x^3 \delta \phi _1' \delta \phi _2'\\\nonumber 
    &&+8 \delta \phi _2 t^{3/2} x^3 \delta \phi _1''-16 t^{3/2} x^3 \phi _3
       \phi _3'-8 \delta \phi _2 t^{3/2} x^2 \delta \phi _1'+8 \delta \phi _1
       t^{3/2} x^2 \delta \phi _2'-8 \delta \phi _1 t^{3/2} x^2 \delta \phi
       _1''\\\nonumber 
    &&+12 t^{3/2} x^2 \phi _3^2+16 \delta \phi _1 t^{3/2} x \delta \phi _1'-16
       \delta \phi _1 \delta \phi _2 t^{3/2} x-16 \delta{p}_1 t^3 x^6-32
       \delta{p}_2 t^3 x^5-8 \delta{p}_1 \delta \phi _1 t^2 x^4\\\nonumber 
    &&-8 \delta{p}_2 \delta \phi _2 t^2 x^4-\delta \phi _1 \delta \phi _2 \sqrt{t} x^3\\
    0&=&-2 \delta{p}_1 \delta \phi _1'+2 \delta \phi _2 \delta{p}_2'+2 \phi _3
         p_3'+\frac{\delta \phi _1'}{\sqrt{t}}-\frac{\delta \phi
         _2}{\sqrt{t}}+4 t x \delta{p}_2'-4 \delta{p}_1 t x \label{deltaphi1prime}\\ 
    0&=&32 \delta{p}_1 p_3 t^{7/2} x^5+32 \delta{p}_2 p_3 t^{7/2} x^4-32 p_3
         t^3 x^5-8 \delta \phi _2 p_3 t^2 x^4+2 t^{3/2} x^5 \phi _3+16 t^{5/2}
         x^4 \phi _3'\\\nonumber 
    &&-16 t^{3/2} x^3 \phi _3 \delta \phi _2'-16 \delta \phi _2 t^{3/2} x^3
       \phi _3'+8 t^{3/2} x^3 \delta \phi _1' \phi _3'+8 t^{3/2} x^3 \phi _3
       \delta \phi _1''-8 t^{3/2} x^2 \phi _3 \delta \phi _1'\\\nonumber 
     &&+8 \delta \phi _1 t^{3/2} x^2 \phi _3'+24 \delta \phi _2 t^{3/2} x^2
        \phi _3-16 \delta \phi _1 t^{3/2} x \phi _3-8 \delta{p}_2 t^2 x^4 \phi
        _3-\delta \phi _1 \sqrt{t} x^3 \phi _3\\ 
    \label{eq:one_pair_simp_con4}
    0&=& 2 \phi _3 \delta{p}_2'+2 \delta \phi _2 p_3'+4 t x p_3'-\frac{\phi _3}{\sqrt{t}}\, 
\end{eqnarray}
and the evolution equations
\begin{eqnarray}
    \delta \dot{\phi}_1&=&\frac{\delta \phi _1}{2 t}-\frac{\delta{p}_2 \delta
                            \phi _1}{\sqrt{t} x}-2 \delta{p}_2 \sqrt{t} x\\ 
    \delta\dot{p}_1&=&\frac{p_3 \phi _3}{4 t^2 x^2}-\frac{p_3^2}{2 \sqrt{t}
                         x^2}-\frac{3 \delta \phi _1 \delta \phi _1'}{4
                         t^{5/2} x^5}+\frac{\delta \phi _1 \delta \phi _2}{2
                         t^{5/2} x^5}-\frac{\delta \phi _2 \delta \phi _1'}{4
                         t^{5/2} x^4}+\frac{\delta \phi _1 \delta \phi _1''}{4
                         t^{5/2} x^4}+\frac{\delta \phi _1}{2 t^{3/2}
                         x^4}-\frac{\phi _3^2}{8 t^{5/2} x^4}\\\nonumber 
    &&+\frac{\delta \phi _1'}{2 t^{3/2} x^3}+\frac{\delta \phi _2 \delta \phi
       _1''}{4 t^{5/2} x^3}+\frac{3 \delta \phi _1 \delta \phi _2}{32 t^{7/2}
       x^3}-\frac{\delta \phi _1''}{2 t^{3/2} x^2}+\frac{\delta \phi _1}{16
       t^{5/2} x^2}-\frac{\delta \phi _2}{16 t^{5/2} x}-\frac{1}{8
       t^{3/2}}-\frac{\delta{p}_2 \delta \phi _1}{2 t^2 x^3}\\\nonumber 
    &&+\frac{\delta{p}_1 \delta \phi _1}{4 t^2 x^2}+\frac{\delta{p}_2 \delta
       \phi _2}{4 t^2 x^2}-\frac{\delta{p}_1}{2 t}-\frac{1}{\sqrt{t}
       x^2}+\frac{\delta{p}_1 \delta{p}_2}{\sqrt{t} x}\\ 
    \delta \dot{\phi}_2&=&-2 {K} \phi _3 \phi _3'-{K'} \phi _3^2-\frac{p_3
                            \phi _3}{\sqrt{t} x}-\frac{\delta \phi _1 \delta
                            \phi _2}{4 t^2 x^2}-2 \delta{p}_2
                            \sqrt{t}+\frac{\delta \phi _2}{2
                            t}+\frac{\delta{p}_2 \delta \phi _1}{\sqrt{t}
                            x^2}-\frac{\delta{p}_1 \delta \phi _1}{\sqrt{t}
                            x}\\\nonumber 
    &&-\frac{\delta{p}_2 \delta \phi _2}{\sqrt{t} x}-2 \delta{p}_1 \sqrt{t} x\\
    \delta\dot{p}_2&=&-{K} \phi _3 p_3'+\frac{p_3^2}{2 \sqrt{t}
                         x}+\frac{\delta \phi _1 \delta \phi _1'}{4 t^{5/2}
                         x^4}-\frac{\delta \phi _1 \delta \phi _2}{4 t^{5/2}
                         x^4}+\frac{\delta \phi _2 \delta \phi _1'}{2 t^{5/2}
                         x^3}-\frac{3 \phi _3^2}{8 t^{5/2} x^3}-\frac{\delta
                         \phi _1'}{2 t^{3/2} x^2}\\\nonumber 
    &&+\frac{\delta \phi _2}{2 t^{3/2} x^2}-\frac{\delta{p}_2 \sqrt{t x}}{2
      t^{3/2} \sqrt{x}}-\frac{\delta \phi _1}{16 t^{5/2} x}-\frac{x}{8
      t^{3/2}}+\frac{\delta{p}_2 \delta \phi _1}{4 t^2 x^2}\\ 
    \dot{\phi}_3&=&-2 {I} p_3 t-\frac{{I} \phi _3}{2 \sqrt{t}}-2 {I'} p_3 t
                    x+4 J p_3 x-\frac{J x \phi _3}{8 t^{3/2}}-\frac{J \phi
                    _3'}{\sqrt{t}}-{K} \phi _3 \delta \phi _2'-\delta \phi _2
                    {K} \phi _3'\\\nonumber 
    &&-2 {K} t x \phi _3'-2 {K} t \phi _3-\delta \phi _2 {K'} \phi _3-2 {K'} t x \phi _3+2 L x \phi _3+\frac{\delta \phi _1 p_3}{\sqrt{t} x^2}-\frac{\delta \phi _2 p_3}{\sqrt{t} x}-2 p_3 \sqrt{t}\\\nonumber
    &&-\frac{\delta \phi _1 \phi _3}{4 t^2 x^2}-\frac{\delta{p}_2 \phi
       _3}{\sqrt{t} x}+\frac{\phi _3}{2 t}\\ 
    \dot{p}_3&=&\frac{{I} p_3}{2 \sqrt{t}}-\frac{p_3 J'}{\sqrt{t}}+\frac{J p_3
                 x}{8 t^{3/2}}-\frac{J p_3'}{\sqrt{t}}-2 {K} \phi _3
                 \delta{p}_2'-\delta \phi _2 {K} p_3'-2 {K} t x p_3'+\frac{{K}
                 \phi _3}{\sqrt{t}}-\frac{\phi _3 L'}{\sqrt{t}}\\\nonumber 
    &&-2 L p_3 x+\frac{L x \phi _3 \sqrt{t x}}{4 \sqrt{2} t^2}+\frac{\delta
       \phi _1 p_3}{4 t^2 x^2}+\frac{\delta{p}_2 p_3}{\sqrt{t} x}-\frac{p_3}{2
       t}+\frac{\delta \phi _1 \phi _3}{t^{5/2} x^4}+\frac{3 \phi _3 \delta
       \phi _1'}{4 t^{5/2} x^3}-\frac{\delta \phi _1 \phi _3'}{2 t^{5/2}
       x^3}+\frac{\delta \phi _1 \phi _3}{2 t^{5/2} x^3}\\\nonumber 
    &&-\frac{11 \delta \phi _2 \phi _3}{4 t^{5/2} x^3}+\frac{\phi _3 \delta
       \phi _1'}{2 t^{5/2} x^2}+\frac{5 \phi _3 \delta \phi _2'}{4 t^{5/2}
       x^2}+\frac{3 \delta \phi _2 \phi _3'}{2 t^{5/2} x^2}-\frac{\delta \phi
       _1' \phi _3'}{2 t^{5/2} x^2}-\frac{\phi _3 \delta \phi _1''}{2 t^{5/2}
       x^2}-\frac{3 \delta \phi _2 \phi _3}{2 t^{5/2} x^2}+\frac{\phi
       _3}{t^{3/2} x^2}\\\nonumber 
    &&-\frac{\phi _3'}{t^{3/2} x}+\frac{\phi _3}{t^{3/2} x}\, .
\end{eqnarray}

A final step to make these equations solvable is to assume values for the
functions $I$, $J$, $K$, and $L$, which come from the higher order
constraints. These currently have no physical interpretation and thus no
special motivations for our choices. As such, we will assume them all to have
the small constant value $0.1$ everywhere in spacetime, which can be
interpreted as choosing a gauge consistent with the assumed smallness of
moments.

A physical interpretation of these functions would require an
extension of the classical space-time geometry to a higher-dimensional line
element that also includes moment directions. At present, however, it remains
unknown whether such an extension is possible because higher-order constraints
do not follow the form of the classical brackets of hypersurface deformations
in space-time. Alternatively, one could impose gauge conditions on the
fluctuation fields $\phi_3$ or $p_3$ and derive conditions on $I$, $J$, $K$,
and $L$ by requiring that the gauge conditions be preserved in time. However,
this approach would require some knowledge of the underlying state.

\subsubsection{Results}

We now have a complete set of equations. Unfortunately, even with this
perturbative approach, neither the constraint nor evolution equations are
simple enough to yield analytic solutions. As such, we proceed numerically,
making use of the ``NDSolve'' function in Wolfram \textit{Mathematica}.

\begin{figure}[!ht]
    \centering
    \includegraphics[scale=1]{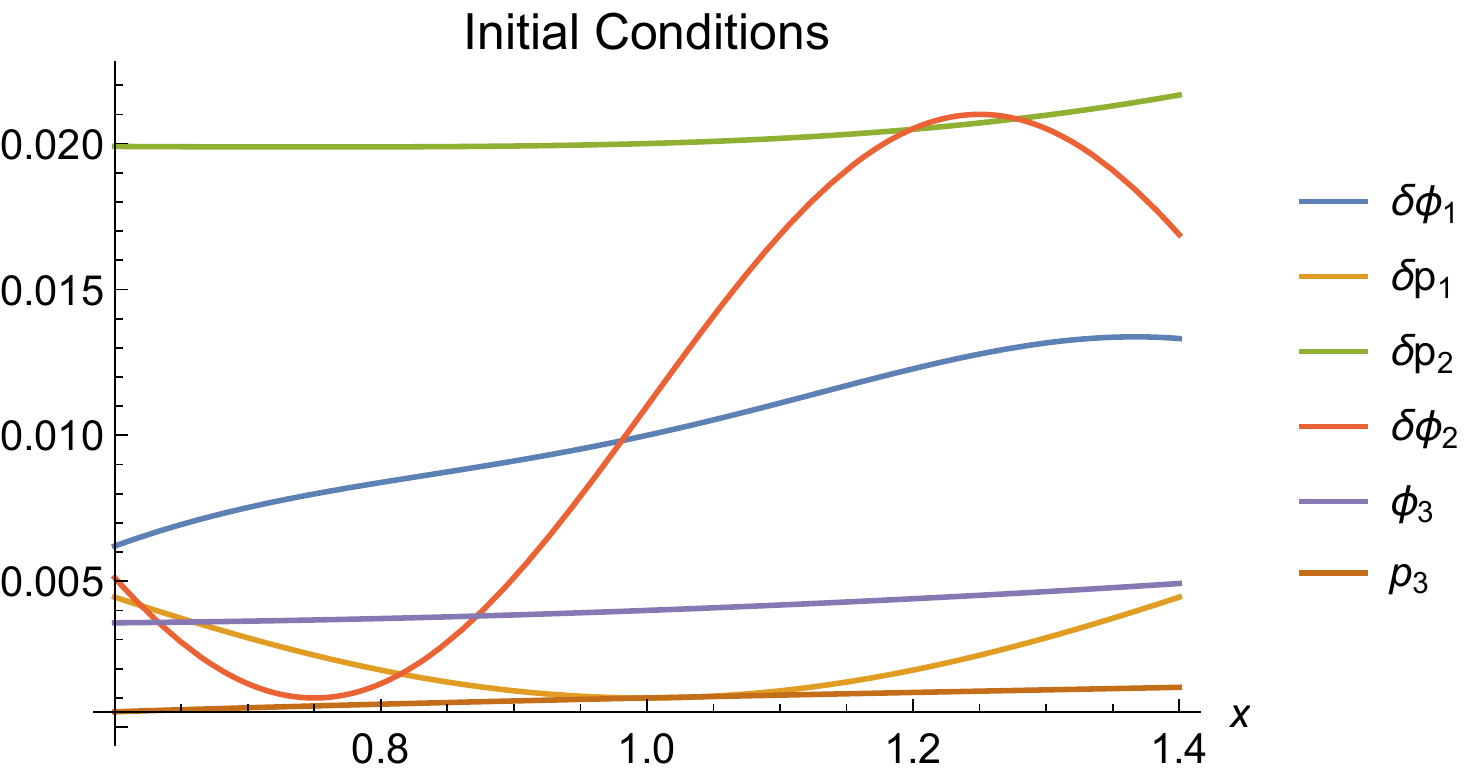}
    \caption[Initial conditions for first model]{Plot of the initial
      conditions used. Specific choices for the two fields $\delta\phi_2$ and
      $\delta p_1$ determine solutions for the remaining fields via the
      constraints. For $x\sim 1$ and greater in this range, the quantum
      corrections $\phi_3$ and $p_3$ of $(\phi_2,p_2)$ are smaller than the
      inhomogeneous perturbations, $\delta\phi_2$ and $\delta p_2$ of the same
      fields.}
    \label{fig:one_pair_init_conds}
\end{figure}

The constraint equations give four equations for the six fields, representing
our initial values. We are therefore free to select any functions for two of
the fields, with the only constraint being that the perturbations remain
small. Thus, we set $\delta\phi_2$ and $\delta p_1$ to sinusoids with
amplitudes on the order of $0.01$. Solving the constraint equations
numerically yielded the initial conditions shown in Figure
\ref{fig:one_pair_init_conds}. All fields remained small except
$\delta\phi_1$, which grows toward both sides of the $x$-range used here. A
linearization of equation (\ref{deltaphi1prime}),
\begin{equation}
   \delta \phi _1'=\delta \phi_2-4 t^{3/2} x (\delta{p}_2'-\delta{p}_1)\,,
\end{equation}
suggests that this growth is generically
quadratic, integrating terms linear in $x$ and modulated only by the more slowly
changing $\delta p_1$ and $\delta p_2'$. The quadratic growth may be avoided
by initial values that imply small $\delta p_1$ and $\delta p_2'$, as in the
example of Figure \ref{fig:one_pair_init_conds}.

\begin{figure}
    \centering
    \includegraphics[scale=0.7]{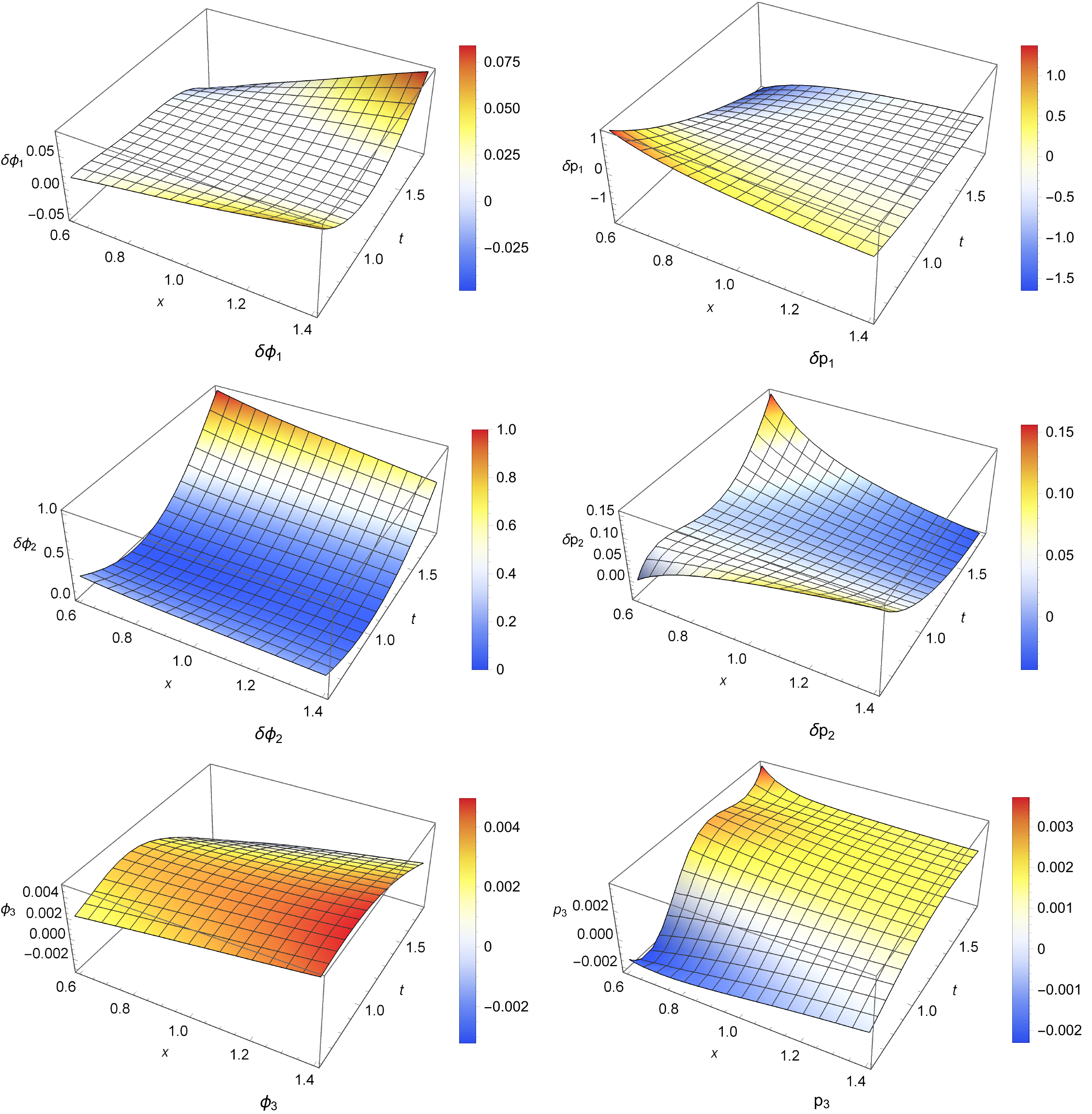}
    \caption[Solutions for first model]{Plots of the solutions found using the initial conditions in Figure \ref{fig:one_pair_init_conds} imposed at $t=1$.}
    \label{fig:one_pair_solns}
\end{figure}

Imposing the initial conditions at $t=1$ and evolving them using the evolution
equations, we found the solutions shown in Figure
\ref{fig:one_pair_solns}. Aside from some of the fields growing to order
unity, these solutions appear to fit our requirements for acceptable
solutions at least in a restricted region. However, this conclusion turns out
to be misleading because it does not take into account additional consistency
conditions for constrained systems.

\begin{figure}[!ht]
    \centering
    \includegraphics[scale=0.7]{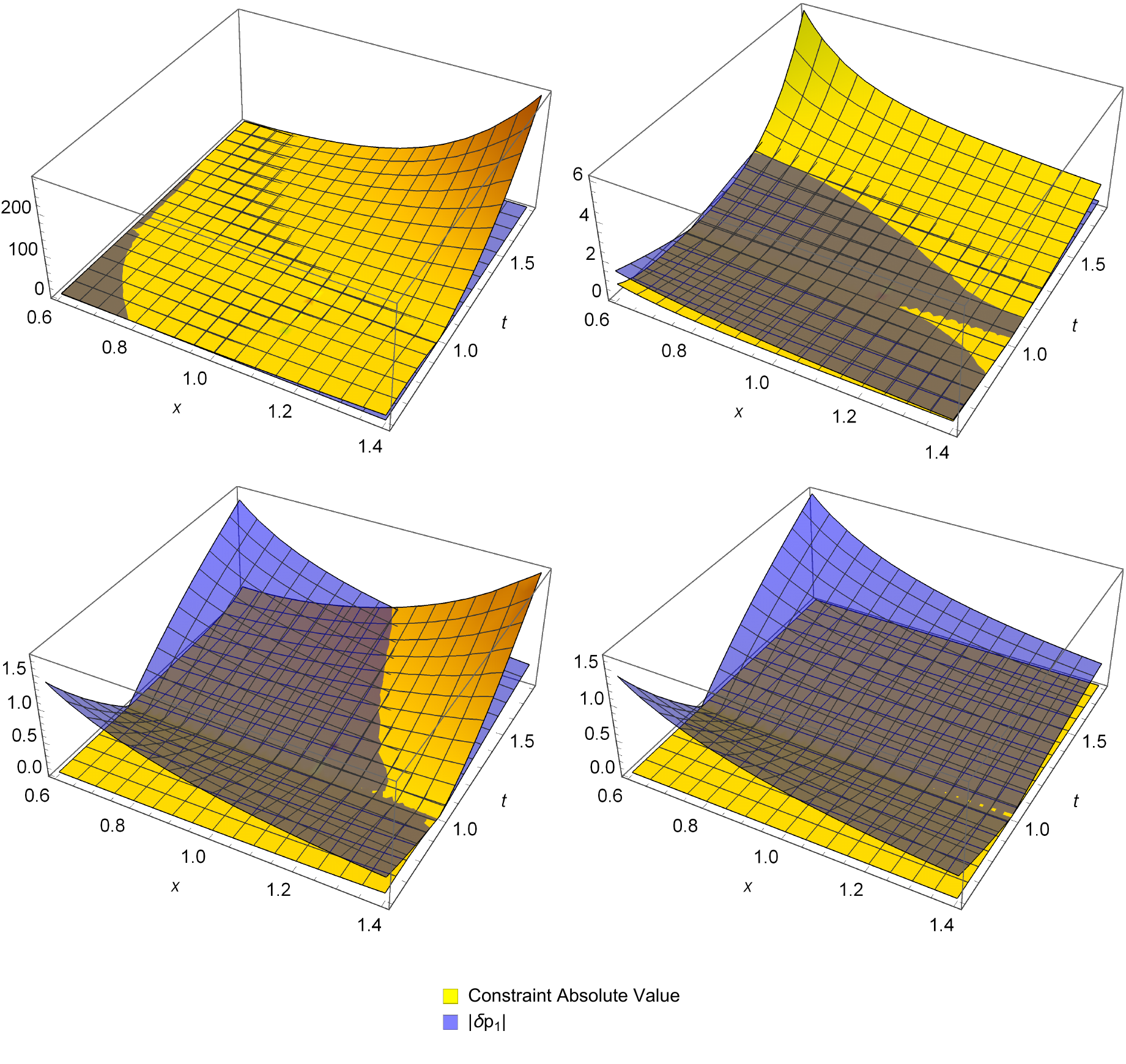}
    \caption[Non-satisfaction of first model's constraints]{Each constraint's
      absolute divergence from zero plotted alongside the absolute value of
      $\delta p_1$, the field with the largest magnitude. The constraints are
      clearly not preserved.} 
    \label{fig:one_pair_cons_test}
\end{figure}

Plotted in Figure \ref{fig:one_pair_cons_test} is a comparison of the absolute
value of $\delta p_1$ with the absolute value of the right-hand side of each
constraint equation
(\ref{eq:one_pair_simp_con1})--(\ref{eq:one_pair_simp_con4}) evaluated for
these solutions. Some divergence from zero in the constraints is expected due
to our approximate solution procedure; we solved this system perturbatively
and applied additional truncations beyond just considering moment order, so we
should expect the constraints to be only approximately preserved. In addition,
since these are numerical solutions, we can expect there is some numerical
error. However, the magnitude of these divergences is much larger than either
of these effects should produce. On this spacetime region, the first
constraint's divergence reaches two orders of magnitude greater than the
maximum value of the largest-valued field, $\delta p_1$. This shows that the
constraints are not in general preserved for a spherically symmetric system
expanded to second order in one canonical pair. Even for this small region,
there is nowhere where we can treat the constraints as preserved.

We conclude that the additional truncation of the moment system used in this
section is not valid beyond static solutions.  In this section, we
showed that an expansion to second order in $\phi_2$ and $p_2$ is insufficient
to produce a consistent system. In order to produce solutions that preserve
the constraints, we will have to increase in complexity.

\subsection{Expansion in Both Canonical Pairs}

For our next attempt to produce a consistent system, we add an additional
canonical pair corresponding to fluctuations in $\phi_1$ and $p_1$. The
relationship between moments and canonical fields is defined in much the same
way as before: We now have two quantum fluctuation canonical pairs
$(\phi_3,p_3)$ and $(\phi_4,p_4)$ given by
\begin{eqnarray}
    \label{eq:two_pair_quantum_vars_def_1}
    \Delta(\phi_1^2) = \phi_3^2 \quad , \quad
    \Delta(\phi_1p_1) &=& \phi_3p_3 \quad , \quad
    \Delta(p_1^2) = p_3^2 + \frac{U_1}{\phi_3^2}\\
    \label{eq:two_pair_quantum_vars_def_2}
    \Delta(\phi_2^2) = \phi_4^2 \quad , \quad
    \Delta(\phi_2p_2) &=& \phi_4p_4 \quad , \quad
    \Delta(p_2^2) = p_4^2 + \frac{U_2}{\phi_4^2}\, .
\end{eqnarray}
Note that $\phi_3$ and $p_3$ now relate to fluctuations in $\phi_1$ and $p_1$,
whereas in the previous section they related to fluctuations in $\phi_2$ and
$p_2$. The effective Hamiltonian and diffeomorphism constraints can be derived
in the same manner as before but with an expansion to second-order in moments
of both canonical pairs rather than just $(\phi_2,p_2)$. This yields an
effective Hamiltonian constraint still of the form
\begin{equation}
    \bar{H}[N] = H[N] + H_2[N]
\end{equation}
given by
\begin{eqnarray}
    \label{eq:2_pair_clas_ham_con}
    H[N] &=& -\int {\rm d}x N \left(\frac{\phi_2p_2^2}{2\sqrt{\phi_1}} +
             2\sqrt{\phi_1}p_1p_2 + \left(1 -
             \left(\frac{\phi_1'}{\phi_2}\right)^2\right)\frac{\phi_2}{2\sqrt{\phi_1}}
             - 2\left(\frac{\phi_1'}{\phi_2}\right)'\sqrt{\phi_1}\right) \\ 
    H_2[N] &=& -\int {\rm d}x {N} \biggl(\frac{p_4^2 \phi _2}{2 \sqrt{\phi
               _1}}+\frac{p_2 p_4 \phi _4}{\sqrt{\phi _1}}+\frac{3 p_2^2 \phi
               _2 \phi _3^2}{16 \phi _1^{5/2}}+\frac{p_2 p_3 \phi
               _3}{\sqrt{\phi _1}}-\frac{p_1 p_2 \phi _3^2}{4 \phi
               _1^{3/2}}
             % +\frac{U_2 \phi _2}{2 \sqrt{\phi _1} \phi_4^2}
  \\\nonumber 
    &&+\frac{6 \sqrt{\phi _1} \phi _4^2 \phi _1' \phi _2'}{\phi
       _2^4}+\frac{\phi _3 \phi _1' \phi _3'}{2 \phi _1^{3/2} \phi
       _2}+\frac{\phi _3 \phi _2' \phi _3'}{\sqrt{\phi _1} \phi
       _2^2}-\frac{\left(\phi _3'\right){}^2}{2 \sqrt{\phi _1} \phi
       _2}-\frac{3 \phi _3^2 \left(\phi _1'\right){}^2}{16 \phi _1^{5/2} \phi
       _2}-\frac{\phi _3^2 \phi _1' \phi _2'}{4 \phi _1^{3/2} \phi
       _2^2}\\\nonumber 
    &&-\frac{4 \sqrt{\phi _1} \phi _4 \phi _1' \phi _4'}{\phi _2^3}-\frac{\phi
       _4^2 \left(\phi _1'\right){}^2}{2 \sqrt{\phi _1} \phi _2^3}+\frac{\phi
       _3^2 \phi _1''}{4 \phi _1^{3/2} \phi _2}-\frac{\phi _3 \phi
       _3''}{\sqrt{\phi _1} \phi _2}-\frac{2 \sqrt{\phi _1} \phi _4^2 \phi
       _1''}{\phi _2^3}+\frac{3 \phi _2 \phi _3^2}{16 \phi _1^{5/2}}\biggr) 
\end{eqnarray}
and the effective diffeomorphism constraint
\begin{equation}
    \label{eq:2exp_eff_diff_con}
    \bar{D}[M] = \int {\rm d}x M (-\phi_1'p_1 + p_2'\phi_2 -p_3 \phi _3'+ \phi _4 p_4')\, .
\end{equation}
As with the number of quantum fluctuation fields, the number of higher-order
constraints doubles. Derived in the same way as before, we now have the
higher-order constraints
\begin{eqnarray}
    H_{\phi_1}[L] &=& -\int {\rm d}x L\biggl( \frac{p_2^2 \phi _3 \phi _4}{2
                     \sqrt{\phi _1}}-\frac{p_2^2 \phi _2 \phi _3^2}{4 \phi
                     _1^{3/2}}+\frac{p_1 p_2 \phi _3^2}{\sqrt{\phi _1}}+2 p_3
                     p_2 \sqrt{\phi _1} \phi _3+\frac{p_4 p_2 \phi _2 \phi
                     _3}{\sqrt{\phi _1}}\\\nonumber 
    &&+2 p_1 p_4 \sqrt{\phi _1} \phi _3+\frac{\phi _3^2 \left(\phi
       _1'\right){}^2}{4 \phi _1^{3/2} \phi _2}+\frac{\phi _3 \phi _4
       \left(\phi _1'\right){}^2}{2 \sqrt{\phi _1} \phi _2^2}+\frac{\phi _3^2
       \phi _1' \phi _2'}{\sqrt{\phi _1} \phi _2^2}+\frac{2 \sqrt{\phi _1}
       \phi _3 \phi _2' \phi _3'}{\phi _2^2}+\frac{2 \sqrt{\phi _1} \phi _3
       \phi _1' \phi _4'}{\phi _2^2}\\\nonumber 
    &&-\frac{\phi _3 \phi _1' \phi _3'}{\sqrt{\phi _1} \phi _2}-\frac{4
       \sqrt{\phi _1} \phi _3 \phi _4 \phi _1' \phi _2'}{\phi _2^3}+\frac{2
       \sqrt{\phi _1} \phi _3 \phi _4 \phi _1''}{\phi _2^2}-\frac{2 \sqrt{\phi
       _1} \phi _3 \phi _3''}{\phi _2}-\frac{\phi _3^2 \phi _1''}{\sqrt{\phi
       _1} \phi _2}+\frac{\phi _3 \phi _4}{2 \sqrt{\phi _1}}-\frac{\phi _2
       \phi _3^2}{4 \phi _1^{3/2}} \biggr)\\ 
    D_{\phi_1}[K] &=& \int {\rm d}x K\left( \phi _3 \phi _4 p_2'+\phi _2 \phi
                      _3 p_4'-p_3 \phi _3 \phi _1'-p_1 \phi _3 \phi _3'
                      \right)
\end{eqnarray}
for the $\phi_1$-constraints,
\begin{eqnarray}
    H_{p_1}[J] &=& -\int {\rm d}x J\biggl(\frac{p_3 \phi _3 \left(\phi
                   _1'\right){}^2}{4 \phi _1^{3/2} \phi _2}+\frac{p_3 \phi _4
                   \left(\phi _1'\right){}^2}{2 \sqrt{\phi _1} \phi
                   _2^2}+\frac{p_3 \phi _3 \phi _1' \phi _2'}{\sqrt{\phi _1}
                   \phi _2^2}+\frac{2 p_3 \sqrt{\phi _1} \phi _2' \phi
                   _3'}{\phi _2^2}\\\nonumber 
    &&+\frac{2 p_3 \sqrt{\phi _1} \phi _1' \phi _4'}{\phi _2^2}-\frac{p_3 \phi
       _1' \phi _3'}{\sqrt{\phi _1} \phi _2}-\frac{4 p_3 \sqrt{\phi _1} \phi
       _4 \phi _1' \phi _2'}{\phi _2^3}+\frac{2 p_3 \sqrt{\phi _1} \phi _4
       \phi _1''}{\phi _2^2}-\frac{2 p_3 \sqrt{\phi _1} \phi _3''}{\phi
       _2}-\frac{p_3 \phi _3 \phi _1''}{\sqrt{\phi _1} \phi _2}\\\nonumber 
    &&+\frac{p_3 p_2^2 \phi _4}{2 \sqrt{\phi _1}}-\frac{p_3 p_2^2 \phi _2 \phi
       _3}{4 \phi _1^{3/2}}+2 p_3^2 p_2 \sqrt{\phi _1}+\frac{p_3 p_4 p_2 \phi
       _2}{\sqrt{\phi _1}}+\frac{p_1 p_3 p_2 \phi _3}{\sqrt{\phi _1}}+2 p_1
       p_3 p_4 \sqrt{\phi _1}+\frac{p_3 \phi _4}{2 \sqrt{\phi _1}}\\\nonumber 
    &&-\frac{p_3 \phi _2 \phi _3}{4 \phi _1^{3/2}} \biggr)\\
    D_{p_1}[I] &=& \int {\rm d}x I\left( p_3^2 \left(-\phi _1'\right)+p_3 \phi
                   _4 p_2'+p_3 \phi _2 p_4'-p_1 p_3 \phi _3' \right)
\end{eqnarray}
for the $p_1$-constraints,
\begin{eqnarray}
H_{\phi_2}[H] &=& -\int {\rm d}x H\biggl( \frac{p_2^2 \phi _4^2}{2 \sqrt{\phi
                  _1}}-\frac{p_2^2 \phi _2 \phi _3 \phi _4}{4 \phi _1^{3/2}}+2
                  p_3 p_2 \sqrt{\phi _1} \phi _4+\frac{p_4 p_2 \phi _2 \phi
                  _4}{\sqrt{\phi _1}}+\frac{p_1 p_2 \phi _3 \phi
                  _4}{\sqrt{\phi _1}}\\\nonumber 
    &&+2 p_1 p_4 \sqrt{\phi _1} \phi _4+\frac{\phi _4^2 \left(\phi
       _1'\right){}^2}{2 \sqrt{\phi _1} \phi _2^2}+\frac{\phi _3 \phi _4
       \left(\phi _1'\right){}^2}{4 \phi _1^{3/2} \phi _2}+\frac{\phi _3 \phi
       _4 \phi _1' \phi _2'}{\sqrt{\phi _1} \phi _2^2}+\frac{2 \sqrt{\phi _1}
       \phi _4 \phi _2' \phi _3'}{\phi _2^2}+\frac{2 \sqrt{\phi _1} \phi _4
       \phi _1' \phi _4'}{\phi _2^2}\\\nonumber 
    &&-\frac{\phi _4 \phi _1' \phi _3'}{\sqrt{\phi _1} \phi _2}-\frac{4
       \sqrt{\phi _1} \phi _4^2 \phi _1' \phi _2'}{\phi _2^3}+\frac{2
       \sqrt{\phi _1} \phi _4^2 \phi _1''}{\phi _2^2}-\frac{2 \sqrt{\phi _1}
       \phi _4 \phi _3''}{\phi _2}-\frac{\phi _3 \phi _4 \phi _1''}{\sqrt{\phi
       _1} \phi _2}+\frac{\phi _4^2}{2 \sqrt{\phi _1}}-\frac{\phi _2 \phi _3
       \phi _4}{4 \phi _1^{3/2}}\biggr)\\ 
    D_{\phi_2}[G] &=& \int {\rm d}x G\left(\phi _4^2 p_2'+\phi _2 \phi _4
                      p_4'-p_3 \phi _4 \phi _1'-p_1 \phi _4 \phi _3' \right)
\end{eqnarray}
for the $\phi_2$-constraints, and
\begin{eqnarray}
    H_{p_2}[F] &=& -\int {\rm d}x F\biggl( \frac{p_4 \phi _3 \left(\phi
                   _1'\right){}^2}{4 \phi _1^{3/2} \phi _2}+\frac{p_4 \phi _4
                   \left(\phi _1'\right){}^2}{2 \sqrt{\phi _1} \phi
                   _2^2}+\frac{p_4 \phi _3 \phi _1' \phi _2'}{\sqrt{\phi _1}
                   \phi _2^2}+\frac{2 p_4 \sqrt{\phi _1} \phi _2' \phi
                   _3'}{\phi _2^2}\\\nonumber 
    &&+\frac{2 p_4 \sqrt{\phi _1} \phi _1' \phi _4'}{\phi _2^2}-\frac{p_4 \phi
       _1' \phi _3'}{\sqrt{\phi _1} \phi _2}-\frac{4 p_4 \sqrt{\phi _1} \phi
       _4 \phi _1' \phi _2'}{\phi _2^3}+\frac{2 p_4 \sqrt{\phi _1} \phi _4
       \phi _1''}{\phi _2^2}-\frac{2 p_4 \sqrt{\phi _1} \phi _3''}{\phi
       _2}-\frac{p_4 \phi _3 \phi _1''}{\sqrt{\phi _1} \phi _2}\\\nonumber 
    &&+\frac{p_4 p_2^2 \phi _4}{2 \sqrt{\phi _1}}-\frac{p_4 p_2^2 \phi _2 \phi
       _3}{4 \phi _1^{3/2}}+2 p_3 p_4 p_2 \sqrt{\phi _1}+\frac{p_4^2 p_2 \phi
       _2}{\sqrt{\phi _1}}+\frac{p_1 p_4 p_2 \phi _3}{\sqrt{\phi _1}}+2 p_1
       p_4^2 \sqrt{\phi _1}+\frac{p_4 \phi _4}{2 \sqrt{\phi _1}}\\\nonumber 
    &&-\frac{p_4 \phi _2 \phi _3}{4 \phi _1^{3/2}}\biggr)\\
    \label{eq:2_pair_high_ord_con_8}
    D_{p_2}[E] &=& \int {\rm d}x E\left( p_4 \phi _4 p_2'+p_4 \phi _2 p_4'-p_3 p_4 \phi _1'-p_1 p_4 \phi _3' \right)
\end{eqnarray}
for the $p_2$-constraints.

Note that the unadorned $H$ in the $H_{\phi_2}$
constraint is simply a free function and not the Hamiltonian constraint. The
total Hamiltonian is now the sum of ten constraints. Setting each equal to zero
yields the reduced set of four equations constraining our initial conditions:
\begin{eqnarray}
    \label{2_pairs_coneq1}
    0&=&8 p_2^2 \phi _1^2 \phi _2^5+8 p_4^2 \phi _1^2 \phi _2^5+3 p_2^2 \phi
         _3^2 \phi _2^5+32 p_1 p_2 \phi _1^3 \phi _2^4-4 p_1 p_2 \phi _1 \phi
         _3^2 \phi _2^4+16 p_2 p_3 \phi _1^2 \phi _3 \phi _2^4\\\nonumber 
    &&+16 p_2 p_4 \phi _1^2 \phi _4 \phi _2^4-8 \phi _1^2 \phi _2^3 \left(\phi
       _1'\right){}^2-3 \phi _3^2 \phi _2^3 \left(\phi _1'\right){}^2-8 \phi
       _1^2 \phi _2^3 \left(\phi _3'\right){}^2+8 \phi _1 \phi _3 \phi _2^3
       \phi _1' \phi _3'+32 \phi _1^3 \phi _2^2 \phi _1' \phi _2'\\\nonumber 
    &&-4 \phi _1 \phi _3^2 \phi _2^2 \phi _1' \phi _2'+16 \phi _1^2 \phi _3
      \phi _2^2 \phi _2' \phi _3'-8 \phi _1^2 \phi _4^2 \phi _2 \left(\phi
      _1'\right){}^2-64 \phi _1^3 \phi _4 \phi _2 \phi _1' \phi _4'+96 \phi
      _1^3 \phi _4^2 \phi _1' \phi _2'\\\nonumber 
    &&-32 \phi _1^3 \phi _2^3 \phi _1''+4 \phi _1 \phi _3^2 \phi _2^3 \phi
       _1''-16 \phi _1^2 \phi _3 \phi _2^3 \phi _3''-32 \phi _1^3 \phi _4^2
       \phi _2 \phi _1''+8 \phi _1^2 \phi _2^5+3 \phi _3^2 \phi _2^5\\ 
    \label{2_pairs_coneq2}
    0&=&\phi _2 p_2'+\phi _4 p_4'-p_1 \phi _1'-p_3 \phi _3'\\
    \label{2_pairs_coneq3}
    0&=&4 p_2 p_4 \phi _1 \phi _2^4-p_2^2 \phi _3 \phi _2^4+8 p_2 p_3 \phi
         _1^2 \phi _2^3+8 p_1 p_4 \phi _1^2 \phi _2^3+4 p_1 p_2 \phi _1 \phi
         _3 \phi _2^3+2 p_2^2 \phi _1 \phi _4 \phi _2^3\\\nonumber 
    &&+\phi _3 \phi _2^2 \left(\phi _1'\right){}^2-4 \phi _1 \phi _2^2 \phi
       _1' \phi _3'+2 \phi _1 \phi _4 \phi _2 \left(\phi _1'\right){}^2+4 \phi
       _1 \phi _3 \phi _2 \phi _1' \phi _2'+8 \phi _1^2 \phi _2 \phi _2' \phi
       _3'+8 \phi _1^2 \phi _2 \phi _1' \phi _4'\\\nonumber 
    &&-16 \phi _1^2 \phi _4 \phi _1' \phi _2'-4 \phi _1 \phi _3 \phi _2^2 \phi
       _1''-8 \phi _1^2 \phi _2^2 \phi _3''+8 \phi _1^2 \phi _4 \phi _2 \phi
       _1''-\phi _3 \phi _2^4+2 \phi _1 \phi _4 \phi _2^3\\ 
    \label{2_pairs_coneq4}
    0&=&\phi _4 p_2'+\phi _2 p_4'-p_3 \phi _1'-p_1 \phi _3'\, .
\end{eqnarray}
With these ingredients, we can now derive a set of evolution equations for the
system. Due to the increased complexity of the Hamiltonian, these equations
are too long to state in their full form. We direct readers to
Appendix~\ref{section:app_b} to see them explicitly.  With these, we can now
proceed in the same manner as last section. We will simplify these equations
via Taylor expansion and then solve them numerically.

\subsubsection{Simplified Equations}

Noticing that none of the constraints contain spatial derivatives of $p_1$ or
$p_3$, we can simplify the process of finding initial conditions by
eliminating these fields algebraically. Solving (\ref{2_pairs_coneq1}) and
(\ref{2_pairs_coneq3}) for $p_1$ and $p_3$, we obtain
\begin{eqnarray}
    p_1 &=& -\frac{-\phi _2 p_2' \phi _1'-\phi _4 p_4' \phi _1'+\phi _4 p_2'
            \phi _3'+\phi _2 p_4' \phi _3'}{\left(\phi
            _1'\right){}^2-\left(\phi _3'\right){}^2}\\ 
    p_3 &=& -\frac{\phi _4 p_2' \phi _1'+\phi _2 p_4' \phi _1'-\phi _2 p_2'
            \phi _3'-\phi _4 p_4' \phi _3'}{\left(\phi
            _3'\right){}^2-\left(\phi _1'\right){}^2}\, . 
\end{eqnarray}
These give the reduced set of two constraint equations
\begin{eqnarray}
    0&=& 8 p_2^2 \phi _1^2 \phi _2^5+8 p_4^2 \phi _1^2 \phi _2^5+8 \phi _1^2
         \phi _2^5+3 p_2^2 \phi _3^2 \phi _2^5+3 \phi _3^2 \phi _2^5+\frac{32
         p_2 \phi _1^3 p_2' \phi _1' \phi _2^5}{\left(\phi
         _1'\right){}^2-\left(\phi _3'\right){}^2}\\\nonumber 
    &&-\frac{4 p_2 \phi _1 \phi _3^2 p_2' \phi _1' \phi _2^5}{\left(\phi
       _1'\right){}^2-\left(\phi _3'\right){}^2}-\frac{32 p_2 \phi _1^3 p_4'
       \phi _3' \phi _2^5}{\left(\phi _1'\right){}^2-\left(\phi
       _3'\right){}^2}+\frac{4 p_2 \phi _1 \phi _3^2 p_4' \phi _3' \phi
       _2^5}{\left(\phi _1'\right){}^2-\left(\phi _3'\right){}^2}-\frac{16 p_2
       \phi _1^2 \phi _3 p_4' \phi _1' \phi _2^5}{\left(\phi
       _3'\right){}^2-\left(\phi _1'\right){}^2}\\\nonumber 
    &&+\frac{16 p_2 \phi _1^2 \phi _3 p_2' \phi _3' \phi _2^5}{\left(\phi
       _3'\right){}^2-\left(\phi _1'\right){}^2}+16 p_2 p_4 \phi _1^2 \phi _4
       \phi _2^4+\frac{32 p_2 \phi _1^3 \phi _4 p_4' \phi _1' \phi
       _2^4}{\left(\phi _1'\right){}^2-\left(\phi _3'\right){}^2}-\frac{4 p_2
       \phi _1 \phi _3^2 \phi _4 p_4' \phi _1' \phi _2^4}{\left(\phi
       _1'\right){}^2-\left(\phi _3'\right){}^2}\\\nonumber 
    &&-\frac{32 p_2 \phi _1^3 \phi _4 p_2' \phi _3' \phi _2^4}{\left(\phi
       _1'\right){}^2-\left(\phi _3'\right){}^2}+\frac{4 p_2 \phi _1 \phi _3^2
       \phi _4 p_2' \phi _3' \phi _2^4}{\left(\phi _1'\right){}^2-\left(\phi
       _3'\right){}^2}-\frac{16 p_2 \phi _1^2 \phi _3 \phi _4 p_2' \phi _1'
       \phi _2^4}{\left(\phi _3'\right){}^2-\left(\phi
       _1'\right){}^2}+\frac{16 p_2 \phi _1^2 \phi _3 \phi _4 p_4' \phi _3'
       \phi _2^4}{\left(\phi _3'\right){}^2-\left(\phi
       _1'\right){}^2}\\\nonumber 
    &&-8 \phi _1^2 \left(\phi _1'\right){}^2 \phi _2^3-3 \phi _3^2 \left(\phi
       _1'\right){}^2 \phi _2^3-8 \phi _1^2 \left(\phi _3'\right){}^2 \phi
       _2^3+8 \phi _1 \phi _3 \phi _1' \phi _3' \phi _2^3-32 \phi _1^3 \phi
       _1'' \phi _2^3+4 \phi _1 \phi _3^2 \phi _1'' \phi _2^3\\\nonumber 
    &&-16 \phi _1^2 \phi _3 \phi _3'' \phi _2^3+32 \phi _1^3 \phi _1' \phi _2'
      \phi _2^2-4 \phi _1 \phi _3^2 \phi _1' \phi _2' \phi _2^2+16 \phi _1^2
      \phi _3 \phi _2' \phi _3' \phi _2^2-8 \phi _1^2 \phi _4^2 \left(\phi
      _1'\right){}^2 \phi _2\\\nonumber 
    &&-64 \phi _1^3 \phi _4 \phi _1' \phi _4' \phi _2-32 \phi _1^3 \phi _4^2
       \phi _1'' \phi _2+96 \phi _1^3 \phi _4^2 \phi _1' \phi _2'\\ 
    0&=& 4 p_2 p_4 \phi _1 \phi _2^4-p_2^2 \phi _3 \phi _2^4-\phi _3 \phi
         _2^4+\frac{8 p_4 \phi _1^2 p_2' \phi _1' \phi _2^4}{\left(\phi
         _1'\right){}^2-\left(\phi _3'\right){}^2}+\frac{4 p_2 \phi _1 \phi _3
         p_2' \phi _1' \phi _2^4}{\left(\phi _1'\right){}^2-\left(\phi
         _3'\right){}^2}-\frac{8 p_4 \phi _1^2 p_4' \phi _3' \phi
         _2^4}{\left(\phi _1'\right){}^2-\left(\phi _3'\right){}^2}\\\nonumber 
    &&-\frac{4 p_2 \phi _1 \phi _3 p_4' \phi _3' \phi _2^4}{\left(\phi
       _1'\right){}^2-\left(\phi _3'\right){}^2}-\frac{8 p_2 \phi _1^2 p_4'
       \phi _1' \phi _2^4}{\left(\phi _3'\right){}^2-\left(\phi
       _1'\right){}^2}+\frac{8 p_2 \phi _1^2 p_2' \phi _3' \phi
       _2^4}{\left(\phi _3'\right){}^2-\left(\phi _1'\right){}^2}+2 p_2^2 \phi
       _1 \phi _4 \phi _2^3+2 \phi _1 \phi _4 \phi _2^3\\\nonumber 
    &&+\frac{8 p_4 \phi _1^2 \phi _4 p_4' \phi _1' \phi _2^3}{\left(\phi
       _1'\right){}^2-\left(\phi _3'\right){}^2}+\frac{4 p_2 \phi _1 \phi _3
       \phi _4 p_4' \phi _1' \phi _2^3}{\left(\phi _1'\right){}^2-\left(\phi
       _3'\right){}^2}-\frac{8 p_4 \phi _1^2 \phi _4 p_2' \phi _3' \phi
       _2^3}{\left(\phi _1'\right){}^2-\left(\phi _3'\right){}^2}-\frac{4 p_2
       \phi _1 \phi _3 \phi _4 p_2' \phi _3' \phi _2^3}{\left(\phi
       _1'\right){}^2-\left(\phi _3'\right){}^2}\\\nonumber 
    &&-\frac{8 p_2 \phi _1^2 \phi _4 p_2' \phi _1' \phi _2^3}{\left(\phi
       _3'\right){}^2-\left(\phi _1'\right){}^2}+\frac{8 p_2 \phi _1^2 \phi _4
       p_4' \phi _3' \phi _2^3}{\left(\phi _3'\right){}^2-\left(\phi
       _1'\right){}^2}+\phi _3 \left(\phi _1'\right){}^2 \phi _2^2-4 \phi _1
       \phi _1' \phi _3' \phi _2^2-4 \phi _1 \phi _3 \phi _1'' \phi _2^2-8
       \phi _1^2 \phi _3'' \phi _2^2\\\nonumber 
    &&+2 \phi _1 \phi _4 \left(\phi _1'\right){}^2 \phi _2+4 \phi _1 \phi _3
      \phi _1' \phi _2' \phi _2+8 \phi _1^2 \phi _2' \phi _3' \phi _2+8 \phi
      _1^2 \phi _1' \phi _4' \phi _2+8 \phi _1^2 \phi _4 \phi _1'' \phi _2-16
      \phi _1^2 \phi _4 \phi _1' \phi _2'\, . 
\end{eqnarray}
As before, we assume the classical fields are given by perturbations on a FRW
background. We then simplify our constraints and evolution equations by
performing a Taylor expansion in all small quantities and removing
terms of greater than second order. This gives the new set of constraint
equations
\begin{eqnarray}
    0&=& -32 p_4^2 a_0{}^4 t^3 x^3+16 a_0{}^3 t^{3/2} x^4 \phi _4 p_4'-16
         a_0{}^3 t^{3/2} x^4 p_4' \phi _3'+16 a_0{}^3 t^{3/2} x^3 \phi _3
         p_4'\\\nonumber 
    &&+16 p_4 a_0{}^3 t^{3/2} x^3 \phi _4+32 a_0{}^5
       t^{5/2} x^5 {\delta p}_2'+64 {\delta p}_2 a_0{}^5 t^{5/2} x^4-64
       {\delta p}_2 a_0{}^4 t^3 x^4 {\delta p}_2'\\\nonumber 
    &&-32 {\delta p}_2^2 a_0{}^4 t^3 x^3+8 a_0{}^4 t x^4 \delta \phi _1'-60
       \delta \phi _2 a_0{}^4 t x^4-64 \delta \phi _1 a_0{}^4 t x^3+80 \delta
       \phi _2 a_0{}^3 t^{3/2} x^4 {\delta p}_2'\\\nonumber 
    &&-16 a_0{}^3 t^{3/2} x^4 {\delta p}_2' \delta \phi _1'+96 \delta \phi _1
       a_0{}^3 t^{3/2} x^3 {\delta p}_2'-16 {\delta p}_2 a_0{}^3 t^{3/2} x^3
       \delta \phi _1'\\\nonumber 
    &&+160 {\delta p}_2 \delta \phi _2 a_0{}^3 t^{3/2} x^3+160 {\delta p}_2
       \delta \phi _1 a_0{}^3 t^{3/2} x^2-32 a_0{}^2 t^2 x^3 \delta \phi
       _2'+32 a_0{}^2 t^2 x^3 \delta \phi _1''\\\nonumber 
    &&+20 \delta \phi _2 a_0{}^2 x^3 \delta \phi _1'-4 a_0{}^2 x^3 \left(\phi
       _3'\right){}^2+4 a_0{}^2 x^3 \phi _4 \phi _3'+24 \delta \phi _1 a_0{}^2
       x^2 \delta \phi _1'-160 \delta \phi _1 \delta \phi _2 a_0{}^2
       x^2\\\nonumber 
    &&+4 a_0{}^2 x^2 \phi _3 \phi _3'-4 a_0{}^2 x^2 \phi _3 \phi _4-a_0{}^2 x
       \phi _3^2-32 \delta \phi _1 t \delta \phi _1'+32 \delta \phi _1 \delta
       \phi _2 t-32 \delta \phi _2 t x^2 \delta \phi _2'\\\nonumber 
    &&-16 t x^2 \delta \phi _1' \delta \phi _2'+48 \delta \phi _2 t x^2 \delta
       \phi _1''+32 t x^2 \phi _4 \phi _4'+16 \delta \phi _2 t x \delta \phi
       _1'-96 \delta \phi _1 t x \delta \phi _2'+96 \delta \phi _1 t x \delta
       \phi _1''\\\nonumber 
    &&+8 t x \left(\phi _3'\right){}^2+16 t x \phi _3 \phi _3''-24 t x \phi _4^2-32 t \phi _3 \phi _3'\\
    0&=&48 a_0{}^3 x^2 p_4 \delta \phi _1 t^{3/2}+64 a_0{}^3 x^3 p_4 \delta
         \phi _2 t^{3/2}+16 a_0{}^3 x^3 {\delta p}_2 \phi _4 t^{3/2}+32
         a_0{}^3 x^3 \delta \phi _1 p_4' t^{3/2}\\\nonumber 
    &&+32 a_0{}^3 x^4 \delta \phi _2 p_4' t^{3/2}+8 a_0{}^3 x^3 \phi _3
       {\delta p}_2' t^{3/2}+8 a_0{}^3 x^4 \phi _4 {\delta p}_2' t^{3/2}-8
       a_0{}^3 x^3 p_4 \delta \phi _1' t^{3/2}\\\nonumber 
    &&-8 a_0{}^3 x^4 p_4' \delta \phi _1' t^{3/2}-8 a_0{}^3 x^3 {\delta p}_2
       \phi _3' t^{3/2}-8 a_0{}^3 x^4 {\delta p}_2' \phi _3' t^{3/2}+32
       a_0{}^5 x^4 p_4 t^{5/2}+16 a_0{}^5 x^5 p_4' t^{5/2}\\\nonumber 
    &&-32 a_0{}^4 x^3 p_4 {\delta p}_2 t^3-32 a_0{}^4 x^4 {\delta p}_2 p_4'
       t^3-32 a_0{}^4 x^4 p_4 {\delta p}_2' t^3-16 a_0{}^2 x^3 \phi _4' t^2+16
       a_0{}^2 x^3 \phi _3'' t^2\\\nonumber 
    &&-2 a_0{}^4 x^3 \phi _3 t+16 \delta \phi _2 \phi _3 t-6 a_0{}^4 x^4 \phi
       _4 t+16 \delta \phi _1 \phi _4 t-24 x \delta \phi _2 \phi _4 t-16 \phi
       _3 \delta \phi _1' t+8 x \phi _4 \delta \phi _1' t\\\nonumber 
    &&-8 x \phi _3 \delta \phi _2' t+16 x^2 \phi _4 \delta \phi _2' t+4
       a_0{}^4 x^4 \phi _3' t-16 \delta \phi _1 \phi _3' t+8 x \delta \phi _2
       \phi _3' t+8 x \delta \phi _1' \phi _3' t-8 x^2 \delta \phi _2' \phi
       _3' t\\\nonumber 
    &&-32 x \delta \phi _1 \phi _4' t-8 x^2 \delta \phi _2 \phi _4' t-8 x^2
       \delta \phi _1' \phi _4' t+8 x \phi _3 \delta \phi _1'' t-8 x^2 \phi _4
       \delta \phi _1'' t+32 x \delta \phi _1 \phi _3'' t+16 x^2 \delta \phi
       _2 \phi _3'' t\\\nonumber 
    &&-4 a_0{}^2 x \delta \phi _1 \phi _3-4 a_0{}^2 x^2 \delta \phi _2 \phi
       _3-10 a_0{}^2 x^2 \delta \phi _1 \phi _4-9 a_0{}^2 x^3 \delta \phi _2
       \phi _4+2 a_0{}^2 x^2 \phi _3 \delta \phi _1'+2 a_0{}^2 x^3 \phi _4
       \delta \phi _1'\\\nonumber 
    &&+8 a_0{}^2 x^2 \delta \phi _1 \phi _3'+8 a_0{}^2 x^3 \delta \phi _2 \phi
       _3'-4 a_0{}^2 x^3 \delta \phi _1' \phi _3'\,  
\end{eqnarray}
and the evolution equations
\begin{eqnarray}
    \delta \dot{\phi}_1&=&-\frac{\phi _3^2}{8 a_0{}^2 t^2 x^2}-\frac{{\delta
                            p}_2 \delta \phi _1}{a_0 \sqrt{t} x}-2 {\delta
                            p}_2 a_0 \sqrt{t} x+\frac{\delta \phi _1}{2 t}\\ 
    \delta \dot{p}_1&=&\frac{p_4 \phi _4}{4 a_0{}^2 t^2 x^2}+\frac{p_3 \phi
                          _3}{4 a_0{}^2 t^2 x^2}-\frac{p_4^2}{2 a_0 \sqrt{t}
                          x^2}-\frac{\delta \phi _1 \delta \phi _1'}{a_0{}^5
                          t^{5/2} x^5}+\frac{\delta \phi _1 \delta \phi
                          _2}{a_0{}^5 t^{5/2} x^5}-\frac{\phi _3 \phi
                          _3'}{a_0{}^5 t^{5/2} x^5}\\\nonumber 
    &&-\frac{3 \delta \phi _2 \delta \phi _1'}{4 a_0{}^5 t^{5/2}
       x^4}-\frac{\delta \phi _1 \delta \phi _2'}{4 a_0{}^5 t^{5/2}
       x^4}+\frac{\delta \phi _1 \delta \phi _1''}{4 a_0{}^5 t^{5/2}
       x^4}+\frac{\left(\phi _3'\right){}^2}{8 a_0{}^5 t^{5/2} x^4}+\frac{\phi
       _3 \phi _3''}{4 a_0{}^5 t^{5/2} x^4}+\frac{5 \phi _4^2}{8 a_0{}^5
       t^{5/2} x^4}\\\nonumber 
    &&+\frac{\delta \phi _1' \delta \phi _2'}{4 a_0{}^5 t^{5/2}
       x^3}-\frac{\delta \phi _2 \delta \phi _2'}{2 a_0{}^5 t^{5/2}
       x^3}+\frac{\delta \phi _2 \delta \phi _1''}{4 a_0{}^5 t^{5/2}
       x^3}-\frac{\phi _4 \phi _4'}{2 a_0{}^5 t^{5/2} x^3}-\frac{9 \phi
       _3^2}{64 a_0{}^3 t^{7/2} x^4}+\frac{\delta \phi _1'}{a_0{}^3 t^{3/2}
       x^3}\\\nonumber 
    &&+\frac{3 \delta \phi _1 \delta \phi _2}{32 a_0{}^3 t^{7/2}
       x^3}-\frac{\delta \phi _2}{a_0{}^3 t^{3/2} x^3}+\frac{\delta \phi
       _2'}{2 a_0{}^3 t^{3/2} x^2}-\frac{\delta \phi _1''}{2 a_0{}^3 t^{3/2}
       x^2}-\frac{{\delta p}_2 \delta \phi _1}{2 a_0{}^2 t^2
       x^3}+\frac{{\delta p}_1 \delta \phi _1}{4 a_0{}^2 t^2 x^2}\\\nonumber 
    &&+\frac{{\delta p}_2 \delta \phi _2}{4 a_0{}^2 t^2 x^2}+\frac{\delta \phi
       _1}{16 a_0 t^{5/2} x^2}-\frac{\delta \phi _2}{16 a_0 t^{5/2}
       x}-\frac{a_0}{8 t^{3/2}}+\frac{{\delta p}_1 {\delta p}_2}{a_0 \sqrt{t}
       x}-\frac{{\delta p}_1}{2 t}\\ 
    \delta \dot{\phi}_2&=&-\frac{p_3 \phi _3}{a_0 \sqrt{t} x}-\frac{p_4 \phi
                            _4}{a_0 \sqrt{t} x}+\frac{\phi _3^2}{4 a_0{}^2 t^2
                            x^3}-\frac{\delta \phi _1 \delta \phi _2}{4
                            a_0{}^2 t^2 x^2}-2 {\delta p}_2 a_0
                            \sqrt{t}+\frac{{\delta p}_2 \delta \phi _1}{a_0
                            \sqrt{t} x^2}-\frac{{\delta p}_1 \delta \phi
                            _1}{a_0 \sqrt{t} x}\\\nonumber 
    &&-\frac{{\delta p}_2 \delta \phi _2}{a_0 \sqrt{t} x}-2 {\delta p}_1 a_0 \sqrt{t} x+\frac{\delta \phi _2}{2 t}\\
    \delta \dot{p}_2&=&\frac{p_4^2}{2 a_0 \sqrt{t} x}+\frac{\delta \phi _1
                          \delta \phi _1'}{4 a_0{}^5 t^{5/2} x^4}-\frac{\delta
                          \phi _1 \delta \phi _2}{4 a_0{}^5 t^{5/2}
                          x^4}+\frac{\phi _3 \phi _3'}{4 a_0{}^5 t^{5/2}
                          x^4}+\frac{\delta \phi _2 \delta \phi _1'}{2 a_0{}^5
                          t^{5/2} x^3}-\frac{\left(\phi _3'\right){}^2}{8
                          a_0{}^5 t^{5/2} x^3}\\\nonumber 
    &&-\frac{3 \phi _4^2}{8 a_0{}^5 t^{5/2} x^3}+\frac{3 \phi _3^2}{64 a_0{}^3
       t^{7/2} x^3}-\frac{\delta \phi _1'}{2 a_0{}^3 t^{3/2} x^2}+\frac{\delta
       \phi _2}{2 a_0{}^3 t^{3/2} x^2}+\frac{{\delta p}_2 \delta \phi _1}{4
       a_0{}^2 t^2 x^2}-\frac{\delta \phi _1}{16 a_0 t^{5/2} x}-\frac{a_0 x}{8
       t^{3/2}}\\\nonumber 
    &&-\frac{{\delta p}_2}{2 t}
\end{eqnarray}
for perturbations and
\begin{eqnarray}
    \dot{\phi}_3&=&-2 {E} a_0{}^2 t x \phi _4+F a_0{}^2 x^2 \phi _4-2 G
                     a_0{}^2 t x \phi _4+H a_0{}^2 x^2 \phi _4+2 {I} a_0{}^2 t
                     x p_4'-4 {I} p_3 a_0{}^2 t x\\\nonumber 
    &&+\frac{{I} a_0 \phi _3'}{2 \sqrt{t}}-\frac{{I} a_0 \phi _4}{2
       \sqrt{t}}+2 J p_3 a_0{}^2 x^2+2 J p_4 a_0{}^2 x-\frac{J a_0 x \phi
       _4}{8 t^{3/2}}-\frac{J a_0 \phi _3}{8 t^{3/2}}-\frac{J \phi _4'}{a_0
       \sqrt{t}}+\frac{J \phi _3''}{a_0 \sqrt{t}}\\\nonumber 
    &&-2 {K} a_0{}^2 t x \phi _3+L a_0{}^2 x^2 \phi _3-\frac{\delta \phi _1
       \phi _3}{4 a_0{}^2 t^2 x^2}-\frac{{\delta p}_2 \phi _3}{a_0 \sqrt{t}
       x}+\frac{\phi _3}{2 t}\\ 
    \dot{p}_3&=&-L x^2 p_3 a_0{}^2+2 {K} t x p_3 a_0{}^2-2 L x p_4 a_0{}^2-2
                 {K} t x p_4' a_0{}^2+\frac{J p_3 a_0}{8 t^{3/2}}+\frac{L \phi
                 _3 a_0}{4 t^{3/2}}\\\nonumber 
    &&+\frac{L x \phi _4 a_0}{8 t^{3/2}}+\frac{F \phi _4 a_0}{8
       t^{3/2}}+\frac{H \phi _4 a_0}{8 t^{3/2}}+\frac{{K} \phi _4 a_0}{2
       \sqrt{t}}+\frac{\phi _4 {E}' a_0}{2 \sqrt{t}}+\frac{p_3 {I}' a_0}{2
       \sqrt{t}}+\frac{\phi _3 {K}' a_0}{2 \sqrt{t}}+\frac{\phi _4 G' a_0}{2
       \sqrt{t}}\\\nonumber 
    &&+\frac{{I} p_3' a_0}{2 \sqrt{t}}+\frac{{E} \phi _4' a_0}{2
       \sqrt{t}}+\frac{G \phi _4' a_0}{2 \sqrt{t}}+\frac{p_3 {\delta p}_2}{a_0
       \sqrt{t} x}+\frac{p_3 \delta \phi _1}{4 a_0{}^2 t^2 x^2}+\frac{{\delta
       p}_1 \phi _3}{4 a_0{}^2 t^2 x^2}+\frac{3 \delta \phi _2 \phi _3}{32
       a_0{}^3 t^{7/2} x^3}\\\nonumber 
    &&+\frac{\delta \phi _2 \phi _3}{a_0{}^5 t^{5/2} x^5}+\frac{\phi _3}{16
       a_0 t^{5/2} x^2}+\frac{\delta \phi _1' \phi _3'}{4 a_0{}^5 t^{5/2}
       x^4}+\frac{\delta \phi _2' \phi _3'}{4 a_0{}^5 t^{5/2} x^3}+\frac{\phi
       _3'}{a_0{}^3 t^{3/2} x^3}+\frac{L \phi _4'}{a_0 \sqrt{t}}+\frac{\phi _3
       \delta \phi _1''}{4 a_0{}^5 t^{5/2} x^4}\\\nonumber 
    &&+\frac{\delta \phi _1 \phi _3''}{4 a_0{}^5 t^{5/2} x^4}+\frac{\delta
       \phi _2 \phi _3''}{4 a_0{}^5 t^{5/2} x^3}-\frac{p_3}{2 t}-\frac{\phi
       _3''}{2 a_0{}^3 t^{3/2} x^2}-\frac{{\delta p}_2 \phi _3}{2 a_0{}^2 t^2
       x^3}-\frac{\phi _3 \delta \phi _2'}{4 a_0{}^5 t^{5/2} x^4}-\frac{3
       \delta \phi _2 \phi _3'}{4 a_0{}^5 t^{5/2} x^4}\\\nonumber 
    &&-\frac{9 \delta \phi _1 \phi _3}{32 a_0{}^3 t^{7/2} x^4}-\frac{\phi _3
       \delta \phi _1'}{a_0{}^5 t^{5/2} x^5}-\frac{\delta \phi _1 \phi
       _3'}{a_0{}^5 t^{5/2} x^5}-\frac{2 J' p_3'}{a_0 \sqrt{t}}-\frac{2 L'
       \phi _3'}{a_0 \sqrt{t}}-\frac{2 F' \phi _4'}{a_0 \sqrt{t}}-\frac{2 H'
       \phi _4'}{a_0 \sqrt{t}}-\frac{\phi _4 F''}{a_0 \sqrt{t}}\\\nonumber 
    &&-\frac{\phi _4 H''}{a_0 \sqrt{t}}-\frac{p_3 J''}{a_0
       \sqrt{t}}-\frac{\phi _3 L''}{a_0 \sqrt{t}}-\frac{2 L \phi _3''}{a_0
       \sqrt{t}}-\frac{F \phi _4''}{a_0 \sqrt{t}}-\frac{H \phi _4''}{a_0
       \sqrt{t}}-\frac{J {p_3}''}{a_0 \sqrt{t}}\\ 
    \dot{\phi}_4&=&-2 a_0{}^2 t x \phi _4 {E}'-2 {E} a_0{}^2 t x \phi _4'-2
                     {E} a_0{}^2 t \phi _4+2 F a_0{}^2 x \phi _4-2 a_0{}^2 t x
                     \phi _4 G'\\\nonumber 
    &&-2 G a_0{}^2 t x \phi _4'-2 G a_0{}^2 t \phi _4+2 H a_0{}^2 x \phi _4-2
       p_3 a_0{}^2 t x {I}'-2 {I} a_0{}^2 t x p_3'-2 {I} p_3 a_0{}^2
       t\\\nonumber 
    &&+2 J p_3 a_0{}^2 x-2 a_0{}^2 t x \phi _3 {K}'-2 {K} a_0{}^2 t x \phi
       _3'-2 {K} a_0{}^2 t \phi _3+2 L a_0{}^2 x \phi _3+\frac{\delta \phi _1
       p_4}{a_0 \sqrt{t} x^2}-\frac{\delta \phi _2 p_4}{a_0 \sqrt{t}
       x}\\\nonumber 
    &&-2 p_4 a_0 \sqrt{t}-\frac{\delta \phi _1 \phi _4}{4 a_0{}^2 t^2
       x^2}-\frac{{\delta p}_2 \phi _4}{a_0 \sqrt{t} x}+\frac{\phi _4}{2 t}\\ 
    \dot{p}_4&=&-2 {E} a_0{}^2 t x p_4'+2 {E} p_3 a_0{}^2 t x-\frac{{E} a_0
                 \phi _3'}{2 \sqrt{t}}+\frac{{E} a_0 \phi
                 _4}{\sqrt{t}}-\frac{\phi _4 F'}{a_0 \sqrt{t}}-F p_3 a_0{}^2
                 x^2\\\nonumber 
    &&-2 F p_4 a_0{}^2 x+\frac{F a_0 x \phi _4}{4 t^{3/2}}+\frac{F a_0 \phi
       _3}{8 t^{3/2}}-\frac{F \phi _3''}{a_0 \sqrt{t}}-2 G a_0{}^2 t x p_4'+2
       G p_3 a_0{}^2 t x-\frac{G a_0 \phi _3'}{2 \sqrt{t}}\\\nonumber 
    &&+\frac{G a_0 \phi _4}{\sqrt{t}}-\frac{\phi _4 H'}{a_0 \sqrt{t}}-H p_3
       a_0{}^2 x^2-2 H p_4 a_0{}^2 x+\frac{H a_0 x \phi _4}{4 t^{3/2}}+\frac{H
       a_0 \phi _3}{8 t^{3/2}}-\frac{H \phi _3''}{a_0 \sqrt{t}}+\frac{{I} p_3
       a_0}{2 \sqrt{t}}\\\nonumber 
    &&-\frac{p_3 J'}{a_0 \sqrt{t}}+\frac{J p_3 a_0 x}{8 t^{3/2}}-\frac{J
       p_3'}{a_0 \sqrt{t}}+\frac{{K} a_0 \phi _3}{2 \sqrt{t}}-\frac{\phi _3
       L'}{a_0 \sqrt{t}}+\frac{L a_0 x \phi _3}{8 t^{3/2}}-\frac{L \phi
       _3'}{a_0 \sqrt{t}}+\frac{\delta \phi _1 p_4}{4 a_0{}^2 t^2
       x^2}+\frac{{\delta p}_2 p_4}{a_0 \sqrt{t} x}\\\nonumber 
    &&-\frac{p_4}{2 t}-\frac{\delta \phi _1 \phi _4}{4 a_0{}^5 t^{5/2}
       x^4}+\frac{\phi _4 \delta \phi _1'}{2 a_0{}^5 t^{5/2} x^3}-\frac{3
       \delta \phi _2 \phi _4}{4 a_0{}^5 t^{5/2} x^3}+\frac{\phi _4}{2 a_0{}^3
       t^{3/2} x^2} 
\end{eqnarray}
for fluctuations.

\begin{figure}
    \centering
    \includegraphics{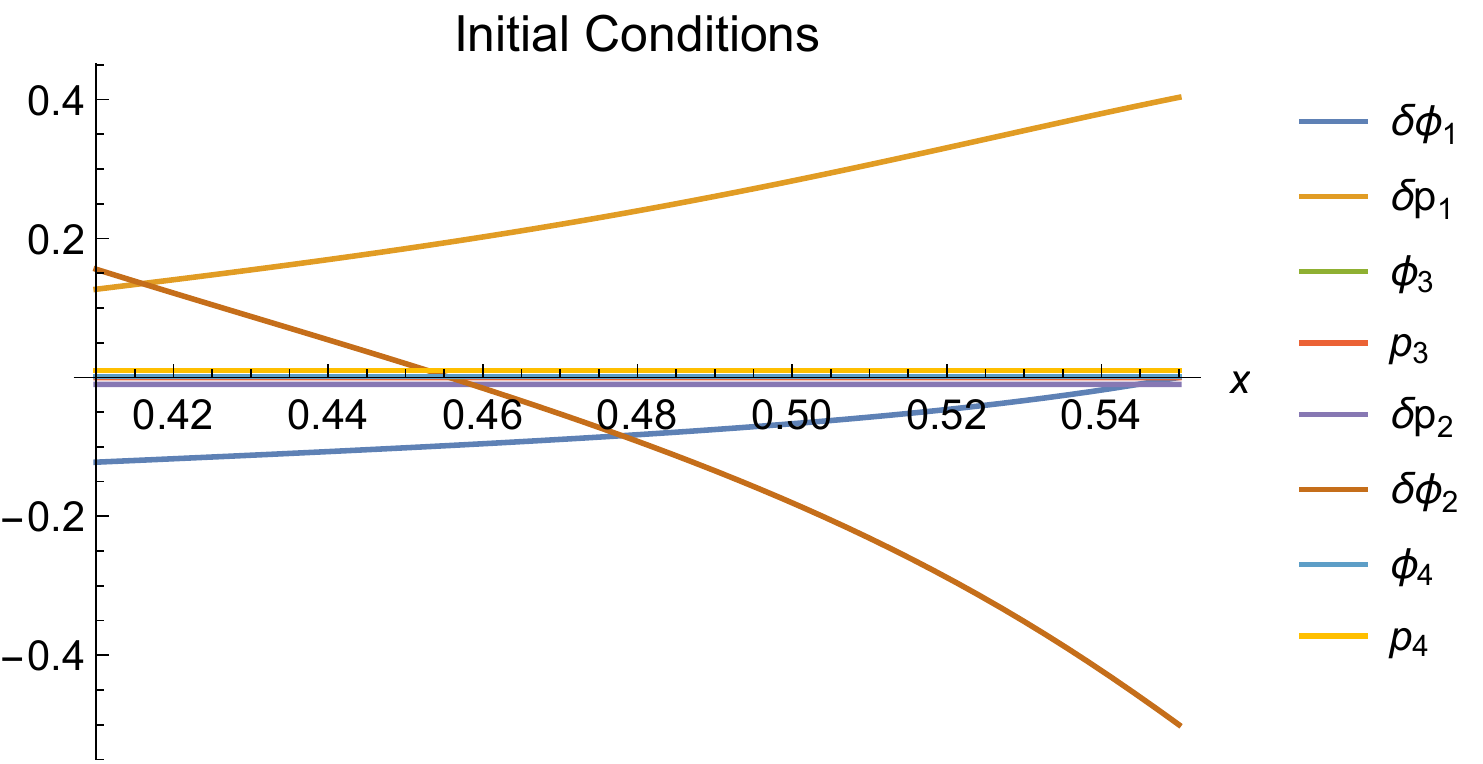}
    \caption[Initial conditions for the second model]{Plot of the initial conditions used. }
    \label{fig:two_pair_init_conds}
\end{figure}

\begin{figure}
    \centering
    \includegraphics[scale=0.7]{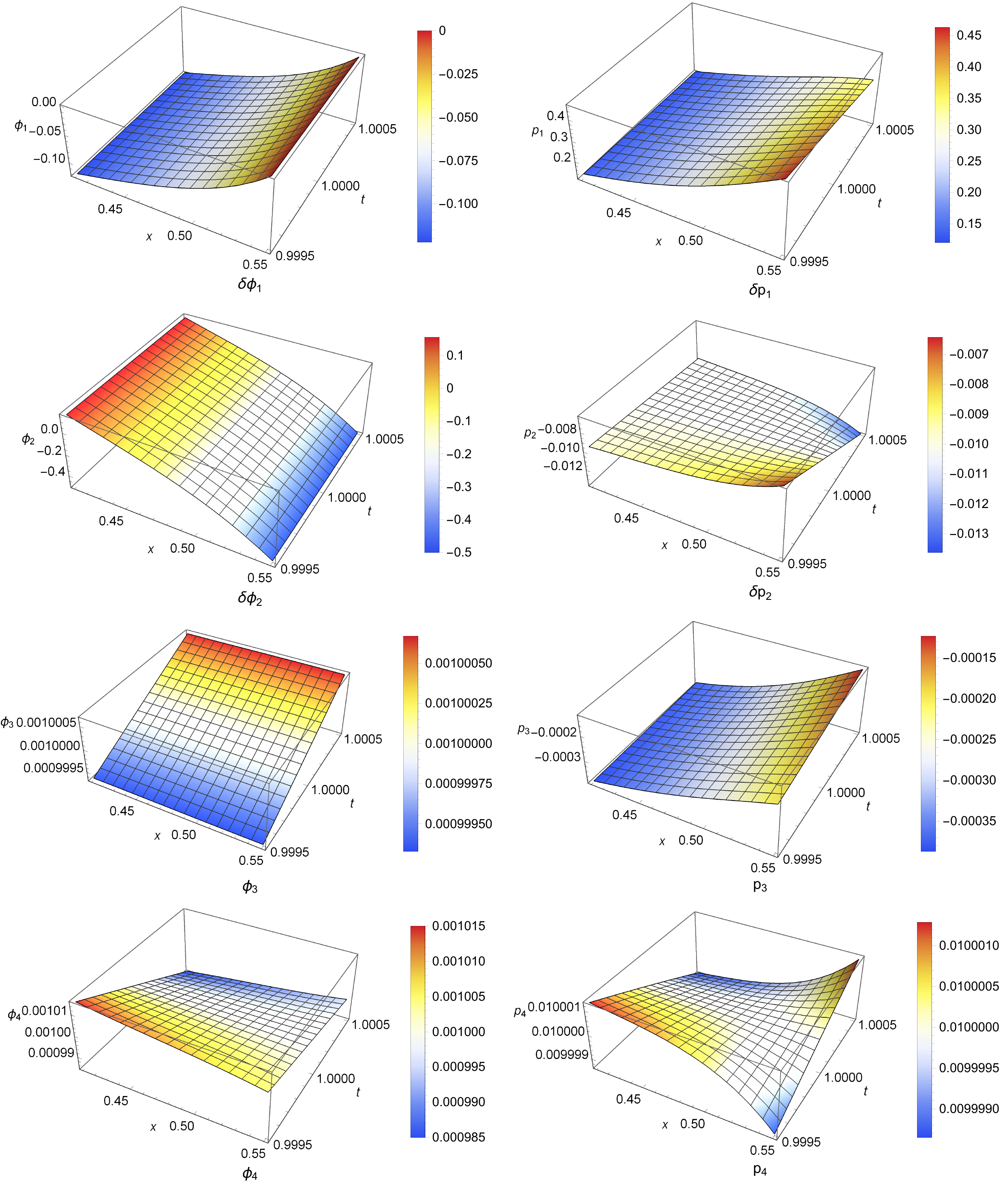}
    \caption[Solutions for the second model]{Plots of the solutions found
      using the initial conditions in Figure \ref{fig:two_pair_init_conds}
      imposed at $t=1$.}
    \label{fig:two_pair_solns}
\end{figure}

\begin{figure}
    \centering
    \includegraphics[scale=0.7]{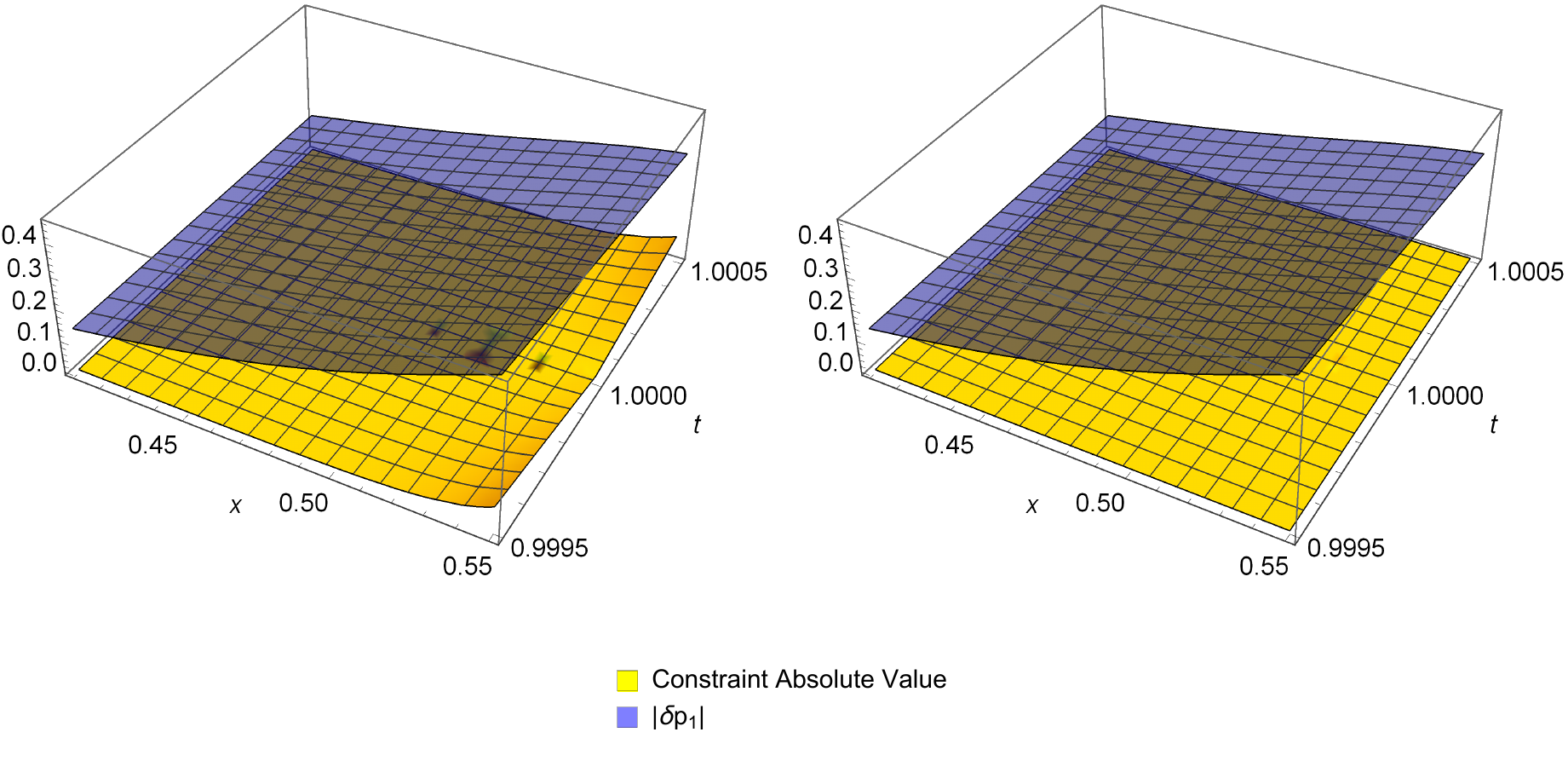}
    \caption[Non-satisfaction of second model's constraints]{Each constraint's
      absolute divergence from zero plotted alongside the absolute value of
      $\delta p_1$, as in Figure \ref{fig:one_pair_cons_test}. Owing to
      numerical instabilities, the available time range here is much smaller
      than in Figure \ref{fig:one_pair_cons_test}.
    \label{fig:two_pair_cons_test}}
\end{figure}

\begin{figure}
    \centering
    \includegraphics[scale=0.7]{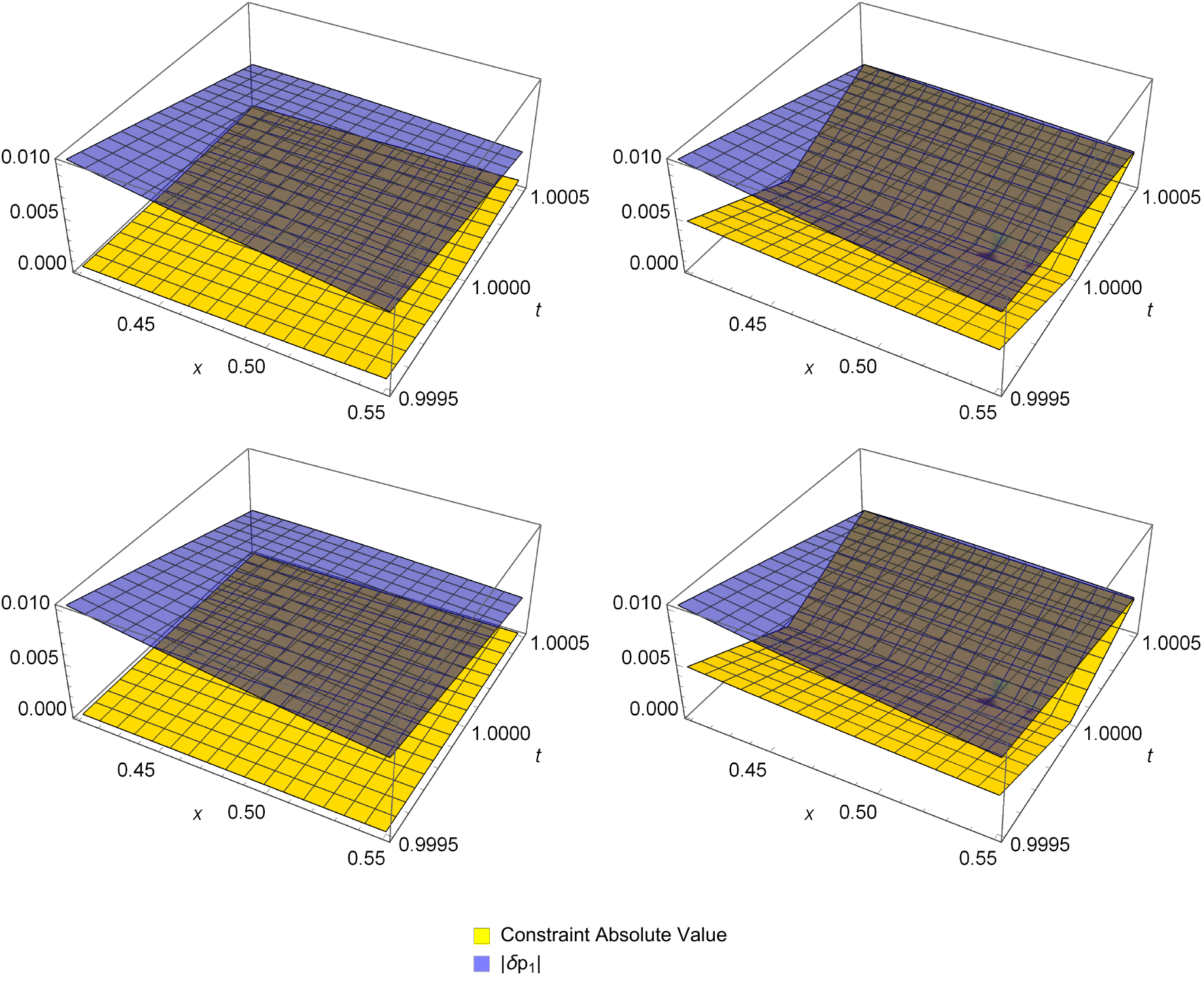}
    \caption[Non-satisfaction of second model's constraints]{Test of
      constraint preservation as in Figure \ref{fig:one_pair_cons_test}, but
      plotted over the ranges of Figure \ref{fig:two_pair_cons_test} to
      facilitate comparisons. The one-pair constraints shown here change more
      quickly than those in the two-pair case of Figure
      \ref{fig:two_pair_cons_test}, relative to the field values.
    \label{fig:one_pair_cons_test_small_region}}
\end{figure}

\subsubsection{Results}

We can now proceed with solving this system numerically. Both the constraint
and evolution equation systems proved highly stiff, making the solution
procedure require considerable fine-tuning. The initial conditions picked are
shown in Fig.~\ref{fig:two_pair_init_conds}. Most choices for the free fields caused
the others to diverge to large values in the initial conditions, but we were
able to find ones that kept all below a magnitude of one for a reasonably
sized region. As the background solutions are of order unity in this region,
this is still considerably larger than we would like the initial perturbations to
be. However, for the purpose of our consistency check, these should suffice.

Imposing these initial conditions at $t=1$, we found the solutions given in
Fig.~\ref{fig:two_pair_solns}. For all initial conditions we tried, the
numerical solutions broke down shortly before and after $t=1$. However,
limiting them to a narrow time region, we found a set of well-behaved
solutions. Even with the large initial conditions, none of the fields showed
exponentially diverging behavior or other instabilities.

The divergences of the constraints from zero for these solutions are shown in
Fig.~\ref{fig:two_pair_cons_test}. Unlike the previous model, the constraints
have a clear region in which they approximately vanish in comparison with the
$\delta p_1$ field. Even though the time range possible here for numerical
reasons is much smaller than in Fig.~\ref{fig:one_pair_cons_test}, it is clear
that the constraints change much more slowly than would be required for
reaching the magnitude of $\delta p_1$, in contrast to the previous
case. Since we used a perturbative treatment, we can only expect the
constraints to be approximately preserved, with smaller perturbations leading
to better preservation. Thus, the large magnitude of the perturbations in this
solution make the slight divergences of the constraints from zero here
acceptable. In addition, since we used a numerical treatment of a stiff
system, numerical error could account for some additional divergence. Both of
these issues are ones that could be treated with additional fine-tuning of the
initial conditions to make all perturbations small. This would hopefully both
assist in preserving the constraints and increase the size of the region on
which reasonable numerical solutions could be found.

\subsection{Non-zero zero-point parameters $U$}

\begin{figure}
    \centering
    \includegraphics[scale=0.7]{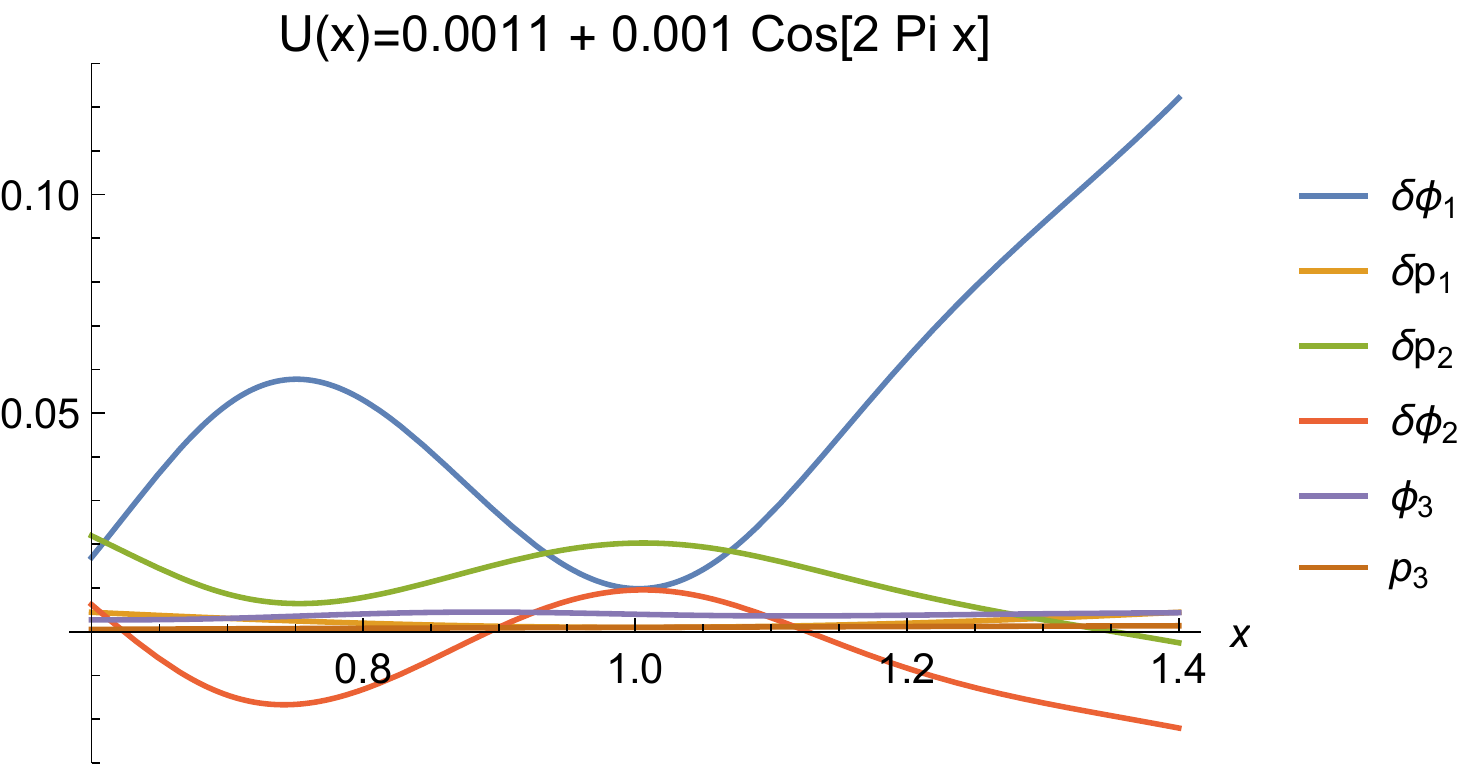}
    \includegraphics[scale=0.7]{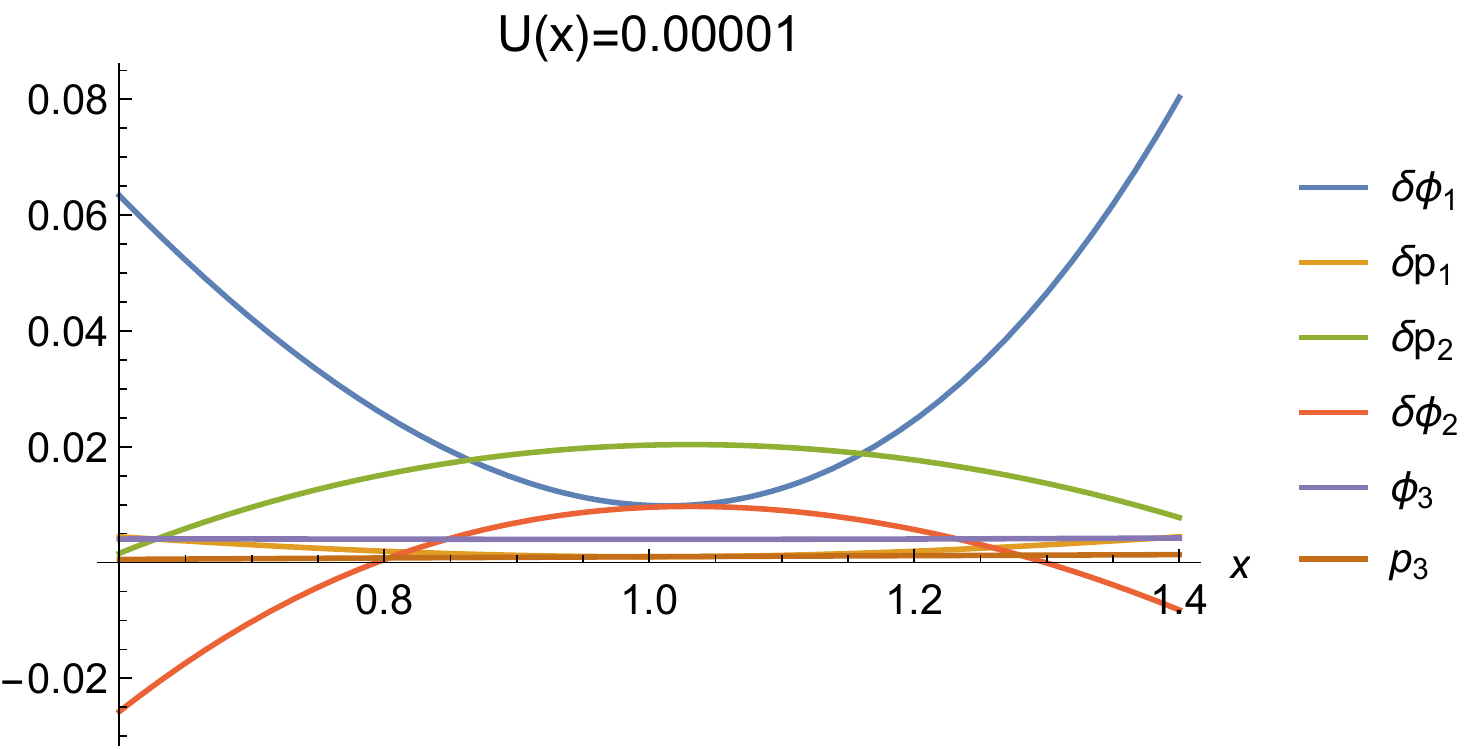}
    \includegraphics[scale=0.7]{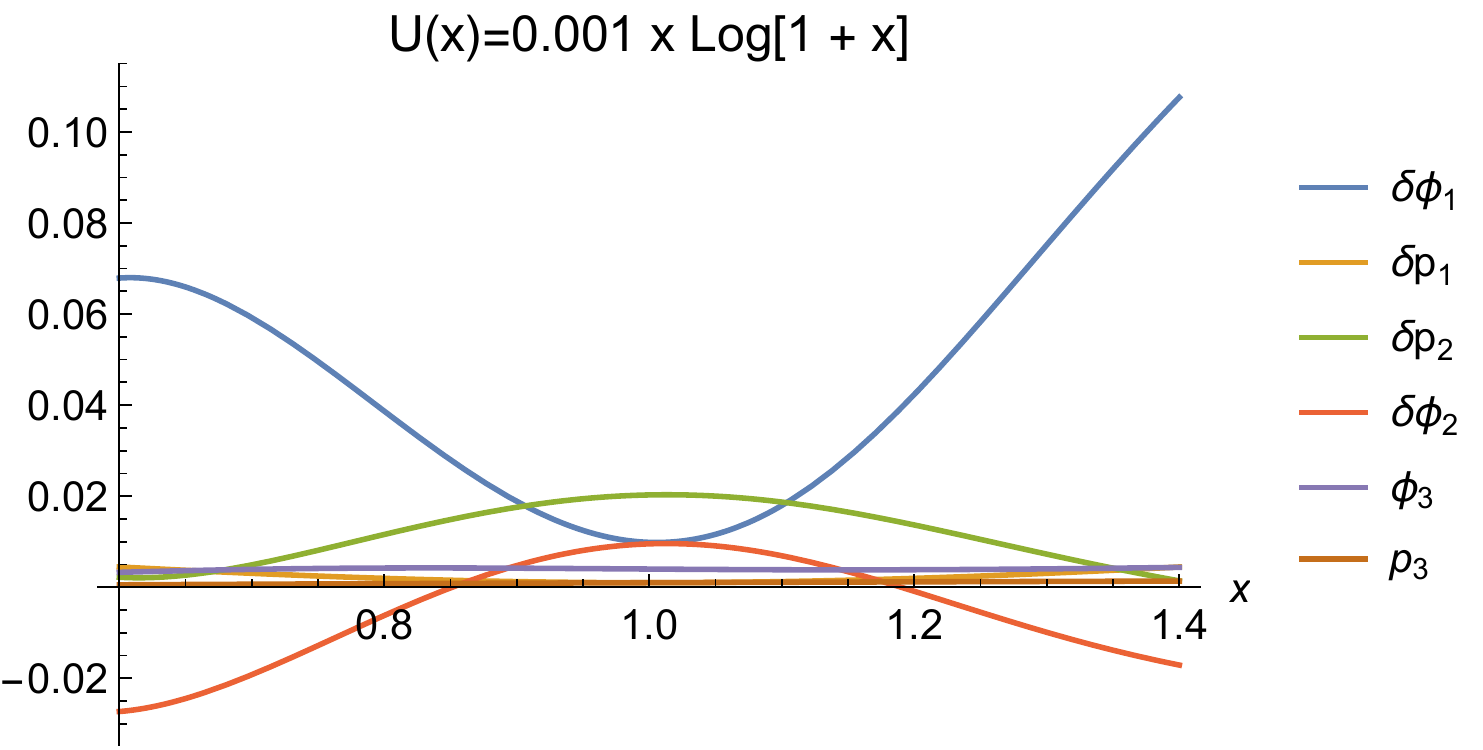}
    \caption[Non-satisfaction of second model's constraints]{Three different
      choices of $U(x)$ in the one-pair model and their implications on
      initial values satisfying the constraints.
    \label{fig:one_pair_init_conds_123}}
\end{figure}

\begin{figure}
    \centering
    \includegraphics[scale=0.7]{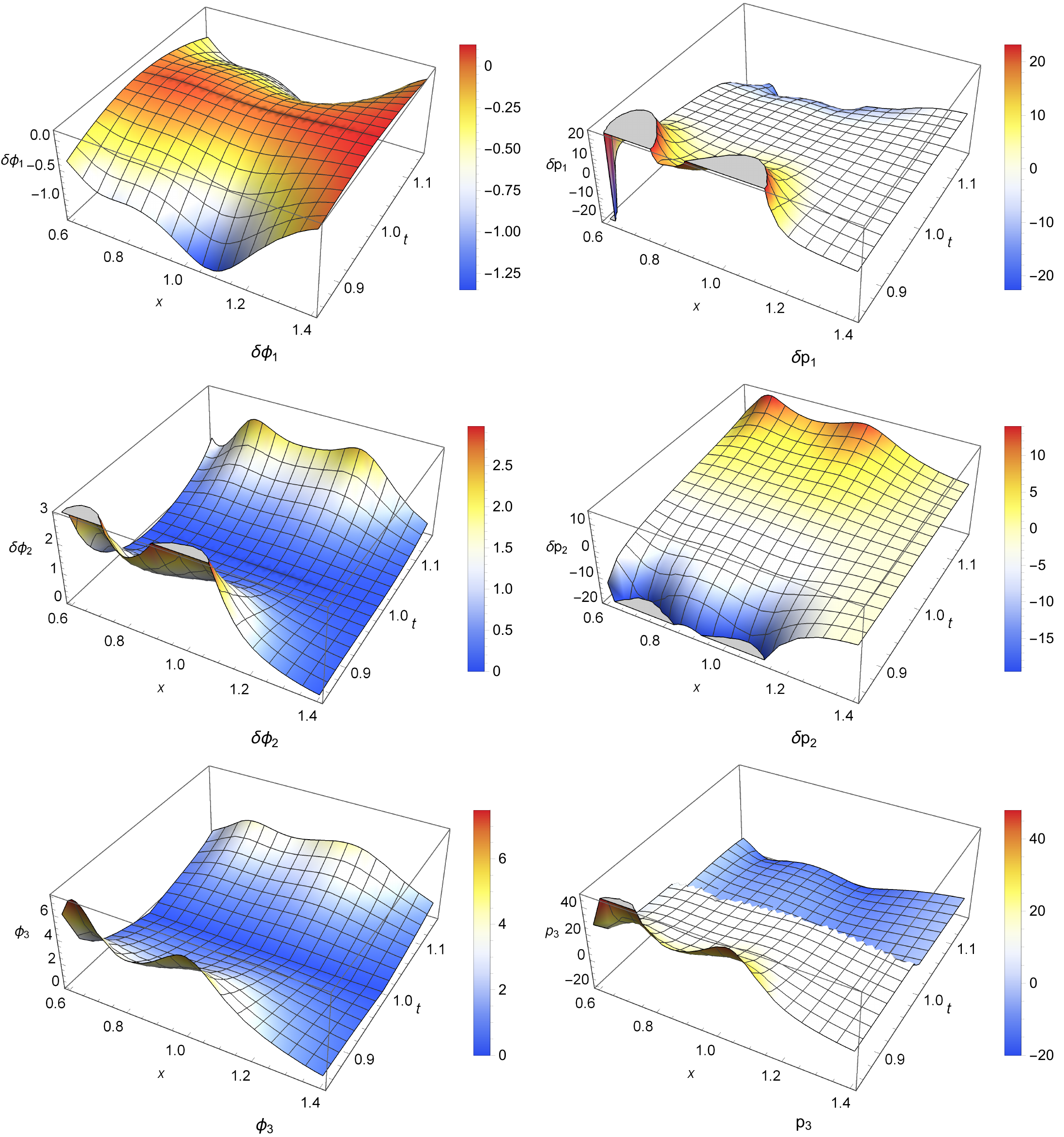}
    \caption[Non-satisfaction of second model's constraints]{Evolution of the
      first choice of initial values in Figure \ref{fig:one_pair_init_conds_123}.
    \label{fig:one_pair_solns_1}}
\end{figure}

\begin{figure}
    \centering
    \includegraphics[scale=0.7]{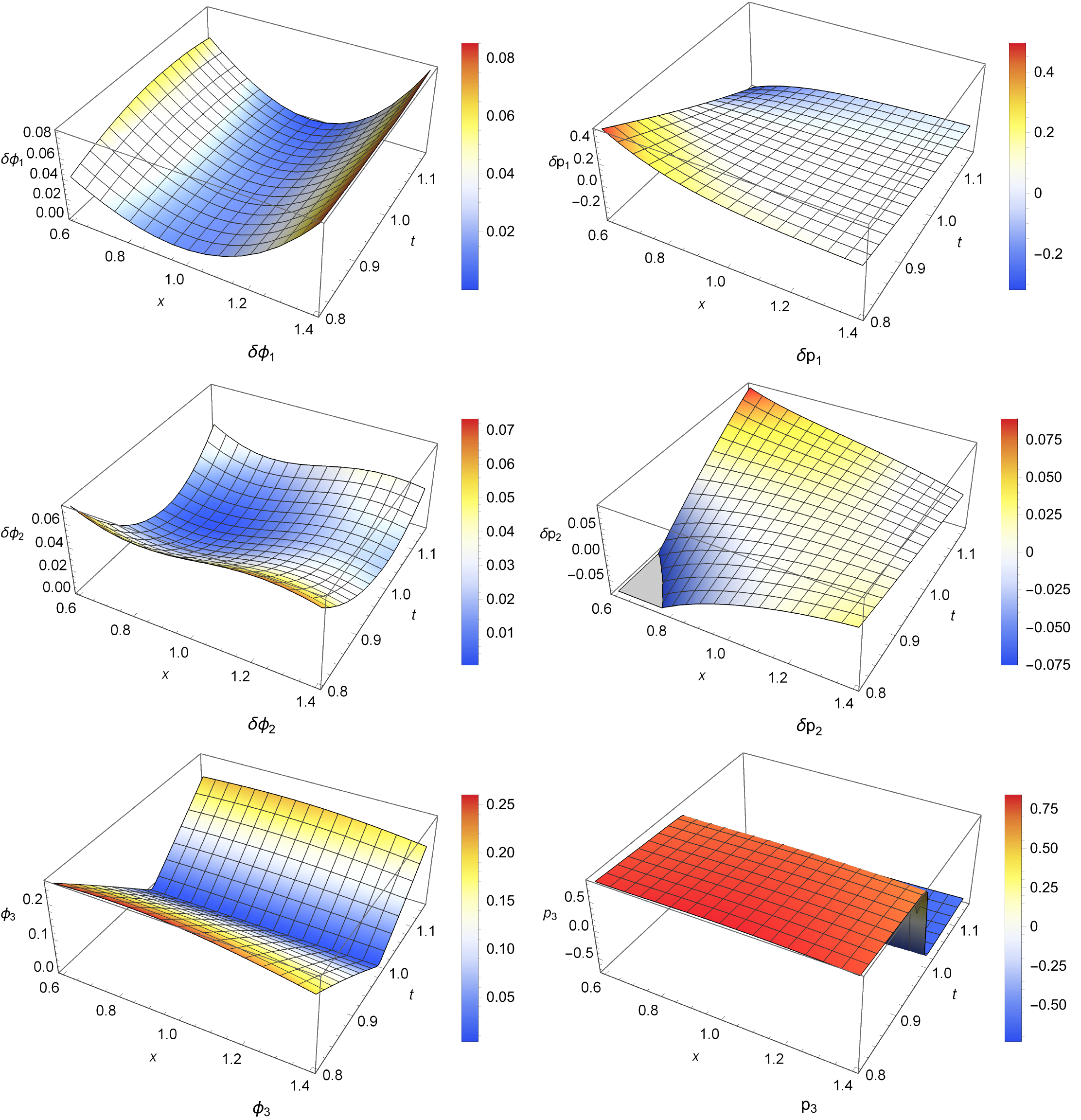}
    \caption[Non-satisfaction of second model's constraints]{Evolution of the
      second choice of initial values in Figure \ref{fig:one_pair_init_conds_123}.
    \label{fig:one_pair_solns_2}}
\end{figure}

\begin{figure}
    \centering
    \includegraphics[scale=0.7]{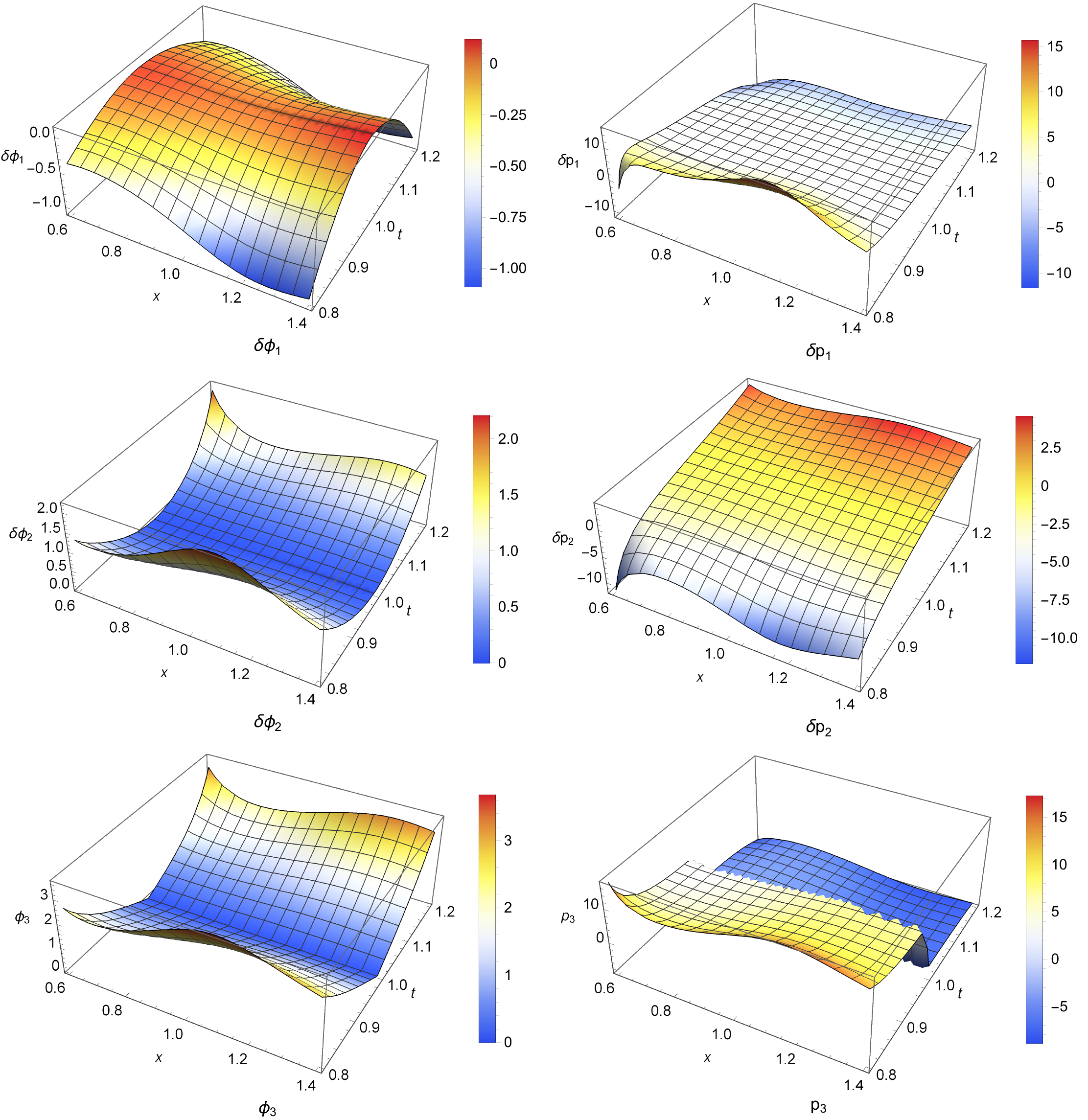}
    \caption[Non-satisfaction of second model's constraints]{Evolution of the
      third choice of initial values in Figure \ref{fig:one_pair_init_conds_123}.
    \label{fig:one_pair_solns_3}}
\end{figure}

Although we did not perform a detailed analysis of quasiclassical models with
$U$-terms included in the constraints and the equations of motion they
generate. Back in the model with a single quantum degree of freedom, we show
the implications on constrained initial values with three different choices of
$U(x)$, possibly with a spatial dependence. There are noticeable changes in
the initial values obtained by solving the constraints in Figure
\ref{fig:one_pair_init_conds_123}. Differences in the resulting evolutions are
even more visible; see Figures \ref{fig:one_pair_solns_1}--\ref{fig:one_pair_solns_3}.

\section{Conclusions}

We have presented the first constraint-based treatment of quasiclassical
cosmological dynamics with due attention paid to the condition of general
covariance. Compared with a previous analysis of static quasiclassical models
of the same symmetry type \cite{SphSymmMoments}, we encountered additional
challenges. In particular, including canonical moment fields of only one of
the two spatial metric or triad components in spherical symmetry does not
result in a consistent system of effective constraints once the staticity
assumption is dropped. We therefore included moment fields of both classical
components, leading to a larger set of fields and corresponding constraint and
evolution equations. The latter turned out to be more challenging from a
computational perspective, but we were able to demonstrate that the
constraints are better preserved by evolution with both types of moments
included. It is encouraging that this degree of consistency did not require us
to include all possible quasiclassical fields, such as cross-correlations
between the two metric components. There is therefore an indication that the
complexity of moments spaces can be kept under control while still maintaining
important consistency conditions such as general covariance.

Physically, the independent fields in our new formulation describe dynamical
properties of a quantum state. While we used perturbative inhomogeneity, we
did not have to perform a derivative expansion as is commonly done in
cosmological effective actions. Instead, the additional fields, formally,
serve as auxiliary fields that allow us to describe non-local quantum dynamics
in terms of a local action. In non-adiabatic regimes, such as at the beginning
or the end of inflation, a derivative expansion would not be reliable, making
it necessary to include additional fields such as our $\phi_3$ and $\phi_4$
and their momenta. As our extended constraints, such as
(\ref{eq:2_pair_clas_ham_con}), demonstrate, these are non-standard scalar
fields. (Nevertheless, they are generally covariant because the effective
constraints are closed, even if there may be no obvious relationship with
higher-curvature terms.) The dynamics therefore differs from two-scalar models
that have been studied extensively in cosmology. Once further studies have
brought the computational challenges observed here under control, our model
will be able to yield potentially observable phenomena.

\section*{Acknowledgements}

This work was supported in part by NSF grants PHY-1912168 and PHY-2206591.

\begin{appendix}

\section{Explicit Form of the Evolution Equations for the First Model}
\label{section:app_a}

\begin{eqnarray}
    \dot{\phi}_1&=&-2 J p_3^2 \sqrt{\phi _1}-2 L p_3 \sqrt{\phi _1} \phi _3-M \phi _1'-2 \text{N} p_2 \sqrt{\phi _1}\\
    \dot{p}_1&=&-\frac{J p_3 \phi _3 \left(\phi _1'\right){}^2}{4 \phi
                 _1^{3/2} \phi _2^2}+\frac{J p_3 \phi _3 \phi _1''}{\sqrt{\phi
                 _1} \phi _2^2}-\frac{J p_3 p_2^2 \phi _3}{4 \phi
                 _1^{3/2}}-\frac{J p_3^2 p_2 \phi _2}{2 \phi _1^{3/2}}+\frac{J
                 p_1 p_3^2}{\sqrt{\phi _1}}-\frac{J p_3 \phi _3}{4 \phi
                 _1^{3/2}}-\frac{L p_2^2 \phi _3^2}{4 \phi
                 _1^{3/2}}\\\nonumber 
    &&-\frac{L p_3 p_2 \phi _2 \phi _3}{2 \phi _1^{3/2}}+\frac{L p_1 p_3 \phi
       _3}{\sqrt{\phi _1}}-\frac{L \phi _3^2 \left(\phi _1'\right){}^2}{4 \phi
       _1^{3/2} \phi _2^2}+\frac{L \phi _3^2 \phi _1''}{\sqrt{\phi _1} \phi
       _2^2}-\frac{L \phi _3^2}{4 \phi _1^{3/2}}-\frac{\text{N} p_2^2 \phi
       _2}{4 \phi _1^{3/2}}+\frac{\text{N} p_1 p_2}{\sqrt{\phi
       _1}}-\frac{\text{N} p_3 p_2 \phi _3}{2 \phi _1^{3/2}}\\\nonumber
                &&-\frac{\text{N} p_3^2 \phi _2}{4 \phi _1^{3/2}}
    %               -\frac{\text{N} U \phi_2}{4 \phi _1^{3/2} \phi _3^2}
       +\frac{\text{N} \phi _3^2 \left(\phi
       _1'\right){}^2}{4 \phi _1^{3/2} \phi _2^3}+\frac{\text{N} \left(\phi
       _1'\right){}^2}{4 \phi _1^{3/2} \phi _2}-\frac{\text{N} \phi
       _1''}{\sqrt{\phi _1} \phi _2}-\frac{\text{N} \phi _3^2 \phi
       _1''}{\sqrt{\phi _1} \phi _2^3}-\frac{\text{N} \phi _2}{4 \phi
       _1^{3/2}}\\
    \dot{\phi}_2 &=& -N\left(\frac{\phi_2p_2}{\sqrt{\phi_1}} + 2\sqrt{\phi_1}
                     p_1 + \frac{\phi_3p_3}{\sqrt{\phi_1}}\right) - (M\phi_2)'
                     - L\left(\frac{p_2\phi_3^2}{\sqrt{\phi_1}} +
                     \frac{\phi_2\phi_3p_3}{\sqrt{\phi_1}}\right) -
                     (K\phi_3^2)' \\\nonumber  
    && - J\left(\frac{p_2p_3\phi_3}{\sqrt{\phi_1}} + \frac{\phi_2p_3^2}{\sqrt{\phi_1}}\right) - (Ip_3\phi_3)'\\
    \dot{p}_2 &=& N\biggl(\frac{p_2^2}{2\sqrt{\phi_1}} +
                  \frac{1}{2\sqrt{\phi_1}} +
                  \frac{(\phi_1')^2}{2\phi_2^2\sqrt{\phi_1}} +
                  2\frac{\sqrt{\phi_1}\phi_1''}{\phi_2^2} -
                  4\frac{\sqrt{\phi_1}\phi_1'\phi_2'}{\phi_2^3} +
                  \frac{p_3^2}{2\sqrt{\phi_1}}\\\nonumber  
    &&- 24 \frac{\sqrt{\phi_1}\phi_1'\phi_3^2\phi_2'}{\phi_2^5} +
       \left(\frac{3}{2}\frac{(\phi_1')^2}{\sqrt{\phi_1}\phi_2^4} +
       6\frac{\phi_1''\sqrt{\phi_1}}{\phi_2^4}\right)\phi_3^2 +
       12\frac{\sqrt{\phi_1}\phi_1'\phi_3\phi_3'}{\phi_2^4} 
 %      +\frac{U}{2\sqrt{\phi_1}\phi_3^2}
       \biggr)\\\nonumber 
    &&- 2\left(\frac{N\sqrt{\phi_1}\phi_1'}{\phi_2^2} +
       3\frac{N\sqrt{\phi_1}\phi_1'\phi_3^2}{\phi_2^4}\right)' - Mp_2' +
       L\biggl(\biggl(-\frac{p_2^2}{4\phi_2^{3/2}} - \frac{(\phi_1')^2 +
       4\phi_1\phi_1''}{\sqrt{\phi_1}\phi_2^3} \\\nonumber 
    &&+ 12\frac{\sqrt{\phi_1}\phi_1'\phi_2'}{\phi_2^4}\biggr)\phi_3^2 -
       4\frac{\sqrt{\phi_1}\phi_1'\phi_3\phi_3'}{\phi_2^3} +
       \frac{p_2\phi_3p_3}{\sqrt{\phi_1}}\biggr) +
       4\left(\frac{L\sqrt{\phi_1}\phi_1'\phi_3^2}{\phi_2^3}\right) -
       K\phi_3p_3' \\\nonumber 
    &&+ J\left(\left(-\frac{(\phi_1')^2}{\sqrt{\phi_1}\phi_2^3} -
       4\frac{\sqrt{\phi_1}\phi_1''}{\phi_2^3} +
       12\frac{\sqrt{\phi_1}\phi_1'\phi_2'}{\phi_2^4}\right)p_3\phi_3 -
       4\frac{\sqrt{\phi_1}\phi_1'p_3\phi_3'}{\phi_2^3} +
       \frac{p_2p_3^2}{\sqrt{\phi_1}}\right) 
    \\\nonumber  
    &&+ 4\left(\frac{J\sqrt{\phi_1}\phi_1'p_3\phi_3}{\phi_2^3}\right)' - Ip_3p_3'\\
    \dot{\phi}_3 &=& -N\left(\frac{\phi_2p_3}{\sqrt{\phi_1}} +
                     \frac{\phi_3p_2}{\sqrt{\phi_1}}\right) + (M\phi_3)' -
                     L\left(\frac{\phi_2p_2}{\sqrt{\phi_1}} +
                     2\sqrt{\phi_1p_1}\right)\phi_3 + (K\phi_2\phi_3)'
  \\\nonumber 
    && - J \biggl(\left(\frac{p_2^2}{2\sqrt{\phi_1}} +
       \frac{1}{2\sqrt{\phi_1}} + \frac{(\phi_1')^2}{2\sqrt{\phi_1}\phi_2^2} +
       \frac{2\sqrt{\phi_1}\phi_1''}{\phi_2^2} -
       \frac{4\sqrt{\phi_1}\phi_1'\phi_2'}{\phi_2^3}\right)\phi_3 +
       \frac{2\sqrt{\phi_1}\phi_1'\phi_3'}{\phi_2^2} \\\nonumber 
    && + 2\left(\frac{\phi_2 p_2}{\sqrt{\phi_1}} + 2\sqrt{\phi_1}p_1\right)p_3 \biggr) + I(p_2'\phi_3 + \phi_2p_3') + (I\phi_2p_3)'\\
    \dot{p}_3 &=& N\biggl(\frac{p_2p_3}{\sqrt{\phi_1}} +
                  2\left(5\frac{\sqrt{\phi_1}\phi_1'\phi_2'}{\phi_2^4} -
                  \frac{1}{2}\frac{(\phi_1')^2}{\sqrt{\phi_1}\phi_2^3} -
                  2\frac{\phi_1''\sqrt{\phi_1}}{\phi_2^3}\right)\phi_3 -
                  4\frac{\sqrt{\phi_1}\phi_1'\phi_3'}{\phi_2^3} 
  %               - \frac{U\phi_2}{\sqrt{\phi_1}\phi_3^3}
                  \biggr) \\\nonumber 
    && + 4\left(\frac{N\sqrt{\phi_1}\phi_1'\phi_3}{\phi_2^3}\right) - Mp_3' +
       L\biggl(2\left(\frac{p_2^2}{2\sqrt{\phi_2}} + \frac{1}{2\sqrt{\phi_1}}
       + \frac{(\phi_1')^2 + 4\phi_1\phi_1''}{2\sqrt{\phi_1}\phi_2^2} -
       4\frac{\sqrt{\phi_1}\phi_1'\phi_2'}{\phi_2^3}\right)\phi_3 \\\nonumber 
    && + 2\frac{\sqrt{\phi_1}\phi_1'\phi_3'}{\phi_2^2} +
       \left(\frac{\phi_2p_2}{\sqrt{\phi_1}} +
       2\sqrt{\phi_1}p_1\right)p_3\biggr) -
       2\left(\frac{L\sqrt{\phi_1}\phi_1'\phi_3}{\phi_2^2}\right)' -
       K(2p_2'\phi_3 + \phi_2p_3') \\\nonumber 
    && + J\left(\frac{p_2^2}{2\sqrt{\phi_1}} + \frac{1}{2\sqrt{\phi_1}} +
       \frac{(\phi_1')^2}{2\sqrt{\phi_1}\phi_2^2} +
       \frac{2\sqrt{\phi_1}\phi_1''}{\phi_2^2} -
       \frac{4\sqrt{\phi_1}\phi_1'\phi_2'}{\phi_2^3}\right)p_3 -
       2\left(\frac{J\sqrt{\phi_1}\phi_1'p_3}{\phi_2^2}\right)' \\\nonumber 
    && - Ip_2'p_3\, .
\end{eqnarray}
  
\section{Explicit Form of the Evolution Equations for the Second Model}
\label{section:app_b}

The equations of motion found for the Hamiltonian given by
(\ref{eq:2_pair_clas_ham_con}-\ref{eq:2_pair_high_ord_con_8}) are the
following:
\begin{eqnarray}
    \dot{\phi}_1&=&-E \phi _4 \phi _3'-\frac{F p_2 \phi _4 \phi _3}{\sqrt{\phi
                    _1}}-2 F p_4 \sqrt{\phi _1} \phi _4-G \phi _4 \phi
                    _3'-\frac{H p_2 \phi _4 \phi _3}{\sqrt{\phi _1}}-2 H p_4
                    \sqrt{\phi _1} \phi _4\\\nonumber 
    &&-{I} p_3 \phi _3'-\frac{J p_2 p_3 \phi _3}{\sqrt{\phi _1}}-2 J p_3 p_4
       \sqrt{\phi _1}-{K} \phi _3 \phi _3'-\frac{L p_2 \phi _3^2}{\sqrt{\phi
       _1}}-2 L p_4 \sqrt{\phi _1} \phi _3-M \phi _1'+\frac{N p_2 \phi _3^2}{4
       \phi _1^{3/2}}\\\nonumber 
                &&-2 N p_2 \sqrt{\phi _1}
\end{eqnarray}
\begin{eqnarray}
  \dot{p}_1&=& \frac{3 L \phi _2 \phi _3^2 p_2^2}{8 \phi _1^{5/2}}+\frac{3 J
               p_3 \phi _2 \phi _3 p_2^2}{8 \phi _1^{5/2}}+\frac{3 F \phi _2
               \phi _3 \phi _4 p_2^2}{8 \phi _1^{5/2}}+\frac{3 H \phi _2
               \phi _3 \phi _4 p_2^2}{8 \phi _1^{5/2}}-\frac{F \phi _4^2
               p_2^2}{4 \phi _1^{3/2}}-\frac{H \phi _4^2 p_2^2}{4 \phi
               _1^{3/2}}\\\nonumber 
           &&-\frac{{N} \phi _2 p_2^2}{4 \phi _1^{3/2}}-\frac{J p_3 \phi _4 p_2^2}{4
              \phi _1^{3/2}}-\frac{L \phi _3 \phi _4 p_2^2}{4 \phi _1^{3/2}}-\frac{15
              {N} \phi _2 \phi _3^2 p_2^2}{32 \phi _1^{7/2}}+\frac{J p_3^2
              p_2}{\sqrt{\phi _1}}+\frac{3 {N} p_1 \phi _3^2 p_2}{8 \phi
              _1^{5/2}}+\frac{{N} p_1 p_2}{\sqrt{\phi _1}}\\\nonumber 
           &&+\frac{L p_3 \phi _3 p_2}{\sqrt{\phi _1}}+\frac{F p_3 \phi _4
              p_2}{\sqrt{\phi _1}}+\frac{H p_3 \phi _4 p_2}{\sqrt{\phi _1}}-\frac{L
              p_1 \phi _3^2 p_2}{2 \phi _1^{3/2}}-\frac{J p_3 p_4 \phi _2 p_2}{2 \phi
              _1^{3/2}}-\frac{{N} p_3 \phi _3 p_2}{2 \phi _1^{3/2}}-\frac{J p_1 p_3
              \phi _3 p_2}{2 \phi _1^{3/2}}\\\nonumber 
           &&-\frac{L p_4 \phi _2 \phi _3 p_2}{2 \phi _1^{3/2}}-\frac{{N} p_4 \phi _4
              p_2}{2 \phi _1^{3/2}}-\frac{F p_4 \phi _2 \phi _4 p_2}{2 \phi
              _1^{3/2}}-\frac{H p_4 \phi _2 \phi _4 p_2}{2 \phi _1^{3/2}}-\frac{F p_1
              \phi _3 \phi _4 p_2}{2 \phi _1^{3/2}}-\frac{H p_1 \phi _3 \phi _4
              p_2}{2 \phi _1^{3/2}}\\\nonumber 
           &&+\frac{3 L \phi _2 \phi _3^2}{8 \phi _1^{5/2}}+\frac{15 {N} \phi _3^2
              \left(\phi _1'\right){}^2}{32 \phi _1^{7/2} \phi _2}+\frac{{N} \phi
              _4^2 \left(\phi _1'\right){}^2}{4 \phi _1^{3/2} \phi _2^3}+\frac{{N}
              \left(\phi _1'\right){}^2}{4 \phi _1^{3/2} \phi _2}+\frac{{N}
              \left(\phi _3'\right){}^2}{4 \phi _1^{3/2} \phi _2}+\frac{2 F
              \sqrt{\phi _1} \left(\phi _4'\right){}^2}{\phi _2^2}\\\nonumber 
           &&+\frac{2 H \sqrt{\phi _1} \left(\phi _4'\right){}^2}{\phi _2^2}+\frac{J
              p_1 p_3 p_4}{\sqrt{\phi _1}}+\frac{L p_1 p_4 \phi _3}{\sqrt{\phi
              _1}}+\frac{3 J p_3 \phi _2 \phi _3}{8 \phi _1^{5/2}}+\frac{F p_1 p_4
              \phi _4}{\sqrt{\phi _1}}+\frac{H p_1 p_4 \phi _4}{\sqrt{\phi
              _1}}+\frac{3 F \phi _2 \phi _3 \phi _4}{8 \phi _1^{5/2}}\\\nonumber 
           &&+\frac{3 H \phi _2 \phi _3 \phi _4}{8 \phi _1^{5/2}}-p_3 \phi _4
              {E}'-p_3^2 {I}'-p_3 \phi _3 {K}'-p_3 \phi _4 G'-p_1 M'-M p_1'-2 {I} p_3
              p_3'-{K} \phi _3 p_3'\\\nonumber 
           &&-{E} \phi _4 p_3'-G \phi _4 p_3'+\frac{4 \sqrt{\phi _1} \phi _4 J'
              p_3'}{\phi _2^2}+\frac{\phi _4^2 F' \phi _1'}{\sqrt{\phi _1} \phi
              _2^2}+\frac{\phi _3 \phi _4 F' \phi _1'}{2 \phi _1^{3/2} \phi
              _2}+\frac{\phi _4^2 H' \phi _1'}{\sqrt{\phi _1} \phi _2^2}+\frac{\phi
              _3 \phi _4 H' \phi _1'}{2 \phi _1^{3/2} \phi _2}\\\nonumber 
           &&+\frac{p_3 \phi _3 J' \phi _1'}{2 \phi _1^{3/2} \phi _2}+\frac{p_3 \phi
              _4 J' \phi _1'}{\sqrt{\phi _1} \phi _2^2}+\frac{\phi _3^2 L' \phi
              _1'}{2 \phi _1^{3/2} \phi _2}+\frac{\phi _3 \phi _4 L' \phi
              _1'}{\sqrt{\phi _1} \phi _2^2}+\frac{J \phi _3 p_3' \phi _1'}{2 \phi
              _1^{3/2} \phi _2}+\frac{J \phi _4 p_3' \phi _1'}{\sqrt{\phi _1} \phi
              _2^2}+\frac{6 \sqrt{\phi _1} \phi _4^2 {N}' \phi _2'}{\phi
              _2^4}\\\nonumber 
           &&+\frac{2 \sqrt{\phi _1} {N}' \phi _2'}{\phi _2^2}+\frac{\phi _3 \phi _4
              F' \phi _2'}{\sqrt{\phi _1} \phi _2^2}+\frac{\phi _3 \phi _4 H' \phi
              _2'}{\sqrt{\phi _1} \phi _2^2}+\frac{p_3 \phi _3 J' \phi
              _2'}{\sqrt{\phi _1} \phi _2^2}+\frac{\phi _3^2 L' \phi _2'}{\sqrt{\phi
              _1} \phi _2^2}+\frac{J \phi _3 p_3' \phi _2'}{\sqrt{\phi _1} \phi
              _2^2}+\frac{3 {N} \phi _3^2 \phi _1' \phi _2'}{8 \phi _1^{5/2} \phi
              _2^2}\\\nonumber 
           &&+\frac{3 {N} \phi _4^2 \phi _1' \phi _2'}{\sqrt{\phi _1} \phi
              _2^4}+\frac{{N} \phi _1' \phi _2'}{\sqrt{\phi _1} \phi _2^2}-{K} p_3
              \phi _3'+\frac{\phi _3 {N}' \phi _3'}{2 \phi _1^{3/2} \phi _2}+\frac{4
              \sqrt{\phi _1} \phi _4 L' \phi _3'}{\phi _2^2}+\frac{L \phi _3 \phi _1'
              \phi _3'}{\phi _1^{3/2} \phi _2}+\frac{F \phi _4 \phi _1' \phi _3'}{2
              \phi _1^{3/2} \phi _2}\\\nonumber 
           &&+\frac{H \phi _4 \phi _1' \phi _3'}{2 \phi _1^{3/2} \phi _2}+\frac{L
              \phi _4 \phi _1' \phi _3'}{\sqrt{\phi _1} \phi _2^2}+\frac{J p_3 \phi
              _1' \phi _3'}{2 \phi _1^{3/2} \phi _2}+\frac{2 L \phi _3 \phi _2' \phi
              _3'}{\sqrt{\phi _1} \phi _2^2}+\frac{F \phi _4 \phi _2' \phi
              _3'}{\sqrt{\phi _1} \phi _2^2}+\frac{H \phi _4 \phi _2' \phi
              _3'}{\sqrt{\phi _1} \phi _2^2}+\frac{J p_3 \phi _2' \phi
              _3'}{\sqrt{\phi _1} \phi _2^2}\\\nonumber 
           &&-{E} p_3 \phi _4'-G p_3 \phi _4'+\frac{6 \sqrt{\phi _1} \phi _4 F' \phi
              _4'}{\phi _2^2}+\frac{6 \sqrt{\phi _1} \phi _4 H' \phi _4'}{\phi
              _2^2}+\frac{2 \sqrt{\phi _1} p_3 J' \phi _4'}{\phi _2^2}+\frac{2
              \sqrt{\phi _1} \phi _3 L' \phi _4'}{\phi _2^2}\\\nonumber 
           &&+\frac{2 J \sqrt{\phi _1} p_3' \phi _4'}{\phi _2^2}+\frac{F \phi _3 \phi
              _1' \phi _4'}{2 \phi _1^{3/2} \phi _2}+\frac{H \phi _3 \phi _1' \phi
              _4'}{2 \phi _1^{3/2} \phi _2}+\frac{L \phi _3 \phi _1' \phi
              _4'}{\sqrt{\phi _1} \phi _2^2}+\frac{2 F \phi _4 \phi _1' \phi
              _4'}{\sqrt{\phi _1} \phi _2^2}+\frac{2 H \phi _4 \phi _1' \phi
              _4'}{\sqrt{\phi _1} \phi _2^2}+\frac{J p_3 \phi _1' \phi
              _4'}{\sqrt{\phi _1} \phi _2^2}\\\nonumber 
           &&+\frac{2 L \sqrt{\phi _1} \phi _3' \phi _4'}{\phi _2^2}+\frac{\phi _3^2
              {N}''}{4 \phi _1^{3/2} \phi _2}+\frac{2 \sqrt{\phi _1} \phi _4^2
              F''}{\phi _2^2}+\frac{2 \sqrt{\phi _1} \phi _4^2 H''}{\phi
              _2^2}+\frac{2 \sqrt{\phi _1} p_3 \phi _4 J''}{\phi _2^2}+\frac{2
              \sqrt{\phi _1} \phi _3 \phi _4 L''}{\phi _2^2}\\\nonumber 
           &&+\frac{L \phi _3^2 \phi _1''}{2 \phi _1^{3/2} \phi _2}+\frac{F \phi _4^2
              \phi _1''}{\sqrt{\phi _1} \phi _2^2}+\frac{H \phi _4^2 \phi
              _1''}{\sqrt{\phi _1} \phi _2^2}+\frac{J p_3 \phi _3 \phi _1''}{2 \phi
              _1^{3/2} \phi _2}+\frac{F \phi _3 \phi _4 \phi _1''}{2 \phi _1^{3/2}
              \phi _2}+\frac{H \phi _3 \phi _4 \phi _1''}{2 \phi _1^{3/2} \phi
              _2}+\frac{L \phi _3 \phi _4 \phi _1''}{\sqrt{\phi _1} \phi
              _2^2}+\frac{J p_3 \phi _4 \phi _1''}{\sqrt{\phi _1} \phi
              _2^2}\\\nonumber 
           &&+\frac{{N} \phi _3 \phi _3''}{2 \phi _1^{3/2} \phi _2}+\frac{2 L
              \sqrt{\phi _1} \phi _4 \phi _3''}{\phi _2^2}+\frac{2 F \sqrt{\phi _1}
              \phi _4 \phi _4''}{\phi _2^2}+\frac{2 H \sqrt{\phi _1} \phi _4 \phi
              _4''}{\phi _2^2}+\frac{2 J \sqrt{\phi _1} \phi _4 p_3''}{\phi
              _2^2}-\frac{F \phi _4^2}{4 \phi _1^{3/2}}-\frac{H \phi _4^2}{4 \phi
              _1^{3/2}}\\\nonumber 
           &&-\frac{{N} p_4^2 \phi _2}{4 \phi _1^{3/2}}-\frac{{N} \phi _2}{4 \phi
              _1^{3/2}}-\frac{J p_3 \phi _4}{4 \phi _1^{3/2}}-\frac{L \phi _3 \phi
              _4}{4 \phi _1^{3/2}}-\frac{15 {N} \phi _2 \phi _3^2}{32 \phi
              _1^{7/2}}-\frac{2 \sqrt{\phi _1} {N}''}{\phi _2}-\frac{L \left(\phi
              _3'\right){}^2}{\sqrt{\phi _1} \phi _2}-\frac{2 \phi _3 J'
              p_3'}{\sqrt{\phi _1} \phi _2}
\end{eqnarray}
\begin{eqnarray*}
           &&-\frac{\phi _4 F' \phi _3'}{\sqrt{\phi _1} \phi _2}-\frac{\phi _4 H'
              \phi _3'}{\sqrt{\phi _1} \phi _2}-\frac{p_3 J' \phi _3'}{\sqrt{\phi _1}
              \phi _2}-\frac{3 \phi _3 L' \phi _3'}{\sqrt{\phi _1} \phi _2}-\frac{J
              p_3' \phi _3'}{\sqrt{\phi _1} \phi _2}-\frac{2 \phi _3 F' \phi
              _4'}{\sqrt{\phi _1} \phi _2}-\frac{2 \phi _3 H' \phi _4'}{\sqrt{\phi
              _1} \phi _2}\\\nonumber 
           &&-\frac{F \phi _3' \phi _4'}{\sqrt{\phi _1} \phi _2}-\frac{H \phi _3'
              \phi _4'}{\sqrt{\phi _1} \phi _2}-\frac{\phi _3 \phi _4 F''}{\sqrt{\phi
              _1} \phi _2}-\frac{\phi _3 \phi _4 H''}{\sqrt{\phi _1} \phi
              _2}-\frac{p_3 \phi _3 J''}{\sqrt{\phi _1} \phi _2}-\frac{\phi _3^2
              L''}{\sqrt{\phi _1} \phi _2}-\frac{{N} \phi _1''}{\sqrt{\phi _1} \phi
              _2}-\frac{J p_3 \phi _3''}{\sqrt{\phi _1} \phi _2}\\\nonumber 
           &&-\frac{2 L \phi _3 \phi _3''}{\sqrt{\phi _1} \phi _2}-\frac{F \phi _4
              \phi _3''}{\sqrt{\phi _1} \phi _2}-\frac{H \phi _4 \phi
              _3''}{\sqrt{\phi _1} \phi _2}-\frac{F \phi _3 \phi _4''}{\sqrt{\phi _1}
              \phi _2}-\frac{H \phi _3 \phi _4''}{\sqrt{\phi _1} \phi _2}-\frac{J
              \phi _3 {p_3}''}{\sqrt{\phi _1} \phi _2}-\frac{3 {N} \phi _3 \phi _1'
              \phi _3'}{4 \phi _1^{5/2} \phi _2}\\\nonumber 
           &&-\frac{3 L \phi _3^2 \left(\phi _1'\right){}^2}{8 \phi _1^{5/2} \phi
              _2}-\frac{3 J p_3 \phi _3 \left(\phi _1'\right){}^2}{8 \phi _1^{5/2}
              \phi _2}-\frac{3 F \phi _3 \phi _4 \left(\phi _1'\right){}^2}{8 \phi
              _1^{5/2} \phi _2}-\frac{3 H \phi _3 \phi _4 \left(\phi
              _1'\right){}^2}{8 \phi _1^{5/2} \phi _2}-\frac{3 \phi _3^2 {N}' \phi
              _1'}{8 \phi _1^{5/2} \phi _2}-\frac{3 {N} \phi _3^2 \phi _1''}{8 \phi
              _1^{5/2} \phi _2}\\\nonumber 
           &&-\frac{L \phi _3^2 \phi _1' \phi _2'}{2 \phi _1^{3/2} \phi _2^2}-\frac{J
              p_3 \phi _3 \phi _1' \phi _2'}{2 \phi _1^{3/2} \phi _2^2}-\frac{F \phi
              _3 \phi _4 \phi _1' \phi _2'}{2 \phi _1^{3/2} \phi _2^2}-\frac{H \phi
              _3 \phi _4 \phi _1' \phi _2'}{2 \phi _1^{3/2} \phi _2^2}-\frac{{N} \phi
              _3 \phi _2' \phi _3'}{2 \phi _1^{3/2} \phi _2^2}-\frac{F \phi _4^2
              \left(\phi _1'\right){}^2}{4 \phi _1^{3/2} \phi _2^2}\\\nonumber 
           &&-\frac{H \phi _4^2 \left(\phi _1'\right){}^2}{4 \phi _1^{3/2} \phi
              _2^2}-\frac{J p_3 \phi _4 \left(\phi _1'\right){}^2}{4 \phi _1^{3/2}
              \phi _2^2}-\frac{L \phi _3 \phi _4 \left(\phi _1'\right){}^2}{4 \phi
              _1^{3/2} \phi _2^2}-\frac{\phi _3^2 {N}' \phi _2'}{4 \phi _1^{3/2} \phi
              _2^2}-\frac{4 \sqrt{\phi _1} \phi _4^2 F' \phi _2'}{\phi _2^3}-\frac{4
              \sqrt{\phi _1} \phi _4^2 H' \phi _2'}{\phi _2^3}\\\nonumber 
           &&-\frac{4 p_3 \sqrt{\phi _1} \phi _4 J' \phi _2'}{\phi _2^3}-\frac{4
              \sqrt{\phi _1} \phi _3 \phi _4 L' \phi _2'}{\phi _2^3}-\frac{4 J
              \sqrt{\phi _1} \phi _4 p_3' \phi _2'}{\phi _2^3}-\frac{4 L \sqrt{\phi
              _1} \phi _4 \phi _2' \phi _3'}{\phi _2^3}-\frac{4 \sqrt{\phi _1} \phi
              _4 {N}' \phi _4'}{\phi _2^3}\\\nonumber 
           &&-\frac{4 F \sqrt{\phi _1} \phi _4 \phi _2' \phi _4'}{\phi _2^3}-\frac{4
              H \sqrt{\phi _1} \phi _4 \phi _2' \phi _4'}{\phi _2^3}-\frac{2
              \sqrt{\phi _1} \phi _4^2 {N}''}{\phi _2^3}-\frac{\phi _4^2 {N}' \phi
              _1'}{\sqrt{\phi _1} \phi _2^3}-\frac{2 F \phi _4^2 \phi _1' \phi
              _2'}{\sqrt{\phi _1} \phi _2^3}-\frac{2 H \phi _4^2 \phi _1' \phi
              _2'}{\sqrt{\phi _1} \phi _2^3}\\\nonumber 
           &&-\frac{2 J p_3 \phi _4 \phi _1' \phi _2'}{\sqrt{\phi _1} \phi
              _2^3}-\frac{2 L \phi _3 \phi _4 \phi _1' \phi _2'}{\sqrt{\phi _1} \phi
              _2^3}-\frac{2 {N} \phi _4 \phi _1' \phi _4'}{\sqrt{\phi _1} \phi
              _2^3}-\frac{{N} \phi _4^2 \phi _1''}{\sqrt{\phi _1} \phi
              _2^3}
 %             -\frac{{N} U_2 \phi _2}{4 \phi _1^{3/2} \phi _4^2}
              +\frac{F \phi _3
              \phi _2' \phi _4'}{\sqrt{\phi _1} \phi _2^2}\\\nonumber
           &&+\frac{H \phi _3 \phi _2' \phi _4'}{\sqrt{\phi _1} \phi _2^2}-\frac{{N}'
              \phi _1'}{\sqrt{\phi _1} \phi _2}
\end{eqnarray*}
\begin{eqnarray}
    \dot{\phi}_2&=&-\phi _4^2 {E}'-2 {E} \phi _4 \phi _4'-2 F p_3 \sqrt{\phi
                    _1} \phi _4+\frac{F p_2 \phi _2 \phi _3 \phi _4}{2 \phi
                    _1^{3/2}}-\frac{F p_2 \phi _4^2}{\sqrt{\phi _1}}-\frac{F
                    p_4 \phi _2 \phi _4}{\sqrt{\phi _1}}\\\nonumber 
    &&-\frac{F p_1 \phi _3 \phi _4}{\sqrt{\phi _1}}-\phi _4^2 G'-2 G \phi _4
       \phi _4'-2 H p_3 \sqrt{\phi _1} \phi _4+\frac{H p_2 \phi _2 \phi _3
       \phi _4}{2 \phi _1^{3/2}}-\frac{H p_2 \phi _4^2}{\sqrt{\phi
       _1}}-\frac{H p_4 \phi _2 \phi _4}{\sqrt{\phi _1}}\\\nonumber 
    &&-\frac{H p_1 \phi _3 \phi _4}{\sqrt{\phi _1}}-p_3 \phi _4 {I}'-{I} p_3
       \phi _4'-{I} \phi _4 p_3'-2 J p_3^2 \sqrt{\phi _1}+\frac{J p_2 p_3 \phi
       _2 \phi _3}{2 \phi _1^{3/2}}-\frac{J p_4 p_3 \phi _2}{\sqrt{\phi
       _1}}-\frac{J p_1 p_3 \phi _3}{\sqrt{\phi _1}}\\\nonumber 
    &&-\frac{J p_2 p_3 \phi _4}{\sqrt{\phi _1}}-\phi _3 \phi _4 {K}'-{K} \phi
       _4 \phi _3'-{K} \phi _3 \phi _4'-2 L p_3 \sqrt{\phi _1} \phi _3+\frac{L
       p_2 \phi _2 \phi _3^2}{2 \phi _1^{3/2}}-\frac{L p_1 \phi
       _3^2}{\sqrt{\phi _1}}-\frac{L p_4 \phi _2 \phi _3}{\sqrt{\phi
       _1}}\\\nonumber 
    &&-\frac{L p_2 \phi _3 \phi _4}{\sqrt{\phi _1}}-\phi _2 M'-M \phi
       _2'-\frac{{N} p_3 \phi _3}{\sqrt{\phi _1}}+\frac{{N} p_1 \phi _3^2}{4
       \phi _1^{3/2}}-2 {N} p_1 \sqrt{\phi _1}-\frac{{N} p_2 \phi
       _2}{\sqrt{\phi _1}}-\frac{{N} p_4 \phi _4}{\sqrt{\phi _1}}\\\nonumber 
                &&-\frac{3 {N} p_2 \phi _2 \phi _3^2}{8 \phi _1^{5/2}}
\end{eqnarray}
\begin{eqnarray}
    \dot{p}_2&=&\frac{3 {N} \phi _3^2 p_2^2}{16 \phi _1^{5/2}}+\frac{{N}
                 p_2^2}{2 \sqrt{\phi _1}}-\frac{L \phi _3^2 p_2^2}{4 \phi
                 _1^{3/2}}-\frac{J p_3 \phi _3 p_2^2}{4 \phi _1^{3/2}}-\frac{F
                 \phi _3 \phi _4 p_2^2}{4 \phi _1^{3/2}}-\frac{H \phi _3 \phi
                 _4 p_2^2}{4 \phi _1^{3/2}}+\frac{J p_3 p_4 p_2}{\sqrt{\phi
                 _1}}\\\nonumber 
     &&+\frac{L p_4 \phi _3 p_2}{\sqrt{\phi _1}}+\frac{F p_4 \phi _4
        p_2}{\sqrt{\phi _1}}+\frac{H p_4 \phi _4 p_2}{\sqrt{\phi
        _1}}+\frac{{N} p_4^2}{2 \sqrt{\phi _1}}+\frac{3 {N} \phi _3^2}{16 \phi
        _1^{5/2}}+\frac{L \phi _3^2 \left(\phi _1'\right){}^2}{4 \phi _1^{3/2}
        \phi _2^2}+\frac{F \phi _4^2 \left(\phi _1'\right){}^2}{\sqrt{\phi _1}
        \phi _2^3}\\\nonumber 
    &&+\frac{H \phi _4^2 \left(\phi _1'\right){}^2}{\sqrt{\phi _1} \phi
       _2^3}+\frac{J p_3 \phi _3 \left(\phi _1'\right){}^2}{4 \phi _1^{3/2}
       \phi _2^2}+\frac{F \phi _3 \phi _4 \left(\phi _1'\right){}^2}{4 \phi
       _1^{3/2} \phi _2^2}+\frac{H \phi _3 \phi _4 \left(\phi
       _1'\right){}^2}{4 \phi _1^{3/2} \phi _2^2}+\frac{L \phi _3 \phi _4
       \left(\phi _1'\right){}^2}{\sqrt{\phi _1} \phi _2^3}\\\nonumber 
    &&+\frac{J p_3 \phi _4 \left(\phi _1'\right){}^2}{\sqrt{\phi _1} \phi
       _2^3}-M p_2'-{I} p_3 p_4'-{K} \phi _3 p_4'-{E} \phi _4 p_4'-G \phi _4
       p_4'+\frac{\phi _3^2 {N}' \phi _1'}{4 \phi _1^{3/2} \phi _2^2}+\frac{4
       \sqrt{\phi _1} \phi _4^2 F' \phi _1'}{\phi _2^3}\\\nonumber 
    &&+\frac{4 \sqrt{\phi _1} \phi _4^2 H' \phi _1'}{\phi _2^3}+\frac{4
       \sqrt{\phi _1} p_3 \phi _4 J' \phi _1'}{\phi _2^3}+\frac{4 \sqrt{\phi
       _1} \phi _3 \phi _4 L' \phi _1'}{\phi _2^3}+\frac{4 J \sqrt{\phi _1}
       \phi _4 p_3' \phi _1'}{\phi _2^3}+\frac{{N} \phi _3 \phi _1' \phi
       _3'}{2 \phi _1^{3/2} \phi _2^2}\\\nonumber 
    &&+\frac{4 L \sqrt{\phi _1} \phi _4 \phi _1' \phi _3'}{\phi _2^3}+\frac{4
       F \sqrt{\phi _1} \phi _4 \phi _1' \phi _4'}{\phi _2^3}+\frac{4 H
       \sqrt{\phi _1} \phi _4 \phi _1' \phi _4'}{\phi _2^3}+\frac{{N}}{2
       \sqrt{\phi _1}}-\frac{L \phi _3^2}{4 \phi _1^{3/2}}-\frac{J p_3 \phi
       _3}{4 \phi _1^{3/2}}\\\nonumber 
    &&-\frac{F \phi _3 \phi _4}{4 \phi _1^{3/2}}-\frac{H \phi _3 \phi _4}{4
       \phi _1^{3/2}}-\frac{2 L \sqrt{\phi _1} \left(\phi _3'\right){}^2}{\phi
       _2^2}-\frac{2 \sqrt{\phi _1} {N}' \phi _1'}{\phi _2^2}-\frac{2
       \sqrt{\phi _1} \phi _4 F' \phi _3'}{\phi _2^2}-\frac{2 \sqrt{\phi _1}
       \phi _4 H' \phi _3'}{\phi _2^2}\\\nonumber 
    &&-\frac{2 p_3 \sqrt{\phi _1} J' \phi _3'}{\phi _2^2}-\frac{2 \sqrt{\phi
       _1} \phi _3 L' \phi _3'}{\phi _2^2}-\frac{2 J \sqrt{\phi _1} p_3' \phi
       _3'}{\phi _2^2}-\frac{2 F \sqrt{\phi _1} \phi _3' \phi _4'}{\phi
       _2^2}-\frac{2 H \sqrt{\phi _1} \phi _3' \phi _4'}{\phi _2^2}\\\nonumber 
    &&-\frac{\phi _3 \phi _4 F' \phi _1'}{\sqrt{\phi _1} \phi _2^2}-\frac{\phi
       _3 \phi _4 H' \phi _1'}{\sqrt{\phi _1} \phi _2^2}-\frac{p_3 \phi _3 J'
       \phi _1'}{\sqrt{\phi _1} \phi _2^2}-\frac{\phi _3^2 L' \phi
       _1'}{\sqrt{\phi _1} \phi _2^2}-\frac{J \phi _3 p_3' \phi
       _1'}{\sqrt{\phi _1} \phi _2^2}-\frac{\phi _3 {N}' \phi _3'}{\sqrt{\phi
       _1} \phi _2^2}-\frac{J p_3 \phi _1' \phi _3'}{\sqrt{\phi _1} \phi
       _2^2}\\\nonumber 
    &&-\frac{2 L \phi _3 \phi _1' \phi _3'}{\sqrt{\phi _1} \phi _2^2}-\frac{F
       \phi _4 \phi _1' \phi _3'}{\sqrt{\phi _1} \phi _2^2}-\frac{H \phi _4
       \phi _1' \phi _3'}{\sqrt{\phi _1} \phi _2^2}-\frac{F \phi _3 \phi _1'
       \phi _4'}{\sqrt{\phi _1} \phi _2^2}-\frac{H \phi _3 \phi _1' \phi
       _4'}{\sqrt{\phi _1} \phi _2^2}-\frac{{N} \left(\phi _1'\right){}^2}{2
       \sqrt{\phi _1} \phi _2^2}-\frac{{N} \left(\phi _3'\right){}^2}{2
       \sqrt{\phi _1} \phi _2^2}\\\nonumber 
    &&-\frac{3 {N} \phi _3^2 \left(\phi _1'\right){}^2}{16 \phi _1^{5/2} \phi
       _2^2}-\frac{6 \sqrt{\phi _1} \phi _4^2 {N}' \phi _1'}{\phi
       _2^4}-\frac{3 {N} \phi _4^2 \left(\phi _1'\right){}^2}{2 \sqrt{\phi _1}
       \phi _2^4}+\frac{{N} U_2}{2 \sqrt{\phi _1} \phi _4^2}
\end{eqnarray}
\begin{eqnarray}
    \dot{\phi}_3&=&-{E} \phi _4 \phi _1'-2 F p_2 \sqrt{\phi _1} \phi _4-G \phi
                    _4 \phi _1'-2 H p_2 \sqrt{\phi _1} \phi _4+{I} \phi _4
                    p_2'+{I} \phi _2 p_4'-2 {I} p_3 \phi _1'\\\nonumber 
    &&-{I} p_1 \phi _3'+\frac{J p_2^2 \phi _2 \phi _3}{4 \phi
       _1^{3/2}}-\frac{J p_2^2 \phi _4}{2 \sqrt{\phi _1}}-4 J p_3 p_2
       \sqrt{\phi _1}-\frac{J p_4 p_2 \phi _2}{\sqrt{\phi _1}}-\frac{J p_1 p_2
       \phi _3}{\sqrt{\phi _1}}-2 J p_1 p_4 \sqrt{\phi _1}\\\nonumber 
    &&+\frac{4 J \sqrt{\phi _1} \phi _4 \phi _1' \phi _2'}{\phi _2^3}+\frac{J
       \phi _1' \phi _3'}{\sqrt{\phi _1} \phi _2}-\frac{J \phi _3 \left(\phi
       _1'\right){}^2}{4 \phi _1^{3/2} \phi _2}-\frac{2 J \sqrt{\phi _1} \phi
       _2' \phi _3'}{\phi _2^2}-\frac{2 J \sqrt{\phi _1} \phi _1' \phi
       _4'}{\phi _2^2}-\frac{J \phi _3 \phi _1' \phi _2'}{\sqrt{\phi _1} \phi
       _2^2}\\\nonumber 
    &&-\frac{J \phi _4 \left(\phi _1'\right){}^2}{2 \sqrt{\phi _1} \phi
       _2^2}+\frac{J \phi _3 \phi _1''}{\sqrt{\phi _1} \phi _2}+\frac{2 J
       \sqrt{\phi _1} \phi _3''}{\phi _2}-\frac{2 J \sqrt{\phi _1} \phi _4
       \phi _1''}{\phi _2^2}+\frac{J \phi _2 \phi _3}{4 \phi _1^{3/2}}-\frac{J
       \phi _4}{2 \sqrt{\phi _1}}-{K} \phi _3 \phi _1'\\\nonumber 
    &&-2 L p_2 \sqrt{\phi _1} \phi _3-M \phi _3'-\frac{{N} p_2 \phi
       _3}{\sqrt{\phi _1}}
\end{eqnarray}
\begin{eqnarray}
    \dot{p}_3&=&\frac{3 {N} \phi _2 \phi _3 p_2^2}{8 \phi _1^{5/2}}+\frac{L
                 \phi _4 p_2^2}{2 \sqrt{\phi _1}}-\frac{L \phi _2 \phi _3
                 p_2^2}{2 \phi _1^{3/2}}-\frac{J p_3 \phi _2 p_2^2}{4 \phi
                 _1^{3/2}}-\frac{F \phi _2 \phi _4 p_2^2}{4 \phi
                 _1^{3/2}}-\frac{H \phi _2 \phi _4 p_2^2}{4 \phi _1^{3/2}}+2 L
                 \sqrt{\phi _1} p_3 p_2\\\nonumber 
    &&+\frac{J p_1 p_3 p_2}{\sqrt{\phi _1}}+\frac{{N} p_3 p_2}{\sqrt{\phi
       _1}}+\frac{L p_4 \phi _2 p_2}{\sqrt{\phi _1}}+\frac{2 L p_1 \phi _3
       p_2}{\sqrt{\phi _1}}+\frac{F p_1 \phi _4 p_2}{\sqrt{\phi _1}}+\frac{H
       p_1 \phi _4 p_2}{\sqrt{\phi _1}}-\frac{{N} p_1 \phi _3 p_2}{2 \phi
       _1^{3/2}}\\\nonumber 
    &&+\frac{L \phi _3 \left(\phi _1'\right){}^2}{2 \phi _1^{3/2} \phi
       _2}+\frac{F \phi _4 \left(\phi _1'\right){}^2}{4 \phi _1^{3/2} \phi
       _2}+\frac{H \phi _4 \left(\phi _1'\right){}^2}{4 \phi _1^{3/2} \phi
       _2}+\frac{L \phi _4 \left(\phi _1'\right){}^2}{2 \sqrt{\phi _1} \phi
       _2^2}+\frac{J p_3 \left(\phi _1'\right){}^2}{4 \phi _1^{3/2} \phi _2}+2
       L \sqrt{\phi _1} p_1 p_4\\\nonumber 
    &&+\frac{3 {N} \phi _2 \phi _3}{8 \phi _1^{5/2}}+\frac{L \phi _4}{2
       \sqrt{\phi _1}}-p_1 \phi _4 {E}'-p_1 p_3 {I}'-p_1 \phi _3 {K}'-p_1 \phi
       _4 G'-p_3 M'-{I} p_3 p_1'-{K} \phi _3 p_1'\\\nonumber 
    &&-{E} \phi _4 p_1'-G \phi _4 p_1'-{K} \phi _4 p_2'-M p_3'-{I} p_1
       p_3'-{K} \phi _2 p_4'+{K} p_3 \phi _1'+\frac{\phi _3 {N}' \phi _1'}{2
       \phi _1^{3/2} \phi _2}+\frac{\phi _3 {N}' \phi _2'}{\sqrt{\phi _1} \phi
       _2^2}\\\nonumber 
    &&+\frac{2 \sqrt{\phi _1} \phi _4 F' \phi _2'}{\phi _2^2}+\frac{2
       \sqrt{\phi _1} \phi _4 H' \phi _2'}{\phi _2^2}+\frac{2 \sqrt{\phi _1}
       p_3 J' \phi _2'}{\phi _2^2}+\frac{2 \sqrt{\phi _1} \phi _3 L' \phi
       _2'}{\phi _2^2}+\frac{2 J \sqrt{\phi _1} p_3' \phi _2'}{\phi
       _2^2}\\\nonumber 
    &&+\frac{2 L \phi _3 \phi _1' \phi _2'}{\sqrt{\phi _1} \phi _2^2}+\frac{F
       \phi _4 \phi _1' \phi _2'}{\sqrt{\phi _1} \phi _2^2}+\frac{H \phi _4
       \phi _1' \phi _2'}{\sqrt{\phi _1} \phi _2^2}+\frac{J p_3 \phi _1' \phi
       _2'}{\sqrt{\phi _1} \phi _2^2}+\frac{{N} \phi _1' \phi _3'}{2 \phi
       _1^{3/2} \phi _2}+\frac{4 L \sqrt{\phi _1} \phi _2' \phi _3'}{\phi
       _2^2}+\frac{{N} \phi _2' \phi _3'}{\sqrt{\phi _1} \phi _2^2}\\\nonumber 
    &&-{E} p_1 \phi _4'-G p_1 \phi _4'+\frac{2 L \sqrt{\phi _1} \phi _1' \phi
       _4'}{\phi _2^2}+\frac{2 F \sqrt{\phi _1} \phi _2' \phi _4'}{\phi
       _2^2}+\frac{2 H \sqrt{\phi _1} \phi _2' \phi _4'}{\phi _2^2}+\frac{{N}
       \phi _3 \phi _1''}{2 \phi _1^{3/2} \phi _2}\\\nonumber 
    &&+\frac{2 L \sqrt{\phi _1} \phi _4 \phi _1''}{\phi _2^2}-\frac{L \phi _2
       \phi _3}{2 \phi _1^{3/2}}-\frac{J p_3 \phi _2}{4 \phi _1^{3/2}}-\frac{F
       \phi _2 \phi _4}{4 \phi _1^{3/2}}-\frac{H \phi _2 \phi _4}{4 \phi
       _1^{3/2}}-\frac{4 \sqrt{\phi _1} J' p_3'}{\phi _2}-\frac{4 \sqrt{\phi
       _1} L' \phi _3'}{\phi _2}\\\nonumber 
    &&-\frac{4 \sqrt{\phi _1} F' \phi _4'}{\phi _2}-\frac{4 \sqrt{\phi _1} H'
       \phi _4'}{\phi _2}-\frac{2 \sqrt{\phi _1} \phi _4 F''}{\phi _2}-\frac{2
       \sqrt{\phi _1} \phi _4 H''}{\phi _2}-\frac{2 p_3 \sqrt{\phi _1}
       J''}{\phi _2}-\frac{2 \sqrt{\phi _1} \phi _3 L''}{\phi _2}\\\nonumber 
    &&-\frac{4 L \sqrt{\phi _1} \phi _3''}{\phi _2}-\frac{2 F \sqrt{\phi _1}
       \phi _4''}{\phi _2}-\frac{2 H \sqrt{\phi _1} \phi _4''}{\phi
       _2}-\frac{2 J \sqrt{\phi _1} {p3}^{(0,2)}(t,x)}{\phi _2}-\frac{\phi _4
       F' \phi _1'}{\sqrt{\phi _1} \phi _2}-\frac{\phi _4 H' \phi
       _1'}{\sqrt{\phi _1} \phi _2}\\\nonumber 
    &&-\frac{p_3 J' \phi _1'}{\sqrt{\phi _1} \phi _2}-\frac{\phi _3 L' \phi
       _1'}{\sqrt{\phi _1} \phi _2}-\frac{J p_3' \phi _1'}{\sqrt{\phi _1} \phi
       _2}-\frac{{N}' \phi _3'}{\sqrt{\phi _1} \phi _2}-\frac{2 L \phi _1'
       \phi _3'}{\sqrt{\phi _1} \phi _2}-\frac{F \phi _1' \phi _4'}{\sqrt{\phi
       _1} \phi _2}-\frac{H \phi _1' \phi _4'}{\sqrt{\phi _1} \phi
       _2}-\frac{\phi _3 {N}''}{\sqrt{\phi _1} \phi _2}\\\nonumber  
    &&-\frac{J p_3 \phi _1''}{\sqrt{\phi _1} \phi _2}-\frac{2 L \phi _3 \phi
       _1''}{\sqrt{\phi _1} \phi _2}-\frac{F \phi _4 \phi _1''}{\sqrt{\phi _1}
       \phi _2}-\frac{H \phi _4 \phi _1''}{\sqrt{\phi _1} \phi _2}-\frac{{N}
       \phi _3''}{\sqrt{\phi _1} \phi _2}-\frac{3 {N} \phi _3 \left(\phi
       _1'\right){}^2}{8 \phi _1^{5/2} \phi _2}-\frac{{N} \phi _3 \phi _1'
       \phi _2'}{2 \phi _1^{3/2} \phi _2^2}\\\nonumber 
             &&-\frac{4 L \sqrt{\phi _1} \phi _4 \phi _1' \phi _2'}{\phi _2^3}
\end{eqnarray}
\begin{eqnarray}
    \dot{\phi}_4&=&-\phi _2 \phi _4 {E}'-{E} \phi _4 \phi _2'-{E} \phi _2 \phi
                    _4'-2 F p_1 \sqrt{\phi _1} \phi _4-\frac{F p_2 \phi _2
                    \phi _4}{\sqrt{\phi _1}}-\phi _2 \phi _4 G'-G \phi _4 \phi
                    _2'\\\nonumber 
    &&-G \phi _2 \phi _4'-2 H p_1 \sqrt{\phi _1} \phi _4-\frac{H p_2 \phi _2
       \phi _4}{\sqrt{\phi _1}}-p_3 \phi _2 {I}'-{I} p_3 \phi _2'-{I} \phi _2
       p_3'-2 J p_1 p_3 \sqrt{\phi _1}\\\nonumber 
    &&-\frac{J p_2 p_3 \phi _2}{\sqrt{\phi _1}}-\phi _2 \phi _3 {K}'-{K} \phi
       _3 \phi _2'-{K} \phi _2 \phi _3'-2 L p_1 \sqrt{\phi _1} \phi _3-\frac{L
       p_2 \phi _2 \phi _3}{\sqrt{\phi _1}}-\phi _4 M'-M \phi _4'\\\nonumber 
    &&-\frac{{N} p_4 \phi _2}{\sqrt{\phi _1}}-\frac{{N} p_2 \phi
       _4}{\sqrt{\phi _1}}
\end{eqnarray}
\begin{eqnarray}
    \dot{p}_4&=&\frac{J p_3 p_2^2}{2 \sqrt{\phi _1}}+\frac{L \phi _3 p_2^2}{2
                 \sqrt{\phi _1}}+\frac{F \phi _4 p_2^2}{\sqrt{\phi
                 _1}}+\frac{H \phi _4 p_2^2}{\sqrt{\phi _1}}-\frac{F \phi _2
                 \phi _3 p_2^2}{4 \phi _1^{3/2}}-\frac{H \phi _2 \phi _3
                 p_2^2}{4 \phi _1^{3/2}}+2 F \sqrt{\phi _1} p_3 p_2\\\nonumber 
    &&+2 H \sqrt{\phi _1} p_3 p_2+\frac{{N} p_4 p_2}{\sqrt{\phi _1}}+\frac{F
       p_4 \phi _2 p_2}{\sqrt{\phi _1}}+\frac{H p_4 \phi _2 p_2}{\sqrt{\phi
       _1}}+\frac{F p_1 \phi _3 p_2}{\sqrt{\phi _1}}+\frac{H p_1 \phi _3
       p_2}{\sqrt{\phi _1}}+\frac{F \phi _3 \left(\phi _1'\right){}^2}{4 \phi
       _1^{3/2} \phi _2}\\\nonumber 
    &&+\frac{H \phi _3 \left(\phi _1'\right){}^2}{4 \phi _1^{3/2} \phi
       _2}+\frac{{N} \phi _4 \left(\phi _1'\right){}^2}{\sqrt{\phi _1} \phi
       _2^3}+\frac{J p_3}{2 \sqrt{\phi _1}}+2 F \sqrt{\phi _1} p_1 p_4+2 H
       \sqrt{\phi _1} p_1 p_4+\frac{L \phi _3}{2 \sqrt{\phi _1}}+\frac{F \phi
       _4}{\sqrt{\phi _1}}\\\nonumber 
    &&+\frac{H \phi _4}{\sqrt{\phi _1}}-{I} p_3 p_2'-{K} \phi _3 p_2'-2 {E}
       \phi _4 p_2'-2 G \phi _4 p_2'-M p_4'-{E} \phi _2 p_4'-G \phi _2
       p_4'+{E} p_3 \phi _1'\\\nonumber 
    &&+G p_3 \phi _1'+\frac{4 \sqrt{\phi _1} \phi _4 {N}' \phi _1'}{\phi
       _2^3}+\frac{F \phi _3 \phi _1' \phi _2'}{\sqrt{\phi _1} \phi
       _2^2}+\frac{H \phi _3 \phi _1' \phi _2'}{\sqrt{\phi _1} \phi _2^2}+{E}
       p_1 \phi _3'+G p_1 \phi _3'+\frac{2 F \sqrt{\phi _1} \phi _2' \phi
       _3'}{\phi _2^2}\\\nonumber 
    &&+\frac{2 H \sqrt{\phi _1} \phi _2' \phi _3'}{\phi _2^2}+\frac{2 F
       \sqrt{\phi _1} \phi _4 \phi _1''}{\phi _2^2}+\frac{2 H \sqrt{\phi _1}
       \phi _4 \phi _1''}{\phi _2^2}-\frac{F \phi _2 \phi _3}{4 \phi
       _1^{3/2}}-\frac{H \phi _2 \phi _3}{4 \phi _1^{3/2}}-\frac{2 F
       \sqrt{\phi _1} \phi _3''}{\phi _2}\\\nonumber 
    &&-\frac{2 H \sqrt{\phi _1} \phi _3''}{\phi _2}-\frac{F \phi _1' \phi
       _3'}{\sqrt{\phi _1} \phi _2}-\frac{H \phi _1' \phi _3'}{\sqrt{\phi _1}
       \phi _2}-\frac{F \phi _3 \phi _1''}{\sqrt{\phi _1} \phi _2}-\frac{H
       \phi _3 \phi _1''}{\sqrt{\phi _1} \phi _2}-\frac{2 \sqrt{\phi _1} \phi
       _4 F' \phi _1'}{\phi _2^2}-\frac{2 \sqrt{\phi _1} \phi _4 H' \phi
       _1'}{\phi _2^2}\\\nonumber 
    &&-\frac{2 p_3 \sqrt{\phi _1} J' \phi _1'}{\phi _2^2}-\frac{2 \sqrt{\phi
       _1} \phi _3 L' \phi _1'}{\phi _2^2}-\frac{2 J \sqrt{\phi _1} p_3' \phi
       _1'}{\phi _2^2}-\frac{2 L \sqrt{\phi _1} \phi _1' \phi _3'}{\phi
       _2^2}-\frac{J p_3 \left(\phi _1'\right){}^2}{2 \sqrt{\phi _1} \phi
       _2^2}-\frac{L \phi _3 \left(\phi _1'\right){}^2}{2 \sqrt{\phi _1} \phi
       _2^2}\\\nonumber 
    &&-\frac{4 F \sqrt{\phi _1} \phi _4 \phi _1' \phi _2'}{\phi _2^3}-\frac{4
       H \sqrt{\phi _1} \phi _4 \phi _1' \phi _2'}{\phi _2^3}-\frac{{N} U_2
       \phi _2}{\sqrt{\phi _1} \phi _4^3}\, . 
\end{eqnarray}
  
\end{appendix}

%\bibliographystyle{../../preprint}
%\bibliography{../../Bib/QuantGra,../../Bib/Tunneling}

\end{document}